\documentclass[%
 reprint,
 amsmath,amssymb,
 aps,
]{revtex4-2}

\usepackage{graphicx}% Include figure files
\usepackage{dcolumn}% Align table columns on decimal point
\usepackage{bm}% bold math

\usepackage{float}
\usepackage{array}
\usepackage{amssymb}
\usepackage{hyperref}
\hypersetup{
    colorlinks=true,
    linkcolor=cyan,
    filecolor=red,      
    urlcolor=blue,
}

\usepackage[font=small,labelfont=bf,justification=RaggedRight]{caption}

\usepackage{subcaption}  
\begin{document}

\preprint{APS/123-QED}

\title{Search for continuous gravitational waves from ten H.E.S.S. sources\\ using a hidden Markov model}% Force line breaks with \\
%\thanks{A footnote to the article title}%

\author{Deeksha Beniwal*$^{1,3}$}
\author{Patrick Clearwater$^{2,3}$}
\author{Liam Dunn$^{2,3}$}
\author{Andrew Melatos$^{2,3}$}
\author{David Ottaway$^{1,3}$}
\altaffiliation{}
%\collaboration{MUSO Collaboration}%\noaffiliation
\email{deeksha.beniwal@adelaide.edu.au}
%Lines break automatically or can be forced with \\

\affiliation{1. Department of Physics, University of Adelaide, Adelaide, SA 5001, Australia}%
\affiliation{2. School of Physics, University of Melbourne, Parkville, Victoria 3010, Australia}%
\affiliation{3. Australian Research Council (ARC) Centre of Excellence for Gravitational Wave Discovery (OzGrav), Clayton, Victoria 3800, Australia}

\date{\today}% It is always \today, today,
             %  but any date may be explicitly specified

\begin{abstract}
Isolated neutron stars are prime targets for continuous$-$wave (CW) searches by ground$-$based gravitational$-$wave interferometers. Results are presented from a CW search targeting ten pulsars. The search uses a semi$-$coherent algorithm, which combines the maximum$-$likelihood $\mathcal{F}-$statistic with a hidden Markov model (HMM) to efficiently detect and track quasi$-$monochromatic signals which wander randomly in frequency. The targets, which are associated with TeV sources detected by the High Energy Stereoscopic System (H.E.S.S.), are chosen to test for gravitational radiation from young, energetic pulsars with strong $\mathrm{\gamma}-$ray emission, and take maximum advantage of the frequency tracking capabilities of HMM compared to other CW search algorithms. The search uses data from the second observing run of the Advanced Laser Interferometer Gravitational$-$Wave Observatory (aLIGO). It scans 1$-$Hz sub$-$bands around $f_*$, 4$f_*$/3, and 2$f_*$, where $f_*$ denotes the star's rotation frequency, in order to accommodate a physically plausible frequency mismatch between the electromagnetic and gravitational$-$wave emission. The 24 sub$-$bands searched in this study return 5,256 candidates above the Gaussian threshold with a false alarm probability of 1$\%$ per sub$-$band per target. Only 12 candidates survive the three data quality vetoes which are applied to separate non$-$Gaussian artifacts from true astrophysical signals. CW searches using the data from subsequent observing runs will clarify the status of the remaining candidates.
\end{abstract}

%\keywords{Suggested keywords}%Use showkeys class option if keyword
                              %display desired
\maketitle

%\tableofcontents

\section{\label{sec:Introduction}Introduction\protect\\}

The catalogue of gravitational wave (GW) transients \cite{Abbott_2019} observed during the first two observing runs of the Advanced Laser Interferometer Gravitational$-$Wave Observatory (aLIGO) and advanced Virgo detectors is a testament to the recent advances in the field of GW astronomy \cite{Aasi_2015,Acernese_2015}. The signals observed to date primarily come from the coalescence of binary black holes \cite{BBH_2012,BBH_2016,BBH_2017} and binary neutron stars \cite{NS_2017,NS_2017-2}. An isolated neutron star whose mass is distributed asymmetrically about its rotation axis is also a promising source of persistent, quasi$-$monochromatic continuous gravitational waves (CW) \cite{Riles_2013,Riles_2017}. Emission from such sources is predicted to be at multiples of the star's spin frequency $f_*$ \cite{Mastrano_2013}. For example, a mass quadrupole caused by a thermoelastic or magnetic mountain emits a CW signal at $f_*$ and $2f_*$ \cite{Ushomirsky_2000,Melatos_2005}, r$-$modes emit at roughly $4f_*$/3 \cite{Owen_1998} and a current quadrupole produced by non$-$axisymmetric circulation of neutron super$-$fluid pinned to the crust emits a signal at $f_*$ \cite{Jones_2001}. \\

Extracting a CW signal from noisy detector data often involves matched filtering against a bank of templates. Managing the size of this template bank and thus the computational cost of the search is an important part of designing a practical search \cite{Wette_2018,Jones_2015}. For a coherent search over a total observation time $T_{\rm{obs}}$, the required number of templates grow as $T_{\rm{obs}}^n$, whereas the sensitivity increases as $T_{\rm{obs}}^{1/2}$ \cite{Wette_2012}. The value of the exponent $n$ is determined by the number of unknown parameters. For example, one has $n\gtrsim5$ for an all$-$sky search, which typically involves searching over sky position, frequency, and frequency derivatives \cite{Wette_2012}.\\

In a search directed at a particular object, such as the one described here, $n$ typically has a smaller value as the sky position, and in many cases, some frequency derivatives are known. Nevertheless the computational challenge is acute, and it is common to use a semi$-$coherent approach, which involves dividing the search into $N_{s}$ short blocks of length $T_{\rm{s}}$,  coherently searching for signals in each block, and then incoherently combining these blocks to cover the entire search space. The sensitivity of a semi$-$coherent search than scales as $\sim N_{\rm{s}}^{1/4} T_{\rm{s}}^{1/2}$ \cite{Wette_2012}. Although this makes it less sensitive than coherent matched filtering, it allows more data to be used, which in turn improves the sensitivity of the search while allowing it to remain computationally tractable \cite{Wette_2012}. \\

Phase templates for CW signals from isolated pulsars are often expressed as Taylor expansions about some reference time $t_0$ of form (e.g., see Eq. (18) in Ref.~\cite{Jaranowski_1998})
\begin{equation}
\phi(t) = \phi(t_0) + 2\pi\sum_{k=0}^{k_{\rm max}} {f_0^{k}}(t_0) \frac{(t-t_0)^{k+1}}{(k+1)!},  
\end{equation}
where $\phi(t_0)$ is the phase at the reference time, ${f_0^{k}}(t_0)$ is the $k^{\rm{th}}$ time derivative of $f_0$ evaluated at $t_0$ and $k_{\rm max}$ is the highest$-$order time derivative included in the search. One therefore has $n = k_{\rm max}(k_{\rm max}-1)/2$, which grows rapidly with $k_{\rm max}$ \cite{Prix_2012}.\\

The challenge increases dramatically when the phase of the signal wanders stochastically on timescales of days to weeks. Radio and X$-$ray timing of isolated pulsars show ``spin$-$wandering" to be a widespread phenomenon, often going by the name timing noise \cite{Hobbs_2010, Ashton_2015,Suvorova_2017}. It is attributed to various mechanisms including changes in the star's magnetosphere \cite{Lyne_2010}, spin micro$-$jumps \cite{Janssen_2006}, super$-$fluid dynamics in the stellar interior \cite{Price_2012,Melatos_2014} and fluctuations in the spin$-$down torque \cite{Cheng_1987, Urama_2016}. It is especially pronounced in young pulsars with characteristic ages $\lesssim 10 $kyr \cite{Arzoumanian_1994,Hobbs_2010}, among which include three targets in this paper. While spin$-$wandering could, in principal, be modelled with higher$-$order derivatives, the number of derivatives required would make $k_{\rm{max}}$ impractically large and thus make the search computationally infeasible.\\

In this paper, we use a hidden Markov model (HMM) solved by the classic Viterbi algorithm to search for CW signals from isolated neutron stars. This scheme combines an existing, efficient and thoroughly tested maximum likelihood detection statistic called the $\mathcal{F}-$statistic \cite{Jaranowski_1998} with a HMM. Loosely speaking, the $\mathcal{F}-$statistic coherently searches for a constant$-$frequency signal within a block of data, and the HMM tracks the stochastic wandering of the signal frequency from one block to the next \cite{Quinn_2001}. The method shares some common elements with other semi$-$coherent methods such as StackSlide \cite{Prix_2012,Dreissigacker_2018} and facilitates the use of data from the entire second observing (O2) run while remaining computationally manageable. Finally, it can accommodate the situation where the gravitational$-$wave$-$emitting quadrupole is not phase locked to the stellar crust, even when the ephemeris is accurately measured via electromagnetic observation~\cite{Abbott_2008}. \\

To sharpen the astrophysical focus of the search, we select 10 pulsars whose pulsations and/or pulsar wind nebulae have been observed by very high$-$energy (VHE) $\gamma-$ray surveys such as the High Energy Stereoscopic System (H.E.S.S.) and the Very Energetic Radiation Imaging Telescope Array System (VERITAS). Previous CW searches have searched for signals from eight of these targets \cite{Limits_2005,ULs_2007,Abbott_2008,Searches_SR5_2010, Beating_SD_Vela_2011,GW_pulsars_2014,Narrowband_2015,Pulsar_survey1-2017,Pulsar_survey2-2017,Pulsar_survey1-2019,Pulsar_survey2-2019}. Leptonic models are often used to completely describe the TeV $\gamma-$rays originating from these sources \cite{Wei_2009,Holder_2012,Abdalla_2018_2}. The $\gamma-$ray emission is unlikely to be tied directly to a gravitational$-$wave$-$emitting mass or current quadrupole, as it comes from low$-$density plasma accelerated in magnetosphere vacuum gaps \cite{Grenier_2015,Bai_2010}. However, it is entirely possible that $\gamma-$ray emission is a reliable co$-$signature of gravitational radiation, because it is tied to young pulsars, whose mass and current quadrupoles at birth have had relatively little time to relax via slow dissipative processes \cite{Chugunov_2010,Knispel_2008,Wette_2010}. Moreover such objects are surrounded by active magnetospheric and particle production processes, which may react back on the star. For example, the heating of polar caps by downward$-$moving particles bombarding the stellar surface can induce non$-$axisymmetric variations in the stellar mass density and thus yield a gravitational$-$wave$-$emitting mass quadrupole \cite{Harding_1998,Harding_2002,Levinson_2005,Muslimov_2003}.\\

In anticipation of a future detection, it is crucial to note that the TeV spectrum provides a comprehensive picture of the electron population around the pulsar \cite{Abdalla_2018}. We can therefore obtain better estimates for the age and magnetic field strength around the progenitors, while multi$-$wavelength observations can yield accurate estimates of the distance to the source. Therefore, the sources in this paper offer unusually rich opportunities to do high$-$impact astrophysics in the event of a detection. Likewise, in the absence of a detection, the accurate age, distance, and magnetisation measurements can be converted into accurate GW spin$-$down upper limits \cite{Riles_2013,Pitkin_2011,Pulsar_survey2-2017} to be interpreted in the light of the upper limits from the GW search.\\

This paper is organised as follows. We outline the targets selected for this study and briefly review their GeV and TeV properties in Section \ref{sec:targets}. In Section \ref{sec:previous_searches}, we review the results from previous CW searches targeting the selected pulsars. Section \ref{sec:search_algorithm} explains the search algorithm which includes a review of the signal model, the detection statistic known as the $\mathcal{F}-$statistic and the HMM formulation. In Section \ref{sec:parameters}, we define the search parameter space for each pulsar as well as the detection threshold and data quality vetoes. The results are presented in Section \ref{sec:Results}. We conclude with a brief statement on the astrophysical implications in Section \ref{sec:Conclusion}.

\section{\label{sec:targets}Targets\protect\\} 
The targets for this search are selected from the publicly available TeV catalogue \footnote{\url{http://tevcat2.uchicago.edu/}}. We restrict the analysis to TeV$-$emitting pulsars, sources firmly associated with isolated neutron stars and located within the galactic disk (i.e., up to a distance of 7kpc). With these restrictions in place, the catalogue includes two $\gamma-$ray$-$emitting pulsars and eight H.E.S.S. sources firmly identified as pulsar wind nebulae (PWNe) powered by known pulsars. Table \ref{tab:search_targets} gives the astrophysical parameters for the ten targets. Note that a small discrepancy exists between the position of the H.E.S.S. sources and their associated pulsar due to the pulsar being offset from the centre of the TeV emission region, possibly due to its proper motion, asymmetric crushing of the PWN by the surrounding supernova remnant or by asymmetric pulsar outflow \cite{Abdalla_2018_2,Vigelius_2007}. Below, we briefly discuss the GeV and TeV properties of the H.E.S.S. targets.\\

\begin{table*}[ht]
    \centering
    \setlength{\tabcolsep}{7pt}
    \renewcommand{\arraystretch}{1.2}
    \caption{List of targets. Names of the H.E.S.S. sources and the associated pulsars are listed in columns one and two. The position (RA, DEC), age, distance and spin$-$frequency of each pulsar are taken from \texttt{v1.63} of Australian Telescope National Facility (ATNF) pulsar catalogue \cite{Manchester_2005}. These are reported in columns three through to seven. The minimum and maximum distance estimates for each pulsar are based on the dispersion measurements \cite{Cordes_2002,Lyne_2006}. Distance to PSR J1849-0001 is not reported in the ATNF catalogue so we place it at the same distance as its counterpart IGR J18490$-$0000/HESS J1849$-$000 \cite{Gotthelf_2011}. Quoted error (in parentheses) represents the uncertainty in the last significant figure.} 
    \label{tab:search_targets}
    \begin{tabular}{c|c|l|l|c|c|c}
    \hline
     HESS name & Pulsar name &  RA & DEC &  Age   & Distance & Spin frequency \\
     $[$J2000] & [J2000]  &  &    & [kyrs] &  Min/Max [kpc] & [Hz]  \\\hline \hline
       -   & J0534$+$2200 & $05\mathrm{h}34\mathrm{m}31.973\mathrm{s}$  & $+22^\circ00^{'}52.06^{''}$ & 1.26 & 1.5/2.5    & 29.946923(1)  \\  
      -   & J0835$-$4510 & $08\mathrm{h}35\mathrm{m}20.61149\mathrm{s}$ & $-45^\circ10^{'}34.8751^{''}$ & 11.3 & 0.26/0.30 & 11.1946499395(5)\\ 
       J1018$-$589 B & J1016$-$5857 & $10\mathrm{h}16\mathrm{m}21.16\mathrm{s}$ & $-58^\circ57^{'}12.1^{''}$ & 21    & 3.16/8.01 & 9.31216108951(13)  \\  %Distance (3.16-8.01)
       J1356$-$645   & J1357$-$6429 & $13\mathrm{h}57\mathrm{m}02.43\mathrm{s}$ & $-64^\circ29^{'}30.2^{''}$ &  7.31  & 2.83/3.88  & 6.0201677725(5)  \\
       J1420$-$607 & J1420$-$6048 & $14\mathrm{h}20\mathrm{m}08.237\mathrm{s}$ & $-60^\circ48^{'}16.43^{''}$ & 13 & 5.63   & 14.667084337(5) \\
       J1514$-$591   & J1513$-$5908  & $15\mathrm{h}13\mathrm{m}55.811\mathrm{s}$ & $-59^\circ08^{'}09.60^{''}$ &  1.57     & 3.6/5.7     & 6.59709182778(19)   \\ 
       J1718$-$385   & J1718$-$3825  & $17\mathrm{h}18\mathrm{m}13.565\mathrm{s}$ & $-38^\circ25^{'}18.06^{''}$ &  89.5 & 3.49/3.60 & 13.39227352093(7) \\
      J1825$-$137   & J1826$-$1334  & $18\mathrm{h}26\mathrm{m}13.175\mathrm{s}$ & $-13^\circ34^{'}46.8^{''}$ & 21.4 & 3.61/3.93   & 9.85349875132(2)  \\
       J1831$-$098   & J1831$-$0952  & $18\mathrm{h}31\mathrm{m}34.304\mathrm{s}$ & $-09^\circ52^{'}01.7^{''}$ & 128  & 3.68/4.05 & 14.8661645038(3)  \\ 
      J1849$-$000   & J1849$-$0001  & $18\mathrm{h}49\mathrm{m}01.61\mathrm{s}$ & $-00^\circ01^{'}17.6^{''}$ & 43.1  & 7    & 25.961252072(9)\\ \hline
    \end{tabular}
\end{table*}

\textbf{PSR J0534$+$2200 (Crab pulsar):} Fermi Large Area Telescope (Fermi$-$LAT) observations of the $\gamma-$rays emitted by the Crab pulsar reveal a light curve with two main peaks, labelled P1 and P2, both of which are extremely stable in position across the $\gamma-$ray energy band (See Fig.~\textbf{1} of Ref.~\cite{Abdo_2009}). The first $\gamma-$ray pulse leads the radio pulses by 281$\mu$s \cite{Abdo_2009}. This combined with the measurement of photons up to $\sim$20GeV constraints the production site of non$-$thermal emission in the pulsar magnetosphere \cite{Abdo_2009}. In addition, observations taken by VERITAS between 2007$-$2011 reveal $\gamma-$ray pulses at energies $>$100GeV \cite{Aliu_2011} which are narrower than those recorded by Fermi$-$LAT at 100MeV. Combined analysis of Fermi$-$LAT and VERITAS data yields important information about the shape and location of the particle acceleration site and TeV emission mechanism \cite{Aliu_2011}.\\

\textbf{PSR J0835$-$4510 (Vela pulsar):} Fermi$-$LAT observations of this pulsar reveal two narrow and widely separated pulses (labelled P1 and P2) typically observed in high energy pulsars \cite{Venter_2018}. A complex bridge emission (labelled P3) is also observed between the two main peaks (See Fig.~\textbf{2} of Ref.~\cite{Abdo_2010}). H.E.S.S. II$-$CT5 observations from 2013 to 2015 confirm the characteristics previously known from Fermi$-$LAT studies and reveal a change in the P2 pulse morphology as well as the onset of a new component near P2 at higher energies~\cite{Abdalla_2018_3}. Fermi$-$LAT and H.E.S.S. measurements of the phase$-$averaged spectrum also contains important information about the location and origin of TeV emission~\cite{Abdo_2010}. \\

\textbf{HESS J1018$-$589$-$B:} This is one of two regions of TeV $\gamma-$ray emission around the supernova remnant G284.3$-$1.8. The TeV emission has a radius of $0.15\pm0.03^\circ$ and is likely to be associated with a pulsar wind nebula powered by PSR J1016$-$5857 \cite{Abramowski_2015}.\\

\textbf{HESS J1356-645:} The morphology of this source in radio and X$-$rays, combined with its spectral energy distribution in TeV, as measured by Fermi$-$LAT in 2013, leads to its classification as a clearly identified PWN \cite{Acero_2013}. The emission is most likely associated with the young and energetic pulsar PSR J1357$-$6429, which is located at a distance of $\sim$5pc from the center of the TeV emission \cite{Abramowski_2011}. \\

\textbf{HESS J1420$-$607 (Kookaburra PWN):} This is one of the two TeV sources in a complex of compact and extended radio and X$-$ray sources, called Kookaburra. It is centered north of a young and energetic pulsar PSR J1420$-$6048, believed to be one of the sources powering the complex \cite{Reyes_2012}. Pulsed $\gamma-$ray and extended X$-$ray emissions from the source provide evidence for its PWN nature \cite{Roberts_2001}.\\

\textbf{HESS J1514$-$591 (MSH 15$-$52):}
This composite supernova remnant contains a bright X$-$ray PWN powered by PSR J1513$-$5908 which is surrounded by a shell that dominates the emission in the radio band. Correlation between the H.E.S.S. and X$-$ray data and pulsar jets in the X$-$ray domain lead to its classification as a clearly identified PWN \cite{Aharonian_2005}. The TeV morphology also suggests asymmetric expansion of the supernova remnant, displacing the PWN in the direction away from the high photon density region \cite{Tsirou_2017}. \\

\textbf{HESS J1718$-$385:} Observation of this H.E.S.S. source by the space based X$-$ray Multi$-$Mirror Mission (XMM$-$Newton) reveals a hard X$-$ray source at the position of PSR J1718$-$3825. Diffuse emission in the vicinity of the pulsar suggests the existence of a synchrotron nebula and strengthens its association with the pulsar. However, the relationship between the X$-$ray and $\gamma-$ray emission is not straight forward. The overall asymmetry of the nebula with respect to the pulsar is consistent with the idea of supernova remnant expanding into a non$-$uniform environment \cite{Hinton_2007}. \\

\textbf{HESS J1825-137:} TeV emissions from the PWN extend out to 1.5$^\circ$ from the pulsar's location, which makes it one the most luminous and potentially the largest of all firmly identified PWNs in the Milky Way \cite{Mitchell_2017}. The XMM$-$Newton's observation of the region, including the pulsar, reveals an elongated non$-$thermal$-$emitting nebula with the pulsar located on the northern border and a tail towards the peak of the H.E.S.S. source. This is also consistent with TeV $\gamma-$rays originating from the inverse Compton mechanism within the PWN powered by PSR J1826$-$1256 \cite{Duvidovich_2019}.\\

\textbf{HESS J1831$-$098:} Detected by the H.E.S.S. survey in 2011, this source is proposed to be a member of a new class of TeV emitters known as extended TeV halos \cite{Linden_2017}. A 67ms pulsar, PSR J1831$-$0952, lies at a small angular offset of $\sim$0.05$^\circ$ from the best$-$fit position of the H.E.S.S. source. A conversion efficiency from spin$-$down power to 1$-$20 TeV $\gamma-$rays of $\epsilon = 1\%$ would be required to power this H.E.S.S. source, a value typically inferred for other TeV/PWN systems. This combined with the lack of other plausible counterparts in multi$-$wavelength searches provides evidence for a PWN as the most favorable scenario \cite{Sheidaei_2011}.\\

\textbf{HESS J1849-000:} This extended TeV source is coincident with a hard X$-$ray source IGR J18490$-$0000 \cite{Abeysekara_2017}. Follow$-$up observations of the source by XMM$-$Newton and Rossi X$-$ray Timing Explorer confirm HESS J1849$-$000 as a PWN powered by a young and energetic pulsar, PSR J1849$-$0001 \cite{Gotthelf_2011}. This study also provides clear evidence of a PWN that is asymmetric in X$-$rays~\cite{Calas_2018}.

\section{\label{sec:previous_searches}Previous pulsar searches\protect\\}

Numerous CW searches have previously looked for signals from eight of the targets chosen for this study. In particular, PSR J0534$+$2200 (Crab) has been extensively studied using data from the LIGO science runs (S) 2$-$5 \cite{Limits_2005,ULs_2007,Abbott_2008,Searches_SR5_2010} and Virgo science run (VSR) 4 \cite{Abbott_2008}. Most of these searches exploited a \textit{Bayesian} method, except Refs.~\cite{Searches_SR5_2010} and \cite{Abbott_2008} which used \textit{Markov Chain Monte Carlo (MCMC)} and \textit{5n$-$vector} techniques, respectively. Although none of these searches found strong evidence of a CW signal from the Crab pulsar, they improved the upper limits (ULs) on the GW strain amplitude from $4.1\times10^{-23}$ to $2.4\times10^{-25}$ over a period of five years.\\

Data from VSR2 \cite{Beating_SD_Vela_2011} and VSR4 \cite{Narrowband_2015} was used to search for a CW signal from PSR J0835$-$4510 (Vela). These searches were carried out using three largely independent methods; \textit{Bayesian} and $\mathcal{F}/\mathcal{G}$\textit{-statistic} methods in Ref.~\cite{Beating_SD_Vela_2011} and a \textit{5n-vector} method in Ref.~\cite{Narrowband_2015}. No CW signal was detected and ULs were set on the GW strain amplitude. In 2014, Aasi et al. combined observations from S6, VSR2 and VSR4 to look for CW signals from 179 pulsars \cite{GW_pulsars_2014}, four of which are included in this HMM search. For seven high$-$value targets, the search was performed using three largely independent methods; \textit{Bayesian}, $\mathcal{F}/\mathcal{G}$\textit{-statistic}, and \textit{5n-vector} methods. Only the \textit{Bayesian} method was applied to the remaining 172 pulsars. The search found no credible evidence for GW emission from any pulsar and produced ULs on the emission amplitude. \\

In early 2017, Abbott et al.~\cite{Pulsar_survey2-2017} repeated a similar analysis for 200 pulsars, five of which are considered here, using the data from aLIGO's first observing run (O1). For 11 high$-$value targets, the search was performed using the \textit{Bayesian}, $\mathcal{F}/\mathcal{G}$\textit{-statistic}, and \textit{5n-vector} methods. Only the \textit{Bayesian} method was used for the remaining 189 pulsars. No detections were reported, and ULs were set on the GW strain amplitude and ellipticity of each target. \\

Following the second observing run (O2), Abbott et al.~\cite{Pulsar_survey1-2019} performed a CW search for 33 pulsars using the \textit{5n-vector} pipeline. These included three of the targets chosen for this study. The narrow$-$band search found no evidence for a GW signal from any of the targets, but significantly improved on ULs for pulsars with rotation frequency $>$30Hz. Later that year, Abbott et al. \cite{Pulsar_survey2-2019} combined data from O1 and O2 to search for CW signals from 222 pulsars at $f_0$ and $2f_0$. As in Ref.~\cite{Pulsar_survey1-2017}, they used \textit{Bayesian}, $\mathcal{F}/\mathcal{G}$\textit{-statistic} and \textit{5n-vector} methods for 34 high$-$value targets and only the \textit{Bayesian} method for the remaining 188 pulsars. Although the search found no strong evidence for GW emission from any pulsar, it did allow for updated ULs on the GW amplitude, mass quadrupole moment and ellipticity for 167 pulsars, and first such limits for 55 others.\\

We summarise the ULs set by the aforementioned studies for targets relevant to this search in Table \ref{tab:Upper_limits}. Although these targets have been studied in the past, the HMM enables the examination of a much more inclusive set of spin$-$wandering templates than the Taylor$-$expanded phase model assumed in the above references. Young pulsars with TeV $\gamma-$ray emission, such as the ones considered here, exhibit rapid spin$-$wandering. Hence they derive maximum benefit from being searched again with a HMM.

\begin{table*}[ht]
    \centering
    \setlength{\tabcolsep}{7pt}
    \renewcommand{\arraystretch}{1.2}
    \caption{\small{Upper limits on the GW amplitude (derived from previous studies) for eight out of the 10 targets selected for this search. Names of the pulsars (column~1), relevant observing runs (column~2), search method (column~3), 95$\%$ upper limits on the GW amplitude (column~4) and the relevant references (column~5) are given in the table below. O1 $\&$ O2 = First and second observing runs, SR2$\rightarrow$6 = LIGO Science Runs 2$\rightarrow$6, VSR(2,4) = Virgo Science Runs 2,4, MCMC = Markov Chain Monte Carlo. The Crab pulsar (PSR J0534$+$2200) underwent a small glitch during the O2 run. Thus upper limits on GW amplitude were calculated separately for the period before (BG) and after (AG) the glitch.}} 
    \label{tab:Upper_limits}
    \begin{tabular}{c|c|c|c|c}
    \hline
     Name $[$J2000$]$ & Observing run & Search method(s) &  $h_0^{95\%}$ [$\times 10^{-25}$]   & Ref \\ \hline \hline
    J0534$+$2200  & O1$+$O2 & Bayesian, $\mathcal{F}/\mathcal{G}-$statistic, 5n$-$vector & 0.19(0.15), 0.22(0.13), 0.29(0.29) & \cite{Pulsar_survey2-2019} \\ 
                & O2 & 5n$-$vector & 1.64[BG], 1.31 [AG]  & \cite{Pulsar_survey1-2019} \\
                & O1 & Bayesian, $\mathcal{F}-$statistic, 5n$-$vector & 0.67(0.61), 0.42 (0.24), 0.52 (0.50) & \cite{Pulsar_survey2-2017} \\ 
%               & O1 & $5n-$vector & 1.08 & \cite{Pulsar_survey1-2017} \\
               & SR6, VSR2 $\&$ VSR4 & Bayesian, $\mathcal{F}/\mathcal{G}-$statistic, 5n$-$vector & 1.6(1.4), 2.3 (1.8), 1.8(1.6) & \cite{GW_pulsars_2014}\\
               & VSR4 & 5n$-$vector & 7.0 & \cite{Narrowband_2015} \\
               & S5 & MCMC & 2.4 & \cite{Searches_SR5_2010} \\
               & S5 [1$^{\rm{st}}$ 9 months] & Bayesian/$\mathcal{F}-$statistic & 4.9/3.9 & \cite{Abbott_2008} \\
               & SR3$+$SR4 & Bayesian & 30.9  & \cite{ULs_2007} \\
               & SR2 & Bayesian & 410  & \cite{Limits_2005} \\
               \hline
    J0835-4510  & O1$+$O2 & Bayesian, $\mathcal{F}/\mathcal{G}-$statistic, 5n$-$vector & 1.4(1.2), 2.6(2.0), 2.3(2.4)  & \cite{Pulsar_survey2-2019}\\
                & O2 & 5n$-$vector & 8.82 & \cite{Pulsar_survey1-2019} \\
                & O1 & Bayesian, F$-$statistic, 5n$-$vector & 3.2(2.8), 3.8 (3.3), 2.9 (2.9) &  \cite{Pulsar_survey2-2017} \\
%                & O1 & $5n-$vector & 9.28 & \cite{Pulsar_survey1-2017}  \\
                & SR6, VSR2 $\&$ VSR4 & Bayesian, $\mathcal{F}/\mathcal{G}-$statistic, 5n$-$vector & 11(10), 4.2(9.0), 11(11) & \cite{GW_pulsars_2014}\\ 
                & VSR4 & 5n$-$vector & 32 & \cite{Narrowband_2015} \\
                & VSR2 & Bayesian, $\mathcal{F}/\mathcal{G}-$statistic,5n$-$vector & 21(1), 19(1), 22(1) & \cite{Beating_SD_Vela_2011} \\ \hline
    J1016-5857 & O1 & Bayesian & 14 & \cite{Pulsar_survey2-2017} \\ \hline
    J1420-6048 & O1$+$O2 & Bayesian, 5n$-$vector & 0.41, 0.76 & \cite{Pulsar_survey2-2019}\\ 
               & O2 & 5n$-$vector & 2.75 & \cite{Pulsar_survey1-2019} \\ \hline
    J1718-3825 & O1$+$O2 & Bayesian, 5n$-$vector & 0.87, 0.65  & \cite{Pulsar_survey2-2019} \\ 
               & O1 & Bayesian & 1.7 & \cite{Pulsar_survey2-2017}\\ 
               & SR6, VSR2 $\&$ VSR4 & Bayesian & 0.80 & \cite{GW_pulsars_2014} \\ \hline
    J1826-1334 & O1	& Bayesian & 12 & \cite{Pulsar_survey2-2017} \\ \hline
    J1831-0952 & O1$+$O2 & Bayesian, 5n$-$vector & 0.69, 0.43 & \cite{Pulsar_survey2-2019} \\ 
                & SR6, VSR2 $\&$ VSR4 & Bayesian & 0.65 & \cite{GW_pulsars_2014} \\ \hline
    J1849-0001 & O1$+$O2 & Bayesian, $\mathcal{F}/\mathcal{G}-$statistic, 5n$-$vector & 0.19, 0.28, 0.20 & \cite{Pulsar_survey2-2019} \\ \hline
    \end{tabular}
\end{table*}

\section{\label{sec:search_algorithm}Search Algorithm\protect} 
In this paper, we subdivide the total observation time $T_{\rm{obs}}$ into blocks of duration $T_{\rm{drift}}$, during which the signal frequency is assumed to remain constant to a good approximation. Within each block, we search for a signal predicted to be emitted by a biaxial rotor by coherently combining the detector data. This data is available in the form of short Fourier transforms (SFTs), each with length $T_{\rm{SFT}}$ = 30min. In Sections \ref{subsec:Sig_model} and \ref{subsec:F_statistic}, we review the signal model and the maximum likelihood detection statistic called the $\mathcal{F}-$statistic, as defined in Ref.~\cite{Jaranowski_1998}. Finally, we incoherently combine the output of each block and recover the optimal frequency path through the data using a HMM. We briefly review the HMM framework in Section \ref{subsec:HMM_framework}. 

\subsection{ \label{subsec:Sig_model}Signal model}
In the absence of noise, we model the GW strain from a pulsar as \begin{equation}
    h(t) = \sum_{\mu=1}^4 \mathcal{A}_\mu h_\mu(t),  \label{sig_model}
\end{equation}
where $\mathcal{A}_{\mu}$ denote the amplitudes associated with four linearly independent components, as defined in Eqs.~(28)$-$(31) of Ref.~\cite{Jaranowski_1998}. They depend on parameters such as the characteristic strain ($h_0$), sky location (RA, DEC), source inclination with respect to the line of sight ($\iota$), initial phase ($\psi_0$) and wave polarisation. Here, $h_\mu(t)$ represents four linearly independent components
\begin{align}
    h_{1}(t) &= a(t)\cos\Phi(t),\\ \label{h_start}
    h_{2}(t) &= b(t)\cos\Phi(t),  \\
    h_{3}(t) &= a(t)\sin\Phi(t),\\
    h_{4}(t) &= b(t)\sin\Phi(t), \label{h_end}
\end{align}
where $a(t)$ and $b(t)$ are the antenna$-$pattern functions as defined in Eqs.~(12) and (13) of Ref.~\cite{Jaranowski_1998}. The signal phase, $\Phi(t)$, as measured by a detector on Earth, includes contributions from three terms \cite{ScoX1_2017}, 
\begin{equation}
    \Phi(t) = 2\pi f_0[t  + \Phi_m(t;\alpha,\delta)] +  \Phi_s(t;f_0^{(k)},\alpha,\delta),  
\end{equation}
where $\Phi_m$ represents the time shift produced by the diurnal and annual motion of the detector relative to the solar system barycentre. The term $\Phi_s$ represents the phase shift which results from the intrinsic evolution of the source in its rest frame through the frequency derivatives [$f_0^{(k)} = d^{(k)}f_0/dt^{(k)}$  with $k \geq 1$].\\

The frequency evolution of a CW signal has a secular and a stochastic term. The (slow) secular spin$-$down of the pulsar can be easily modelled by specifying the frequency derivatives $f_*^{(k)}$ measured via electromagnetic observations. However, the stochastic term cannot be easily measured, and it is computationally infeasible to conduct an exhaustive search over physically$-$plausible stochastic paths. Instead, we use a HMM, as outlined in Section \ref{subsec:HMM_framework}, to handle the stochastic evolution of the signal phase and the Viterbi algorithm to efficiently backtrack the optimal pathway in frequency. In other words, in this paper, we set $\dot{f_0}$ according to the spin$-$down rate ($\dot{f}_\ast$) reported in a pulsar catalogue and set higher derivatives to zero within each coherent block. We then let the HMM (instead of the Taylor expansion) do the work of tracking $f_0(t)$, as it evolves from one block to the next. 

\subsection{\label{subsec:F_statistic}$\mathcal{F}$-statistic} 
Following Ref.~\cite{Jaranowski_1998}, we use the maximum likelihood $\mathcal{F}-$statistic to extract a CW signal buried in detector noise. The $\mathcal{F}-$statistic involves maximising the likelihood function ($\Lambda$) with respect to the amplitudes $\mathcal{A}_\mu (1 \leq\mu\leq 4)$. The values of the parameters which maximise $\Lambda$ are known as the maximum likelihood estimators. Below, we briefly review how the $\mathcal{F}-$statistic is defined.  \\

The time$-$dependent output of a single detector can be assumed to take the following form \cite{Jaranowski_1998}: 
\begin{align}
   x(t) &= h(t) + n(t), \label{Detector_data}
\end{align}
where $n(t)$ represents stationary, additive noise and $h(t)$ is the GW signal as defined in Eq. \ref{sig_model}. We can then define a normalised log$-$likelihood function of form \cite{Jaranowski_1998}
\begin{equation}
	\ln\Lambda = (x||h) - \frac{1}{2}(h||h),	\label{log_liklihood}
\end{equation}
where the inner product is given by 
\begin{equation}
	(x||y) = \frac{2}{T_{\rm{obs}}} \int_0^{T_{\rm{obs}}} dt\ x(t)y(t). \label{eq:inner_product}	
\end{equation}
% {\color{red} [Differentiate the log likelihood function with respect to the amplitude parameter and find roots.]}
By maximising $\ln{\Lambda}$ with respect to the four amplitudes ($\mathcal{A}_{\mu}$) over the observation period ($T_{\rm{obs}}$), we find the optimal set of signal parameters. These parameters, also known as the maximum likelihood (ML) estimators, are then used to define a frequency-domain detection statistic called the $\mathcal{F}-$statistic of form %\equiv \ln{\Lambda}_{\mathrm{ML}}
\begin{multline}
	\mathcal{F} \equiv \ln{\Lambda}_{\mathrm{ML}} =  D^{-1}[B (x||h_{1})^2 + A(x||h_{2})^2 \\
        \quad \quad \quad - 2C(x||h_{1}) (x||h_{2}) + B(x||h_{3})^2 \\
+ A(x||h_{4})^2 - 2C(x||h_{3})(x||h_{4})], \label{Fstat_eqn}
\end{multline}
with $A = (a||a)$, $B = (b||b)$, $C = (a||b)$ and $D = AB - C^2$. A full derivation of Eq.~(\ref{Fstat_eqn}) can be found in Sec III.A of Ref.~\cite{Jaranowski_1998}. \\

We claim a detection if the functional $\mathcal{F}$ exceed a predefined threshold, which is calculated to give a desired false alarm probability. Once above threshold, the magnitude of $\mathcal{F}$ indicates the probability of detection \cite{Jaranowski_1998}. In the case of white, Gaussian noise with no signal, the probability density function (PDF) for the $\mathcal{F}-$statistic takes the form of a centralised $\chi^2$ distribution with four degrees of freedom, $p(2\mathcal{F})= \chi^2(2\mathcal{F}; 4, 0)$. When a signal is present, the PDF has a non$-$central $\chi^2$ distribution with four degrees of freedom, $p(2\mathcal{F})= \chi^2(2\mathcal{F}; 4, \rho_0^2)$. The non$-$centrality parameter $\rho_0^2$ is
\begin{equation}
    {\rho_0}^2 =\frac{K{h_0}^2T_{\rm{drift}}}{S_h(f_0)},
\end{equation}
where the constant $K$ depends on the number of detectors, orientation of the source and the sky location, ${h_0}$ is the characteristic strain of the GW signal, and $S_h(f_0)$ represents the one$-$sided noise spectral density of the detector.

\subsection{\label{subsec:HMM_framework}HMM framework}  
A HMM offers a powerful method for detecting and tracking the wandering frequency of a pulsar \cite{Quinn_2001}. This is done by modelling the signal frequency $f_*(t)$ probabilistically as a Markov chain of transitions between unobservable (hidden) states and using the $\mathcal{F}-$statistic as a detection statistic to relate the observed data to the hidden states. HMM tracking plays a critical role in various industrial applications including computational biology \cite{Henderson_1997}, radar target detection systems \cite{Tugac_2012} and biometric recognition \cite{Nefian_1998}. It estimates $f_0(t)$ accurately when the signal$-$to$-$noise ratio is low but the sample size is large, a situation commonly encountered in CW searches \cite{Quinn_2001}. HMMs have been used to search for CW signals from low$-$mass X$-$ray binary systems \cite{ScoX1_2017,ScoX1_2019,Middleton_2020}, supernova remnants \cite{Sun_2018} and the post$-$merger remnant of GW170817 \cite{Sun_2019}.\\

As in Ref.~\cite{ScoX1_2017}, we model the spin parameter of a CW signal as a hidden Markov process where the hidden (unobservable) state variable $q(t)$ transitions between values in a set \{$q_1,....,q_{N_Q}$\} at discrete times \{$t_0,...., t_{N_T}$\}. Meanwhile, the observable state variable $o(t)$ transitions between values from the set \{$o_1,....,o_{N_O}$\}. Under the Markov assumption, the hidden state at time $t_{n+1}$ only depends on the state at time $t_n$. This is described by the transition probability matrix 
\begin{equation}
A_{q_jq_i} = \mathrm{Pr}[q(t_{n+1}) = q_j|q(t_n) = q_i].  \label{transition_prob} 
\end{equation}

The likelihood that the hidden state $q_i$ gives rise to the observation $o_j$ is described by the emission probability matrix
\begin{equation}
L_{o_jq_i} = \mathrm{Pr}[o(t_n) = o_j | q(t_n) = q_i].  
\end{equation}

Finally, the model is completed by specifying the probability of the system occupying each hidden state initially, which is described by the prior vector
\begin{equation}
    \Pi_{q_i} = \mathrm{Pr}[q(t_0) = q_i].
\end{equation}

For a Markov process, the probability that a hidden path $Q = \{q(t_0), .... q(t_{N_T})$\} gives rise to the observed sequence $O = \{o(t_1),...,o(t_{N_T})\}$ can be obtained using the following expression \cite{Suvorova_2016}
\begin{multline}
    Pr(Q|O) \propto L_{o(t_{N_T})q(t_{N_T})}A_{q(t_{N_T})q(t_{N_{T}-1})}...\times\\  L_{o(t_1)q(t_1)}A_{q(t_1)}\Pi_q(t_0). \label{Markov_chain}
\end{multline}

The path which maximises $Pr(Q|O)$ is known as the most probable path $Q^*$ and is defined as
\begin{equation}
    Q^{*}(O) = \mathrm{arg max}\ Pr(Q|O).  \label{optimal_path}
\end{equation}

The Viterbi algorithm, as outlined in Refs \cite{Suvorova_2016, Suvorova_2017} provides a recursive, computationally efficient route to computing $Q^*(O)$ from Eqs.~(\ref{transition_prob}) to (\ref{optimal_path}). It relies on the Principle of Optimality, which states that all sub$-$paths of an optimal path are themselves optimal \cite{Bellman_1966}. To speed up the computation process, we evaluate $\mathcal{L}=\log \left[Pr(Q|O)\right]$, whereby Eq.~(\ref{Markov_chain}) becomes a sum of log$-$likelihoods.\\

Previous HMM searches \cite{Middleton_2020,ScoX1_2017} have modelled timing noise as an unbiased random walk, in which the frequency of the signal ($f_0$) moves by at most one bin up or down during a timescale $T_{\rm{drift}}$. This is represented by the following transition matrix 
\begin{equation}
  A_{q_{i-1}q_i} = A_{q_{i}q_i}  = A_{q_{i+1}q_i} = \frac{1}{3}. \label{eq:TM_old}
\end{equation} 
An unbiased random walk is not appropriate in this paper because, for all targets considered here, the evolution of $f_0(t)$ is dominated by the secular spin$-$down of the pulsar rather than its timing noise [Refer to Section \ref{subsec:Drift_timescale}, Table \ref{tab:Timing_noise_1}]. Consequently, the secular spin$-$down sets $T_{\rm{drift}}$ and $f_0$ moves down by one bin in each segment. Since the stochastic spin$-$wandering timescales are typically longer than the secular timescales, one could argue that an appropriate transition matrix would be the trivial matrix:
\begin{equation}
   A_{q_{i-1}q_i} = 1. \label{eq:TM_new}
\end{equation}
However, the drift timescales given in Table \ref{tab:Timing_noise_1} are conservative estimates, they are only estimates and significant modelling uncertainties remain, particularly because the precise mechanism of GW emission is not yet established. With such a precise prescription as in Eq.~(\ref{eq:TM_new}), any spin$-$wandering leads to a substantial loss of sensitivity \cite{Ashton_2015}. Therefore, it is appropriate to model the system as a biased random walk, with the bias in this case accounting for the secular spin$-$down. A looser prescription also ensures that the HMM can continue to track the signal even if $\dot{f}_*$ is mis-measured or if the emission frequency shifts randomly (e.g. due to internal stellar processes \cite{Melatos_2014}) with respect to $f_\ast$. The resulting transition matrix is
\begin{equation}
  A_{q_{i-2}q_i} = A_{q_{i-1}q_i}  = A_{q_iq_i} = \frac{1}{3}, \label{eq:transition_matrix}
\end{equation} 
with all other terms being zero. For simplicity, we choose equal weighting for the three allowed transitions, which does not have a significant effect on sensitivity of such searches \cite{Sun_2018}.\\

The emission probability takes the following form~\cite{Suvorova_2016,Sun_2018}
\begin{equation}
    L_{o_jq_i} \propto \exp{[\mathcal{F}(f_0)]},
\end{equation}
where the $\mathcal{F}(f_0)$ is computed for each segment of length $T_{\rm{drift}}$ at a frequency resolution of $\triangle f_{\rm{drift}} = 1/(2T_{\rm{drift}})$. Section \ref{subsec:Drift_timescale} outlines the criteria used for setting $T_{\rm{drift}}$.
We choose a uniform prior over the sub$-$band expected to contain the GW signal (see Section \ref{sub:F_range}):  
\begin{equation}
    \Pi_{q_i} = {N_Q}^{-1},
\end{equation}
where $N_Q$ is the total number of frequency bins.

\section{\label{sec:parameters}Searching O2} 
We use publicly available data from the O2 run of the aLIGO detectors, specifically the \texttt{GWOSC$-$4KHz$\_$R1$\_$STRAIN} channel, to search for CW signals from the pulsars listed in Table \ref{tab:search_targets} \cite{Collaboration_2019}. This observing run took place between 30$-$11$-$2016 and 25$-$08$-$2017. The advanced Virgo detector was also operating for the last month of O2. However, we do not use the data from this detector due to its relatively short observation period and lower sensitivity \cite{Collaboration_2019}. During O2, there was a break in the operation of both detectors from 22$-$12$-$2016 23:00:00 UTC to 04$-$01$-$2017 16:00:00 UTC. In addition, a commissioning break took place at the Livingston detector between 08$-$05$-$2017 and 26$-$05$-$2017, while at the Hanford detector it lasted from 08$-$05$-$2017 to 08$-$06$-$2017 \cite{Collaboration_2019}. In order to perform a joint search between the two aLIGO detectors, we choose a common period from 04$-$01$-$2017 to 25$-$08$-$2017, with a total $T_{\rm{obs}}$ $\approx$ 234 days.\\

In Sections \ref{sub:F_range} and \ref{subsec:Drift_timescale}, we outline the procedure for setting the search parameters, namely the frequency range and coherence timescale. In Section \ref{subsec:Glitches}, we briefly discuss whether pulsar glitches have any affect on this search. Section \ref{subsec:Threshold} outlines the procedure for setting the detection threshold. Finally, we describe the vetoes used to distinguish a real signal from a non$-$Gaussian noise feature in Section \ref{subsec:Vetoes}. 

\subsection{\label{sub:F_range}Frequency range} 
All of the targets chosen for this study are observed to pulse electromagnetically, meaning their spin frequency $f_*$ is known to a precision of at least one part in a million. We use {\texttt{v1.63}} of the Australian Telescope National Facility (ATNF) pulsar catalogue \cite{Manchester_2005} to obtain an accurate ephemeris for each pulsar. The search is carried out at simple multiples of the star's spin frequency ($f_*$, $4f_*/3$ and $2f_*$) to account for the different emission mechanisms listed in Section \ref{sec:Introduction} \cite{Jones_2001}. We search a 1$-$Hz sub$-$band around each multiple to accommodate stochastic spin$-$wandering, secular spin$-$down and the possibility that the signal is displaced systematically from the relevant multiple of $f_\ast$, i.e., because the X$-$ray$-$emitting crust and GW$-$emitting quadrupole are not exactly locked together \cite{Bennett_1991,Melatos_2012}. The HMM's speed enables us make the sub$-$bands generously wide, viz. 1Hz, and track a larger set of spectral displacements and spin$-$wandering histories than typical coherent searches based on a Taylor series phase model \cite{Pulsar_survey2-2019,Pulsar_survey2-2017,Abbott_2008,GW_pulsars_2014,Beating_SD_Vela_2011}. We do not search in sub$-$bands where the expected signal frequency falls below 10Hz as the strain data is not calibrated below this frequency. This is primarily because the detector noise rises rapidly below 10Hz and is significantly larger than any plausible GW strain signal. Additionally, the GW open data is aggressively high$-$pass$-$filtered at 8Hz to avoid signal processing issues and thus does not accurately represent either the signal or noise at low frequencies \cite{Cahillane_2017,Collaboration_2019}.
% strain data is not calibrated below 10Hz. 
% the uncertainty in phase and amplitude of the detector response function increases significantly below this limit, which in turn increases calibration uncertainty 

\subsection{\label{subsec:Drift_timescale}Coherence timescale}
The coherence timescale, $T_{\rm{drift}}$, is an important parameter as it sets the frequency resolution ($\triangle f_{\rm{drift}}$) of the search. It is chosen such that the signal frequency falls in one frequency bin during a single coherent step. For each pulsar, we weigh the drift timescale expected due to the secular spin$-$down ($T'_{\rm{drift}}$) against the stochastic spin$-$wandering timescale ($T''_{\rm{drift}}$). We then use the shorter of the two timescales to define $T_{\rm{drift}}$ for this search.

Overestimating the drift timescale means that, in general, the CW signal wanders by more than one search frequency bin (i.e., $\triangle f_{\rm{drift}}$) over a single coherent step of length $T_{\rm{drift}}$. This presents a difficulty for the HMM tracker, which can only ``keep up" with signals that wander down by zero, one or two bins per coherent step, as defined by the transition matrix in Eq.~\ref{eq:transition_matrix}. Once the HMM loses track, the phase mismatch between the signal and model increases monotonically, and the signal-to-noise ratio (SNR) decreases, until it falls below the detection threshold \cite{Jones_2004}.\\ %between each coherent step

In contrast, underestimating $T_{\rm{drift}}$ does not affect the tracking ability of the HMM; the drift condition given by Eq.~{\ref{SD_relation2}} is still satisfied, and the HMM can easily track the signal. However, shorter coherence timescales increase the total number of segments ($N_{\rm{s}}$) which are combined incoherently, reducing the sensitivity of the search, which scales as $\sim N_{\rm{s}}^{-1/4} T_{\rm{obs}}^{1/2}$ \cite{Wette_2012}. For this reason, it is arguably safer to underestimate rather than overestimate $T_{\rm{drift}}$.

% , despite increasing the total number of segments ($N_{\rm{s}}$) which are combined incoherently.

%This reduces the computational cost of the search, but comes at the cost of reduced sensitivity, which scales as $\sim N_{\rm{s}}^{1/4} T_{\rm{drift}}^{1/2}$ \cite{Wette_2012}. For this reason, it is arguably safer to underestimate rather than overestimate $T_{\rm{drift}}$.

%However, shorter coherence timescales increase the total number of segments ($N_{\rm{s}}$) which are combined incoherently, reducing the sensitivity of the search, which scales as $\sim N_{\rm{s}}^{1/4} T_{\rm{drift}}^{1/2}$ \cite{Wette_2012}. 

\subsubsection{Spin$-$down timescale}
%  (i.e., fractional loss of power in the $\mathcal{F}-$statistic due to the discrete sampling of the parameter space)
The spin$-$down rate of a pulsar is governed by the rate at which rotational energy is extracted by magnetic dipole and gravitational quadrupole radiation \cite{Lyne_2015,Ferrari_1969}. The spin$-$down drift timescale ($T'_{\rm{drift}}$) is chosen to satisfy the following relation, satisfying mismatch $m\leq$ 0.2 \cite{Sun_2018}:
\begin{equation}
 \triangle f_{\rm{drift}} = \frac{1}{2T'_{\rm{drift}}}.   \label{SD_relation1}
\end{equation}
Here, $\triangle f_{\rm{drift}}$ represents the frequency resolution of the search. We also require the change in the signal frequency due to the secular spin$-$down over [$t,t+T'_{\rm{drift}}$] to satisfy
\begin{equation}
    \left| \int_t^{t + T'_{\rm{drift}}} dt' \dot{f_0}(t') \right| < \triangle f_{\rm{drift}} \label{SD_relation2}
\end{equation}
for $ 0 < t < T_{\rm{obs}}$ \cite{Sun_2018}. Even though the signal frequency $f_0$ may not be locked to the star's spin frequency $f_\ast$ \cite{Bennett_1991,Melatos_2012}, we can assume that $\dot{f}_0 \approx \dot{f}_\ast$ to a good approximation. Combining relation \ref{SD_relation1} with \ref{SD_relation2}, we arrive at the following equality
\begin{equation}
    |\dot{f}_*|T'_{\rm{drift}} \approx \triangle f_{\rm{drift}} = \frac{1}{2T'_{\rm{drift}}}.
\end{equation} 
Hence, one obtains $T'_{\rm{drift}} = (2|\dot{f}_*|)^{-1/2}$, which depends solely on the spin$-$down rate of the pulsar ($\dot{f}_*$). We obtain the most conservative estimate for $T'_{\rm{drift}}$ by using $|\dot{f_*}|_{\rm{max}}$, where $|\dot{f_*}|_{\rm{max}}$ denotes the maximum allowed value of $\dot{f_*}$ from electromagnetic observations. The spin$-$down drift timescales for the ten targets are presented in column~3 of Table \ref{tab:Timing_noise_1}.

\subsubsection{Stochastic spin$-$wandering timescale}
Pulsar timing measurements reveal stochastic spin$-$wandering as a widespread phenomenon which manifests itself as timing noise \cite{Hobbs_2010, Ashton_2015}, auto$-$correlated on the order of days to weeks \cite{Suvorova_2017}. The timing noise in pulsars can be characterised as a random walk in some combination of rotation phase, frequency and spin$-$down rate. These idealised cases are commonly referred to as phase noise, frequency noise and spin$-$down noise \cite{Phillips_1994,Shannon_2016}. The stochastic spin$-$wandering timescale ($T''_{\rm{drift}}$) can be estimated in terms of two parameters defined below, $\beta$ and $k$, which characterise the power spectral density (PSD) of the timing noise phase residuals \cite{Parthasarathy_2019}. One finds 
\begin{equation}
     T''_{\rm{drift}} = \left[\frac{1}{k(\beta-1)}\right]^{2/(\beta-1)}; \label{eq:Stochastic_drift}
\end{equation}
refer to Appendix \ref{Appendix:SW_derivation} for a full derivation. In equation \ref{eq:Stochastic_drift}, $\beta$ is the exponent of power$-$law PSD of phase residuals, $P_{\rm{red}}(f) = \left(A^2_{\rm{red}}/12\pi^2\right)\left(f/f_{\rm{yr}}\right)^{-\beta}$ \cite{Shannon_2016,Parthasarathy_2019}. Also in Eq. \ref{eq:Stochastic_drift}, we have the parameter $k$ which is defined as follows
\begin{equation}
    k = \frac{\mathrm{\langle\triangle\Phi^2_{EM}\rangle^{1/2}}}{\tau^{(\beta-1)/2}},
\end{equation}
where $\mathrm{\langle\triangle\Phi^2_{EM}\rangle^{1/2}}$ is the rms phase residual accumulated over time$-$gap $\tau$. Depending on which random process best describes the observed timing noise, $\beta$ can be 2, 4 or 6 for phase noise, frequency noise or spin$-$down noise, respectively. The spectral index has only been measured for a handful of the selected targets, namely the Crab pulsar ($\beta$ = 4) \cite{Cordes_1980}, Vela pulsar ($\beta \approx $ 6) \cite{Shannon_2016} and PSR J1513$-$5908 ($\beta \approx$ 6) \cite{Parthasarathy_2019}. For the remaining targets, we experiment with all three scenario (i.e., $\beta = 2,4,6$) and quote the most conservative timescale (i.e., shortest $T''_{\rm{drift}}$) in Table \ref{tab:Timing_noise_1}, obtained using $\beta = 6$.\\

Conservatives estimates for the two drift timescales indicate that $T''_{\rm{drift}} \gg T'_{\rm{drift}}$ on average. We pick $T_{\rm{drift}} = {\rm{min}}(T'_{\rm{drift}},T''_{\rm{drift}})$. Tables \ref{Pulsar_J0534}$-$\ref{Pulsar_J1849} in Appendix \ref{Appendix:search_params} summarise the search parameters for each target. 

\begin{table}[h]
    \setlength{\tabcolsep}{5pt}
    \renewcommand{\arraystretch}{1.2}
    \centering
    \caption{\small{The secular spin$-$down and stochastic spin$-$wandering timescales estimated from electromagnetic observations. The names of pulsars (column~1) and their spin$-$down rates (column~2) are obtained from \texttt{v1.63} of the ATNF catalogue \cite{Manchester_2005}. Quoted errors (in parentheses) represent the uncertainty in the last significant figure. The spin$-$down drift timescales calculated using the maximum allowed value of $\dot{f_\ast}$ are reported in column~3. Phase residuals for each target  (column~4) are calculated using the information provided in Table \ref{tab:Timing_noise_2} of Appendix \ref{Appendix:SW_derivation} and references therein. In column~5, we report the most conservative (i.e., shortest) estimates for the stochastic spin$-$wandering timescales.}}
    \begin{tabular}{c|c|c|c|c}
        \hline
         Pulsar name   & Spin$-$down rate & $T'_{\rm{drift}}$  &  $\mathrm{\langle\triangle\Phi^2_{EM}\rangle^{\frac{1}{2}}}$ & $T''_{\rm{drift}}$ \\ %& Refs \\ 
         $[$J2000$]$      & [$\times10^{-12}$ Hzs$^{-1}$] & [days]   & [Cycles] &  [yrs] \\ \hline \hline % &  \\ \hline
         J0534$+$2200  & 377.535(2)      & 0.41 & 0.003 & 1.902  \\ %&   \cite{Abdo_2010}\\ 
         J0835$-$4510  & 15.666(6)       & 2.06 & 4.9 & 0.609   \\ %&  \cite{Yu_2013}  \\  
         J1016$-$5857  & 7.00965(5)      & 3.08 & 0.0401 & 2.87   \\ % &  \cite{Camilo_2001}\\  
         J1357$-$6429  & 13.05395(5)     & 2.25 & 0.0217 & 1.51   \\ % &  \cite{Zavlin_2007} \\  
         J1420$-$6048  & 17.8912(7)      & 1.94 & 1.02 & 0.57   \\ % &  \cite{Yu_2013}  \\
         J1513$-$5908  & 66.5310558(2.7) & 1.00 & 1.97 & 4.52    \\ % &  \cite{Parthasarathy_2019} \\
         J1718$-$3825  & 2.371346(1.1)   & 5.31 & 0.361 & 1.73  \\ %&  \cite{Yu_2013}  \\
         J1826$-$1334  & 7.3062994(9)    & 3.02 & 22.35                & 1.79   \\ % &  \cite{Hobbs_2010} \\
         J1831$-$0952  & 1.839595(5)     & 6.04 & 0.0297 & 13.04  \\ %&  \cite{Lorimer_2006}  \\
         J1849$-$0001  & 9.59(4)        & 2.63 & 0.0136  & 1.52   \\ \hline %& \cite{Bogdanov_2019} \\ \hline
    \end{tabular} \label{tab:Timing_noise_1}
\end{table}

\subsection{\label{subsec:Glitches}Glitches} 
Glitches are another form of timing irregularity commonly observed in a subset of the isolated pulsar population \cite{Lyne_2006}. They show up as abrupt changes in $f_\ast$, often followed by a relaxation. Glitches offer an opportunity to probe the interior of a neutron star and the properties of matter at super$-$nuclear densities \cite{Espinoza_2011,Haskell_2015}.\\

The Crab pulsar underwent a small glitch on 27$-$03$-$2017 \footnote{\url{http://www.jb.man.ac.uk/pulsar/glitches.html}}. This event changed the star's spin frequency by $\triangle f_\ast=2.14(11)\times 10^{-9}$Hz which is significantly smaller than $\triangle f_{\rm{drift}} =1.3889 \times 10^{-5}$Hz. The Vela pulsar also experienced a glitch on 12$-$12$-$2016, at the start of O2 \cite{Basu_2020,Palfreyman_2018}, with $\triangle f_* = 1.431\times10^{-6}$Hz. This is conveniently outside the chosen observation period and does not affect our analysis. None of the other pulsars are reported to have glitches in the chosen observation period. 

\subsection{\label{subsec:Threshold}Threshold} 
As outlined in Section \ref{sec:search_algorithm}, we apply the $\mathcal{F}-$statistic to the O2 dataset to calculate the probability of a signal being detected for a given set of parameters. We then use the Viterbi algorithm to find a set of optimal paths through the frequency space. Each optimal Viterbi path ends in one of the final states $q_i$ ($1 \leq i \leq N_Q$) and has an associated log$-$likelihood $\mathcal{L}$ as defined in Eq.~\ref{Markov_chain}. A path is classified as a detection candidate if its $\mathcal{L}$ exceeds a pre$-$determined threshold $\mathcal{L}_\mathrm{th}$.\\

We aim to find $\mathcal{L}_\mathrm{th}$ such that the search has a false alarm probability P$_{\mathrm{fa}}=0.01$ per sub$-$band, per target. To do this, we generate 100 Gaussian noise$-$only realisations for each sub$-$band using the \texttt{lalapps$\_$Makefakedata$\_$v4} tool in the LIGO Scientific Collaboration Algorithm Library (LAL) \cite{LALapps_2018}. Each realisation is searched using the procedure outlined in Section \ref{sec:search_algorithm}, and $\mathcal{L}_\mathrm{th}$ is chosen to yield at-most one (false) detection \cite{Middleton_2020}. The thresholds obtained using the above procedure are listed in the third column of Table \ref{tab:Results}.

\subsection{Vetoes} \label{subsec:Vetoes}
Some of the signal templates produce candidates with $\mathcal{L}>\mathcal{L}_{\rm{th}}$. The number of candidates may exceed the number expected from P$_{\mathrm{fa}}$ because of non$-$Gaussian features in the real aLIGO detector noise. Here we lay out three veto criteria used to discard candidates as non$-$astrophysical sources. These are adapted from the vetoes previously used in published HMM searches \cite{ScoX1_2017,ScoX1_2019,Middleton_2020}. 
\begin{enumerate}
    \setlength\itemsep{0.5em}
    \item \textbf{Known lines veto:} Numerous persistent instrumental lines are identified during the aLIGO detector characterisation process \cite{Covas_2018}. These lines are produced by a number of sources, including resonant modes of the suspension system, external environmental causes and interference from equipment around the detector \cite{ScoX1_2019}. If the Viterbi path of a candidate crosses a known noise line over the duration of the observing run, it can be safely vetoed as a known instrumental line. 
    
%    \db{If a known noise line at $f_{\mathrm{line}}$ lies within the full width half maximum of the candidate, it can be safely vetoed as a known instrumental line. NEED TO FIX!}
    %If the optimal HMM path of a candidate crosses a known noise line at $f_{line}$ for any time in the range $0\leq t\leq T_{obs}$, it can be safely vetoed as a known instrumental line.\\
    \item \textbf{Single interferometer (IFO) veto:} A strong noise artifact which is only present in one of the detectors can produce candidates above threshold in the twin$-$detector search. To veto such artifacts, we analyse the data from the Hanford and Livingston detectors separately and compare the log$-$likelihood in each case (i.e., $\mathcal{L}_{\mathrm{H1}}$ and $\mathcal{L}_{\mathrm{L1}}$) with the log$-$likelihood from the twin$-$detector search ($\mathcal{L}_{\mathrm{2ifo}}$). If $\mathcal{L}$ in either interferometer (but not both) exceeds $\mathcal{L}_{\mathrm{2ifo}}$, the candidate can be safely vetoed as an instrumental artifact. Otherwise we keep the candidate for further analysis. 
    \item \textbf{Off$-$target veto:} This veto was first introduced in Ref.~\cite{Middleton_2020} and extensively studied in Refs.~\cite{Jones_2020,Isi_2020,O3SNR_2020}. It involves analysing the detector data in an off$-$target sky position. For this study, we offset the search from the pulsar's sky location by $\mathrm{3hrs}$ in right ascension and $\mathrm{10min}$ in declination, while keeping all other parameters of the search constant. If the off$-$target search returns $\mathcal{L}\geq$ 0.9$\mathcal{L}_{\rm{on-target}}$, where $\mathcal{L}_{\rm{on-target}}$ denotes the log$-$likelihood from the dual interferometer, on$-$target search, we veto the candidate as an instrumental artifact. This veto criteria is based on the outcome of the injection studies presented in Appendix \ref{Appendix:Off-target_veto}. 
\end{enumerate} 
Previous HMM searches have used an additional veto, known as the $T_{\mathrm{obs}}/2$ veto, whereby the data is split in two halves and each half is analysed separately \cite{ScoX1_2017,ScoX1_2019}. A true astrophysical signal should be present in both segments unless it is a transient source like an r$-$mode which happens, by luck, to be off for half of the observation \cite{Caride_2019}. These searches used Viterbi score as a detection statistic, which is defined as the number of standard deviations the log$-$likelihood of the optimal path exceeds the mean log$-$likelihood of all alternative pathways. Due to the normalisation step, which is inherent to the score calculation, the distribution of final $\mathcal{L}$ is (approximately) the same for typical observation lengths. This implies that the detection statistic is independent of $T_{\rm{obs}}$. However, since the distribution of unnormalised $\mathcal{L}$ varies with $T_{\rm{obs}}$, separate thresholds will need to be established for the two data segments in order to adapt the veto to this search. For this reason, we do not use $T_{\mathrm{obs}}/2$ here.

\section{\label{sec:Results}Results}
All of the targets listed in Table \ref{tab:search_targets} return candidates with $\mathcal{L}\geq \mathcal{L}_{\mathrm{th}}$. This is not surprising as the detector noise is not Gaussian. We apply the three data quality vetoes outlined in Section \ref{subsec:Vetoes} to each candidate to separate a real signal from a non$-$Gaussian noise artifact. The results are summarised in Table \ref{tab:Results} while Figs. \ref{Res:PSR_J0534_2200}$-$\ref{Res:PSR_J1849-0001} in Appendix \ref{Appendix:Results_full} show $\mathcal{L}$ against the terminating frequency of the Viterbi paths for each sub$-$band of each target. The search returns 5,256 candidates across 24 sub$-$bands, only 12 of which survive the three data quality vetoes. 

\begin{table}[h!]
    \centering
    \setlength{\tabcolsep}{3pt}
    \renewcommand{\arraystretch}{1.2}
    \caption{\small{Above threshold candidates before and after the data quality vetoes. We omit sub$-$bands below 10Hz. For each pulsar and sub$-$band, we report the threshold log$-$likelihood (column~3), and the number of candidates before (column~4), and after the three data quality vetoes (columns~5).}}\label{tab:Results}
    \begin{tabular}{c|c|c|cc} 
    \hline 
    Pulsar name & Sub$-$band   &$\mathcal{L}_{\mathrm{th}}$ & Candidates & Candidates \\
    $[$J2000]    & [Hz]       &                  & (pre$-$veto) & (post$-$veto)  \\ \hline \hline
    J0534$+$2200 & 29.5$-$30.5  & 4090  & 11 & 1 \\         % & 11 & 2   
                 & 39.5$-$40.5  & 4775  & 9  &  0 \\        % 8 & 2 &
                 & 59.5$-$60.5  & 5789  & 12 & 0 \\ \hline  % 12 & 2 &
    J0835$-$4510 & 11$-$12      & 943   & 64 & 0 \\         % 64 & 10 &     
                 & 14.5$-$15.5  & 1069  & 196 & 0 \\        % 196 & 9 &
                 & 22$-$23      & 1290  & 54 & 0 \\ \hline  % 52 & 4 &
    J1016$-$5857 & 12$-$13      & 756   & 29 & 0 \\       % 29  & 9 &
                 & 18$-$19      & 896   & 10 & 0 \\ \hline  %  10 & 4 &
    J1357$-$6429 & 11.5$-$12.5  & 1187  & 44 & 0 \\ \hline  % & 44 & 11 
    J1420$-$6048 & 14$-$15      & 1006  & 58 & 1 \\         % 58 & 10 & 
                 & 19$-$20      & 1146  & 46 & 1 \\         % 46 & 2 &
                 & 29$-$30      & 1356  & 35 & 1  \\ \hline % 35 & 7 & 
    J1513$-$5908 & 12.5$-$13.5  & 2483  & 23 & 0  \\ \hline % & 23 & 8 
    J1718$-$3825 & 13$-$14      & 413   & 2051 & 0 \\       % & 23 & 8 
                 & 17.5$-$18.5  & 468   & 165 & 1 \\        % & 165 & 4 
                 & 26.5$-$27.5  & 523   & 417 & 0 \\ \hline % & 397 & 12 
    J1826$-$1334 & 12.5$-$13.5  & 760   & 89 & 0  \\        % & 89 & 9     
                 & 19$-$20      & 904   & 63 & 0  \\ \hline % & 63 & 5  
    J1831$-$0953 & 14.5$-$15.5  & 379   & 1413 & 2 \\       % & 1409 & 72
                 & 19.5$-$20.5  & 419   & 263 & 2 \\         %  & 262 & 35 
                 & 29$-$30      & 498   & 81 & 0 \\ \hline   %  & 33 & 1 
    J1849$-$0001 & 25.5$-$26.5  & 761   & 57 & 3 \\         % & 46 & 23 
                 & 34$-$35      & 860   & 61  & 0 \\         % & 8 & 2
                 & 51.5$-$52.5  & 1030  & 5 & 0 \\ \hline    % & 4 & 4 
    \end{tabular}
\end{table}
\subsection{\label{subsec:survivors}Survivors} 
We summarise the properties of the surviving candidates in Table \ref{tab:survivors}. These include the terminating frequency of all surviving Viterbi paths (column~3) and Gaussian thresholds $\mathcal{L_{\rm{th}}}$ for each sub$-$band (column~4). For each candidate, we also report the log$-$likelihood from the dual$-$interferometer, on$-$target search ($\mathcal{L_{\rm{on-target}}}$ or $\mathcal{L_{\rm{2ifo}}}$), single IFO search ($\mathcal{L}_{\rm{H1}}$ and $\mathcal{L}_{\rm{L1}}$) and off$-$target search ($\mathcal{L_{\rm{on-target}}}$) in columns 5 through to 8. Below, we briefly discuss the characteristics of the surviving candidates.\\ %In column~9, we state the outcome of follow$-$up analysis described below.

\begin{table*}
    \centering
    \setlength{\tabcolsep}{5pt}
    \renewcommand{\arraystretch}{1.2}
    \caption{\small{Survivors of the three data quality vetoes. The pulsar name and search sub$-$bands are reported in columns 1 and 2, respectively. For each target and sub$-$band, we report the terminating frequency of the surviving Viterbi path (column~3), Gaussian threshold log$-$likelihood for the sub$-$band (column~4), the dual$-$interferometer, on$-$target $\mathcal{L}$ (column~5), single IFO $\mathcal{L}$s for the Hanford and Livingston detectors (column~6 and 7) as well as $\mathcal{L}$ for the off$-$target search (column~8). We highlight the survivors with either $\mathcal{L}_{\rm{H1}}\approx\mathcal{L}_{\rm{2ifo}}$ or $\mathcal{L}_{\rm{L1}}\approx\mathcal{L}_{\rm{2ifo}}$ and $\mathcal{L}_{\rm{off-target}}=(0.75-0.80)\mathcal{L}_{\rm{on-target}}$ with an asterisk.}} 
    \begin{tabular}{c|c|l|c|cccc} 
    \hline Pulsar & Sub-band & Frequency & $\mathcal{L}_{\rm{th}}$ & $\mathcal{L}_{\rm{on-target}}$ & $\mathcal{L}_{\rm{H1}}$ & $\mathcal{L}_{\rm{L1}}$ & $\mathcal{L}_{\rm{off-target}}$  \\
     $[$J2000]   &  [Hz]   &   [Hz]   & & ($\mathcal{L}_{\rm{2ifo}}$) &   & &  \\ \hline \hline
    J0534$+$2200 & 29.5$-$30.5  & 29.804989 & 4090 & 5159 & 4025 & 4668 & 3662 \\ \hline %\cline{2-9}
%                 & 59.5$-$60.5  & 59.608133 & 5789 & 6029 & 5836 & 4928 & 5746 & $\times$ \\ 
%    J1357$-$6429 &  11.5$-$12.5 & 11.511937 & 1187 & 1315 & 1163 & 1077 & 1175 & $\times$ \\ \hline 
    J1420$-$6048 &  14$-$15     & 14.510879 & 1006 & 1539 & 928 & 1409 & 1045 \\ \cline{2-8}
%                 &              & 14.067361 &  -   & 1037 & 765 & 1033 & 947 & $\times$\\ 
%                 &              & 14.067361 &  -   & 1037 & 765 & 1033 & 947 & $\times$\\ 
%                 &              & 14.772618 &  -   & 1009 & 937 & 870 & 972 & $\times$ \\ 
%                 &              & 14.772866 &  -   & 1021 & 929 & 834 & 972  & $\times$ \\ 
                 & 19$-$20      & 19.514528 & 1146 & 1829 & 1005 & 1664 & 1296  \\ \cline{2-8}
%                 &              & 29.521740 &  & 1397 & 1069 & 1339 & 1263 & $\times$ \\  
                 & 29$-$30      & 29.521799* & 1356    & 1487 & 1098 & 1395 & 1263 \\ \hline
%    J1513$-$5908 & 12.5$-$13.5  & 12.531323 & 2483 & 2551 & 2120 & 2388 & 2402 & $\times$ \\ \hline
    J1718$-$3825 & 17.5$-$18.5  & 17.503407* & 468 & 481 & 438 & 412 & 390  \\ \hline
%    J1826$-$1334 & 19$-$20      & 19.506141 & 904 & 984 & 680 & 983 & 894 & $\times$ \\
%                 &              & 19.506327 &  -  & 909 & 678 & 894 & 873 & $\times$ \\ \hline
    J1831$-$0953 & 14.5$-$15.5  & 14.50181* & 379 & 383 & 244 & 372 & 324  \\
 %                &              & 14.79469  & -   & 383 & 351 & 291  & 375 & $\times$ \\
 %                &              & 15.02224  & -   & 383 & 246 & 382 & 326 & $\times$ \\
 %                &              & 15.02494  & -   & 515 & 236 & 514 & 373 & $\times$ \\
                 &              & 15.401251* & -   & 395 & 229 & 384 & 318  \\ \cline{2-8}
                 & 19.5$-$20.5  & 19.999155 & 419 & 985 & 278 & 962 & 644  \\
                 &              & 19.999173 & -   & 1256 & 296 & 1255 & 664  \\ \hline
    J1849$-$0001 &  25.5$-$26.5 & 26.307929* &  761  & 1141 & 561 & 1136 & 860  \\
                 &              & 26.308067* &  -  & 1064 & 702 & 1061 &  825  \\
                 &              & 26.341044* &  -  & 808 & 573 & 797 & 718  \\ \hline
    \end{tabular}
    \label{tab:survivors}
\end{table*}

\begin{figure}[h]
\begin{subfigure}{0.34\textwidth}
 % \centering
  \includegraphics[trim=0.5cm 0.1cm 0.7cm 0.85cm, clip=true,width=1.0\linewidth]{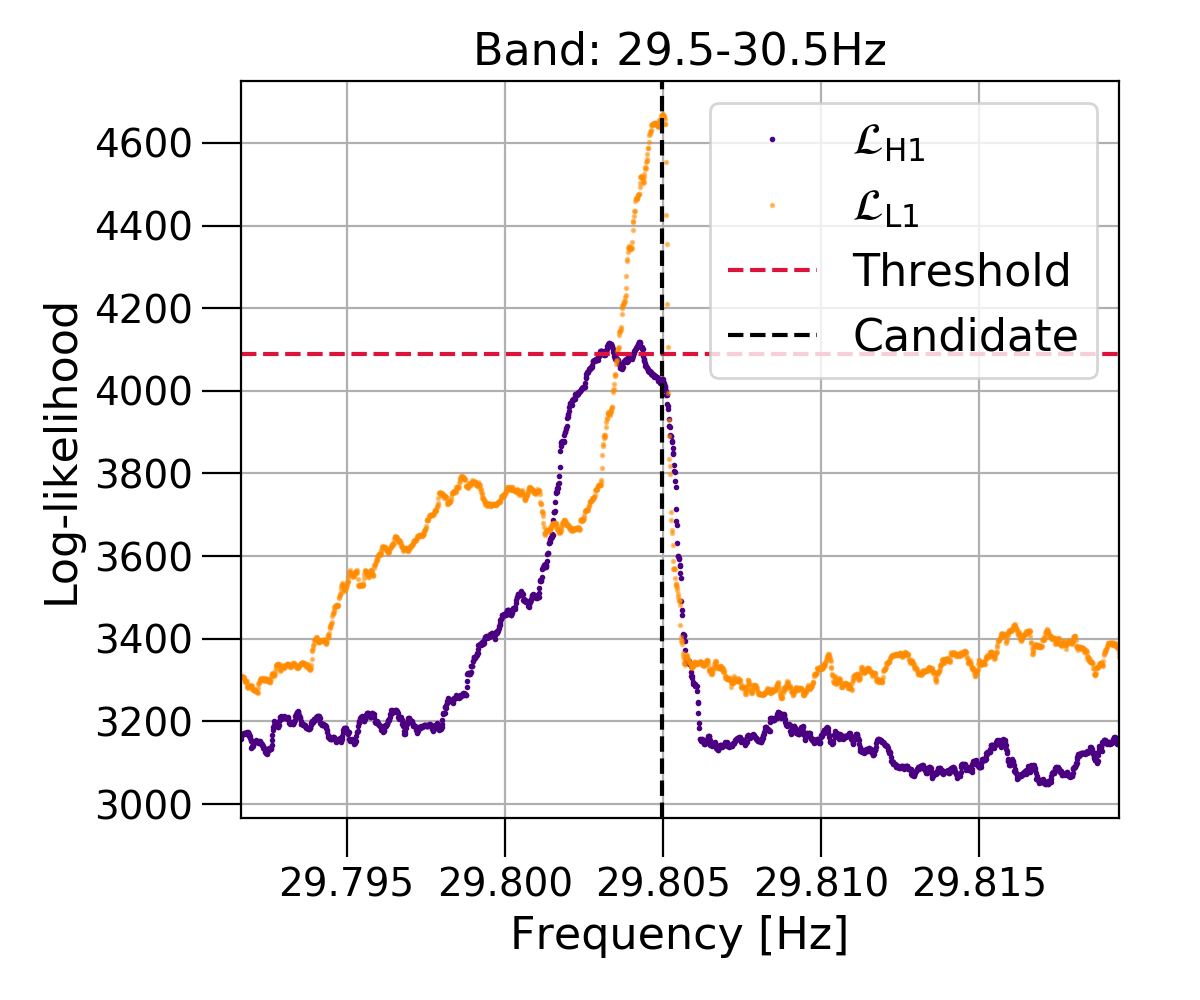}
\end{subfigure}
\begin{subfigure}{0.34\textwidth}
%  \centering
  \includegraphics[trim=0.5cm 0.3cm 0.7cm 0.85cm, clip=true,width=1.0\linewidth]{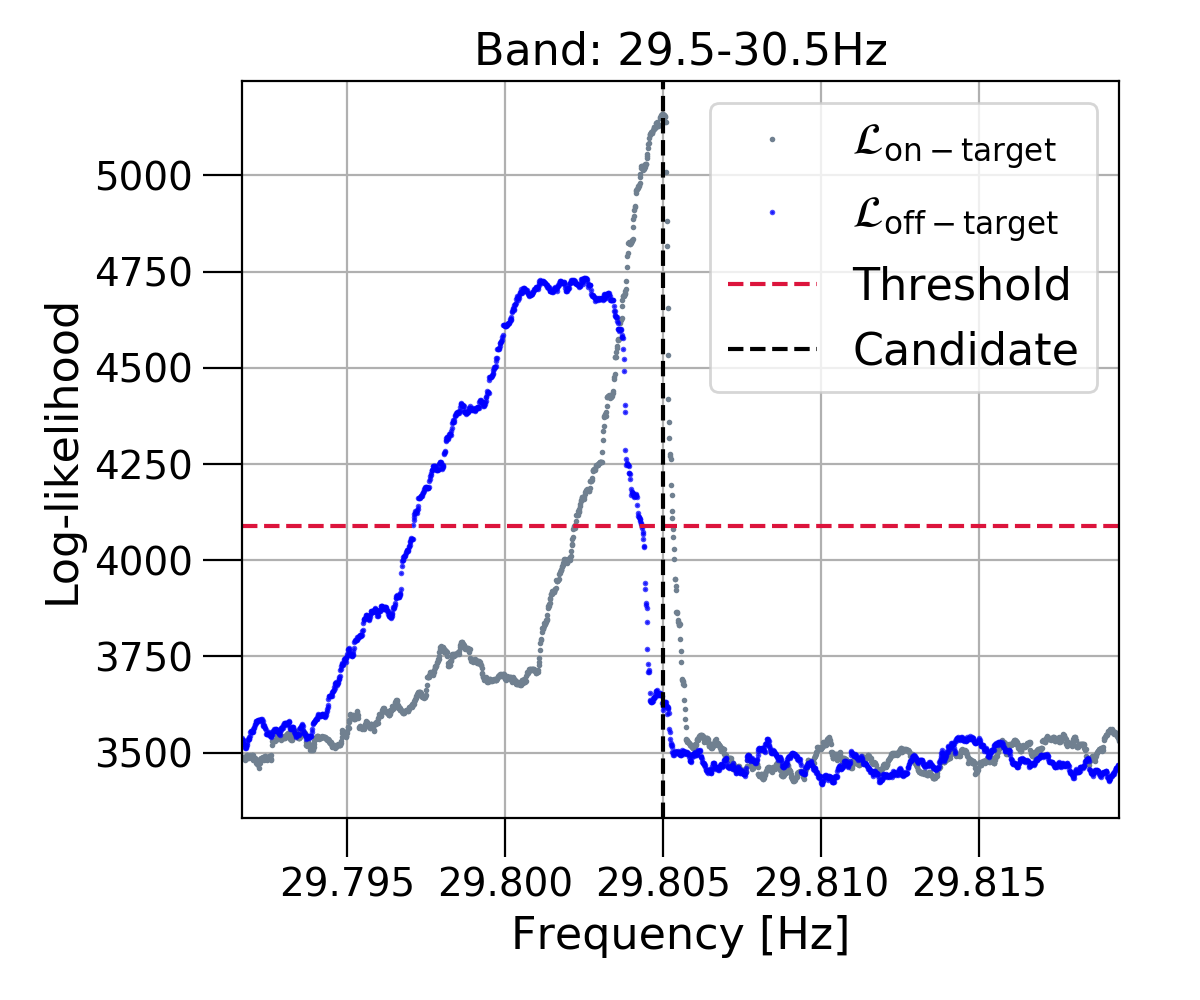}
\end{subfigure}
%\begin{subfigure}{0.23\textwidth}
%  \centering
%  \includegraphics[trim=0.5cm 0.5cm 0.5cm 0.5cm, width=1.0\linewidth]{Results/PSR_J0534+2200_C31-ifo.png}
% \end{subfigure}
% \begin{subfigure}{0.23\textwidth}
%  \centering
%  \includegraphics[trim=0.5cm 0.5cm 0.5cm 0.3cm, clip=true,width=1.0\linewidth]{Results/PSR_J0534+2200_C31-off.png}
% \end{subfigure}
%\captionsetup{justification=raggedright}
\caption{Outcome of the data quality vetoes for the survivor in 29.5-30.5Hz sub$-$band of PSR J0534+2200. The top panel shows the outcome of the single IFO veto with the log$-$likelihoods ($\mathcal{L}$'s) for Hanford (purple) and Livingston (orange) detectors plotted against the terminating frequency of Viterbi paths. In the bottom panel, we plot $\mathcal{L}$'s for the dual$-$interferometer, on$-$target search (gray) and the off$-$target search (blue) against the terminating frequency of Viterbi paths. The survivor frequency is indicated by the black vertical line. The red dashed line shows the Gaussian threshold for the twin$-$detector search ($\mathcal{L}_{\rm{th}}$).}\label{fig:PSR_J0532+2200_candidates}
\end{figure}

\noindent \textbf{PSR J0534+2200:} We search for a signal from this pulsar in the 29.5$-$30.5Hz ($f_*$), 39.5$-$40.5Hz ($4f_*/3$) and 59.5$-$60.5Hz ($2f_*$) sub$-$bands. The search returns 11 candidates at $f_*$, nine at $4f_*/3$ and 12 at $2f_*$. Only one of these candidates survives the data quality vetoes. 
\begin{enumerate}
    \item \textbf{29.5$-$30.5Hz:} The sole survivor in this sub$-$band is located on the peak of a spectral feature in the Livingston data and has $\mathcal{L}_{\rm{L1}} = 4668$, as shown in the top panel of Fig.~\ref{fig:PSR_J0532+2200_candidates}. It also appears as a significant outlier in the Hanford only search with $\mathcal{L}_{\rm{H1}}=4025$ and in the off$-$target search with $\mathcal{L}_{\rm{off-target}}\approx$ 0.7$\mathcal{L}_{\rm{on-target}}$. Since the candidate does not satisfy any of the veto criteria used here, we keep it for a follow$-$up study.
%    \item \textbf{59.5$-$60.5Hz:} The sole survivor in this sub$-$band has $\mathcal{L}_{\rm{H1}}>\mathcal{L}_{\rm{th}}>\mathcal{L}_{\rm{L1}}$ in the single IFO search, as shown in the bottom panel of Fig.~\ref{fig:PSR_J0532+2200_candidates}. It also shows up as a significant outlier in the off$-$target search with $\mathcal{L}_{\rm{off-target}}>$ 0.9$\mathcal{L}_{\rm{on-target}}$, despite being sub$-$threshold. Therefore, the candidate can be rejected as an instrumental artifact in the Hanford detector.\\
\end{enumerate}  

\begin{figure}[h!]
\begin{subfigure}{0.32\textwidth}
  \centering
  \includegraphics[trim=0.5cm 0.1cm 0.5cm 0.85cm, clip=true,width=1.0\linewidth]{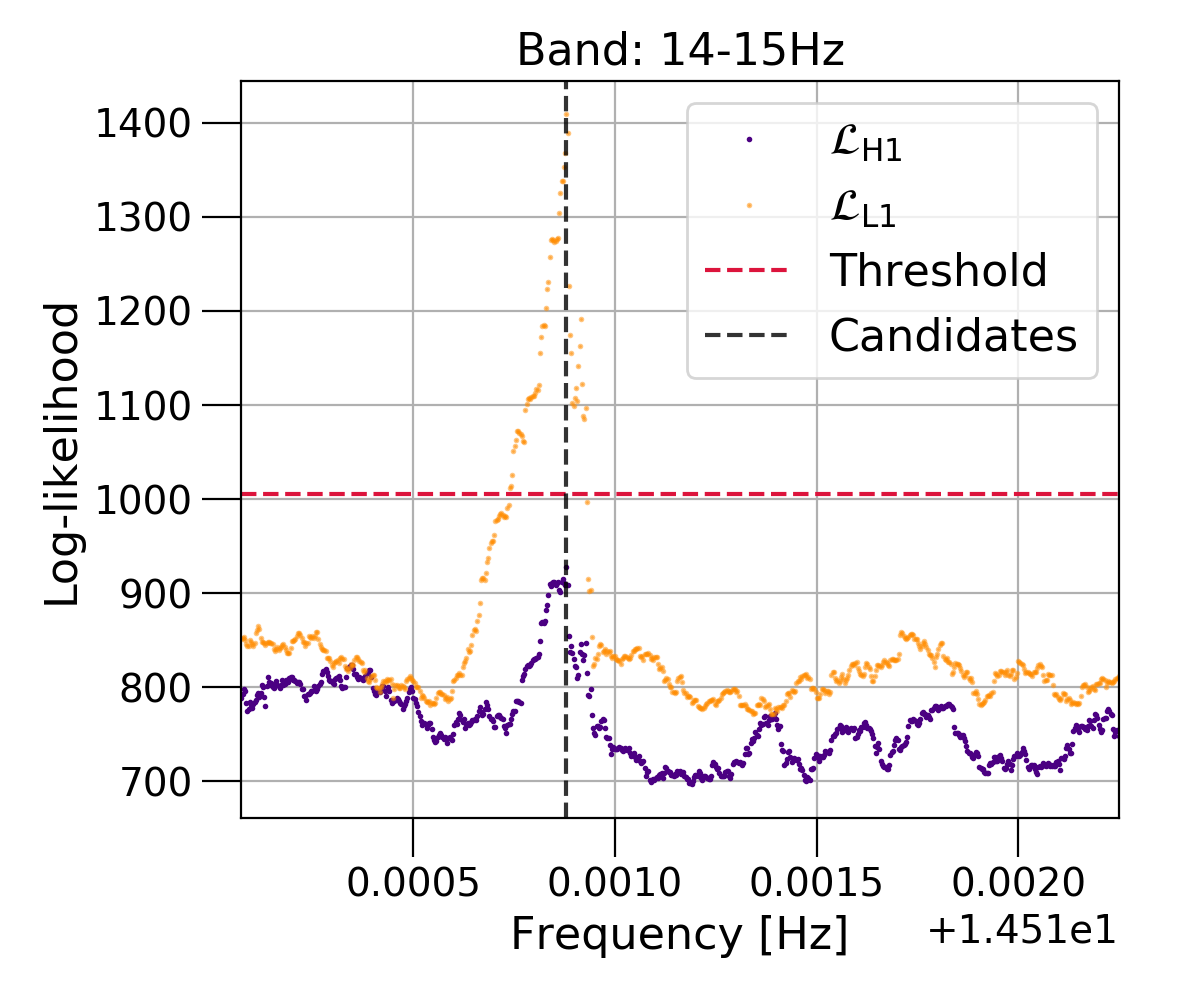}
\end{subfigure}
\begin{subfigure}{0.32\textwidth}
  \centering
  \includegraphics[trim=0.5cm 0.1cm 0.5cm 0.85cm, clip=true,width=1.0\linewidth]{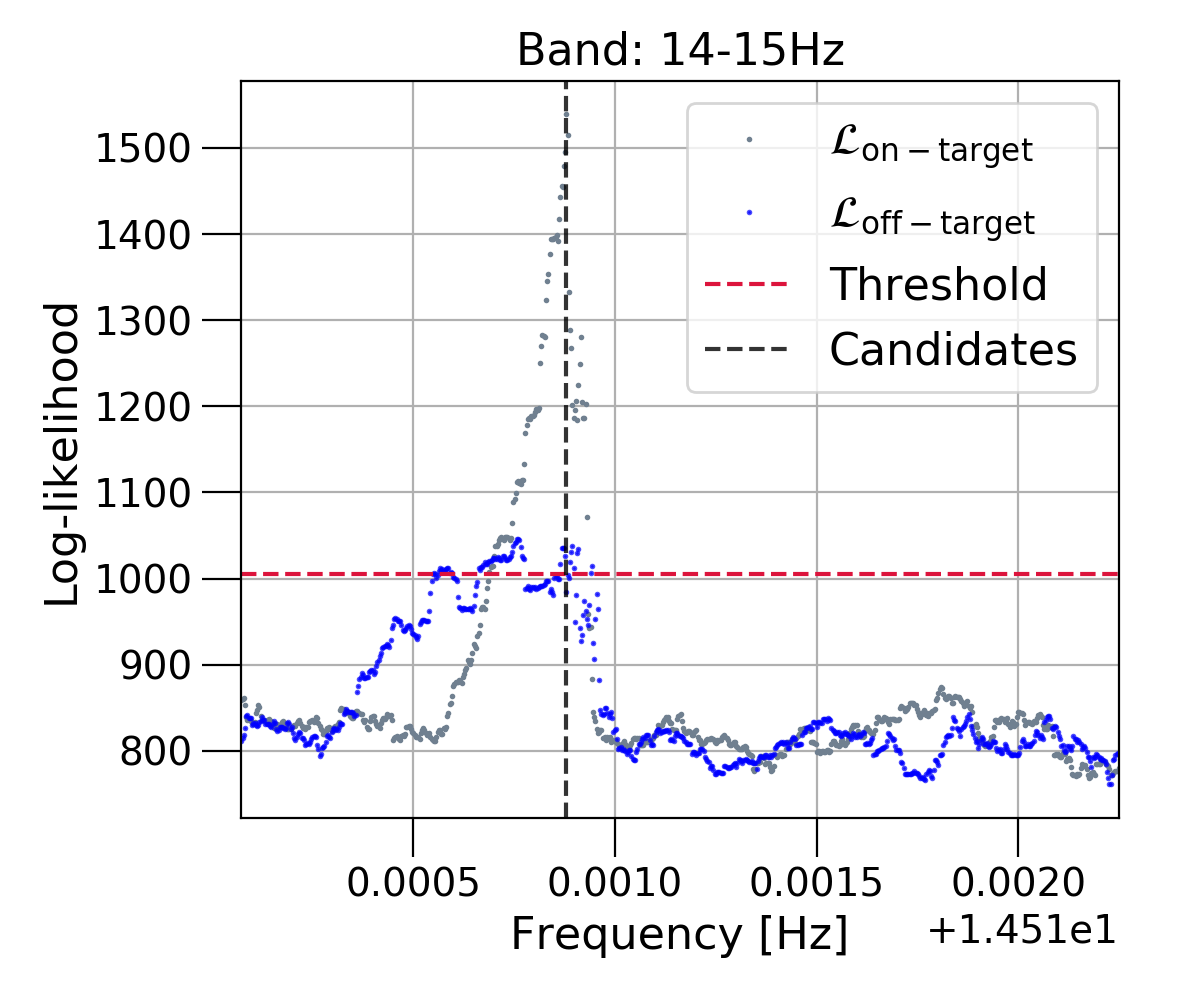}
\end{subfigure}
\caption{\small{Outcome of the single IFO (top) and off$-$target (bottom) vetoes for the survivor in 14$-$15Hz sub$-$band of PSR J1420-6048, laid out as in Fig.~\ref{fig:PSR_J0532+2200_candidates}.}}
\label{fig:J1420-6048_candidates_1}
\end{figure}

\begin{figure}[h!]
\begin{subfigure}{0.32\textwidth}
  \centering
  \includegraphics[trim=0.5cm 0cm 0.5cm 0.85cm, clip=true,width=1.0\linewidth]{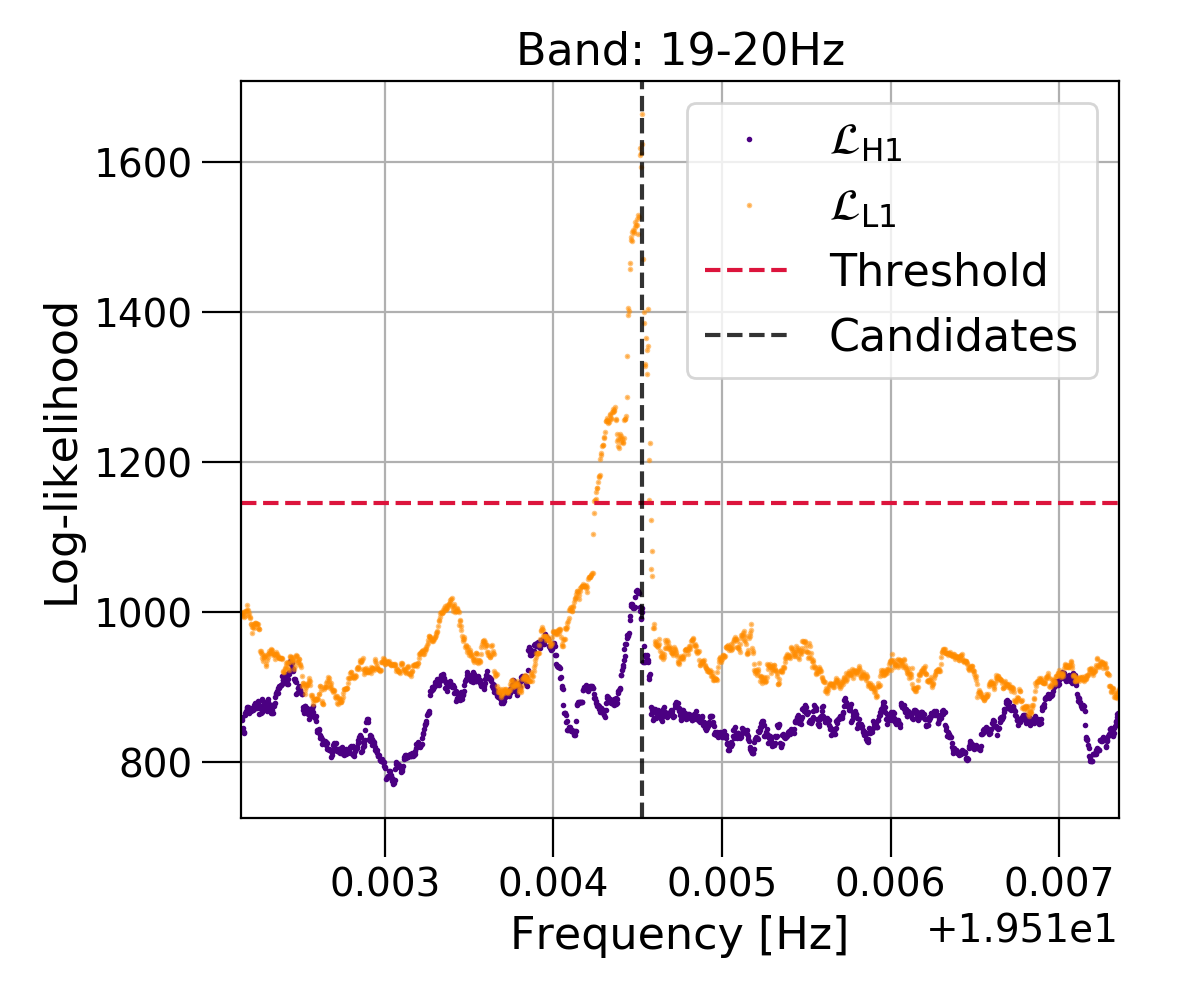}
\end{subfigure}
\begin{subfigure}{0.32\textwidth}
  \centering
  \includegraphics[trim=0.5cm 0cm 0.5cm 0.85cm, clip=true,width=1.0\linewidth]{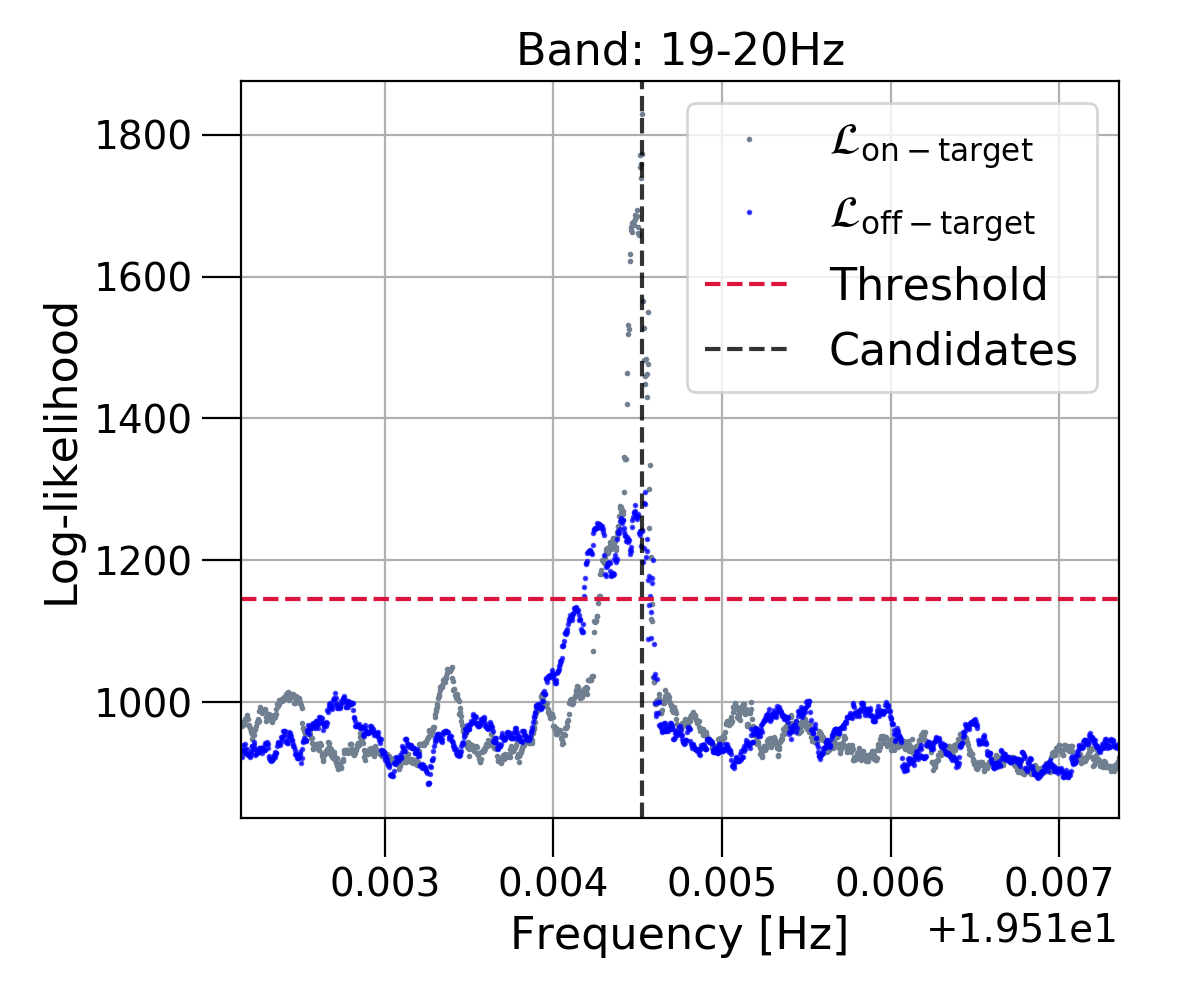}
\end{subfigure}
\caption{\small{Outcome of the single IFO (top) and off$-$target (bottom) vetoes for the survivor in 19$-$20Hz sub$-$band of PSR J1420-6048, laid out as in Fig.~\ref{fig:PSR_J0532+2200_candidates}.}}
\label{fig:J1420-6048_candidates_2}
\end{figure}

\begin{figure}[h!]
\begin{subfigure}{0.33\textwidth}
  \centering
  \includegraphics[trim=0.5cm 0.1cm 0.5cm 0.85cm, clip=true, width=1.0\linewidth]{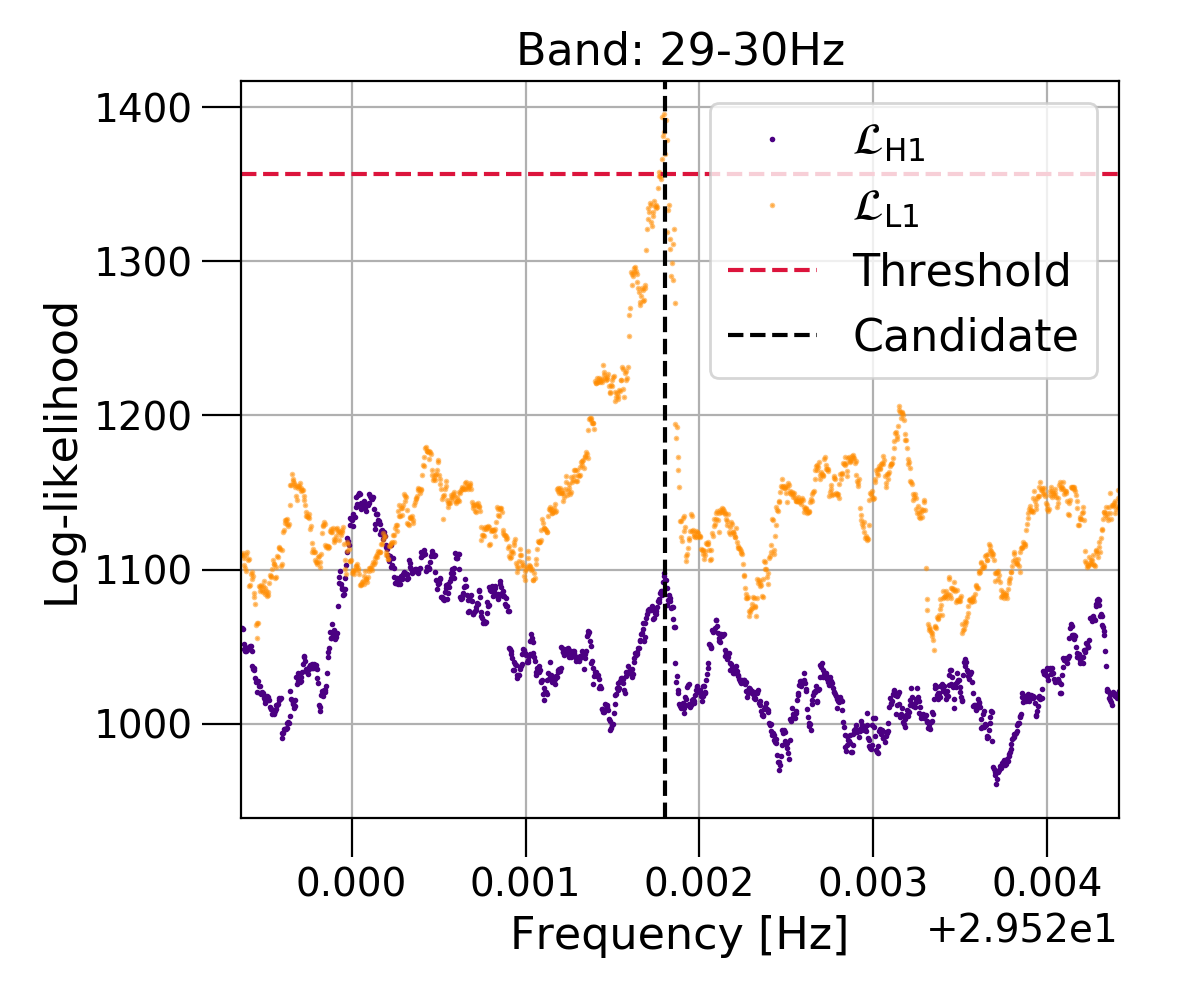}
\end{subfigure}
\begin{subfigure}{0.33\textwidth}
  \centering
  \includegraphics[trim=0.5cm 0.1cm 0.5cm 0.85cm, clip=true, width=1.0\linewidth]{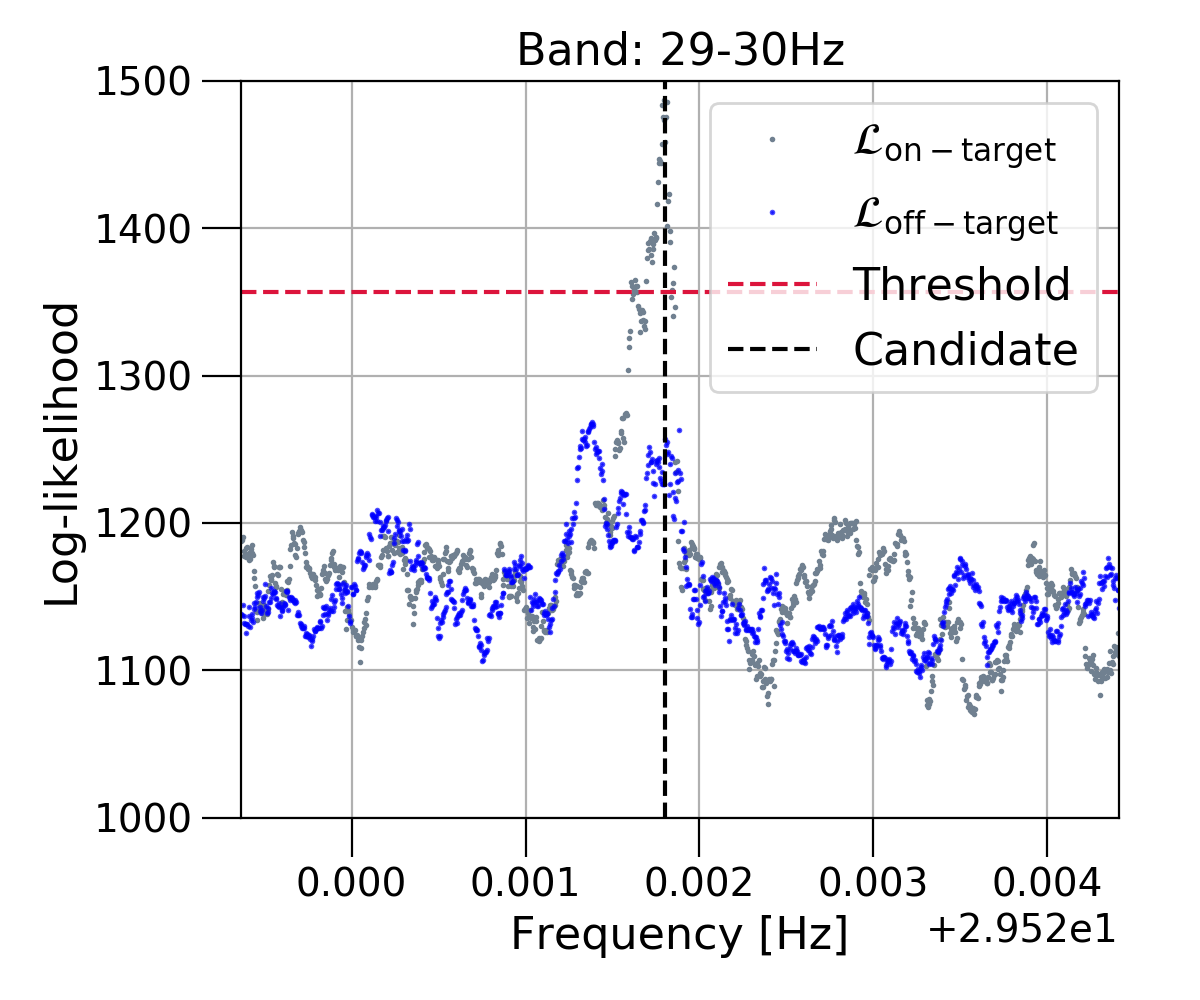}
\end{subfigure} 
\caption{\small{Outcome of the single IFO (top) and off$-$target (bottom) vetoes for the survivor in 29$-$30Hz sub$-$band of PSR J1420-6048, laid out as in Fig.~\ref{fig:PSR_J0532+2200_candidates}.}}
\label{fig:J1420-6048_candidates_3}
\end{figure}

\noindent \textbf{PSR J1420$-$6048:} We search for a signal from this pulsar in the 14$-$15Hz ($f_*$), 19$-$20Hz ($4f_*/3$) and 29$-$30Hz (2$f_*$) sub$-$bands. The search returns 58 candidates at $f_*$, 46 at $4f_*/3$ and 35 at $2f_*$. Only one candidate survives the data quality vetoes in each sub$-$band.
\begin{enumerate}
    \item \textbf{14$-$15Hz:} The candidate at 14.510Hz is more prominent in the Livingston data ($\mathcal{L}_{\rm{L1}} = 1409$) than at Hanford ($\mathcal{L}_{\rm{H1}} = 928$), as shown in the top panel of Fig.~\ref{fig:J1420-6048_candidates_1}. It also resurfaces in the off$-$target search with $\mathcal{L}_{\mathrm{off-target}}\approx~0.7\mathcal{L}_{\mathrm{on-target}}$. Since the candidate does not satisfy any of the veto criteria used here, we keep it for a follow$-$up study. 
    
    \item \textbf{19$-$20Hz:} The sole survivor in this sub$-$band is more prominent in the Livingston data ($\mathcal{L}_{\rm{L1}}=1664$) than at Hanford ($\mathcal{L}_{\rm{H1}}=1005$). It also resurfaces as a significant outlier in the off$-$target search with $\mathcal{L}_{\mathrm{off-target}}\approx~0.7\mathcal{L}_{\mathrm{on-target}}$, as shown in the bottom panel of Fig.~\ref{fig:J1420-6048_candidates_2}. We keep the candidate for a follow$-$up study as it does not satisfy any of the veto criteria used here. 
    
    \item \textbf{29$-$30Hz:} The survivor at 29.5218Hz returns $\mathcal{L}_{\rm{L1}}>~\mathcal{L}_{\rm{th}}>\mathcal{L}_{\rm{H1}}$ in the single IFO search and is shown in the top panel of Fig.~\ref{fig:J1420-6048_candidates_3}. It also resurfaces as a significant outlier in the off$-$target search with $\mathcal{L}_{\mathrm{off-target}}\approx~0.85\mathcal{L}_{\mathrm{on-target}}$. Although it does not strictly satisfy any of the veto criteria used here, the relatively large $\mathcal{L}$ in only one of the detectors, combined with a large $\mathcal{L}_{\mathrm{off-target}}$ points towards a non$-$astrophysical origin.  
\end{enumerate}

% -----------------------------------------------------------------

\begin{figure}[h!]
\begin{subfigure}{0.32\textwidth}
  \centering
  \includegraphics[trim=0.5cm 0.1cm 0.5cm 0.9cm,clip=true, width=1.0\linewidth]{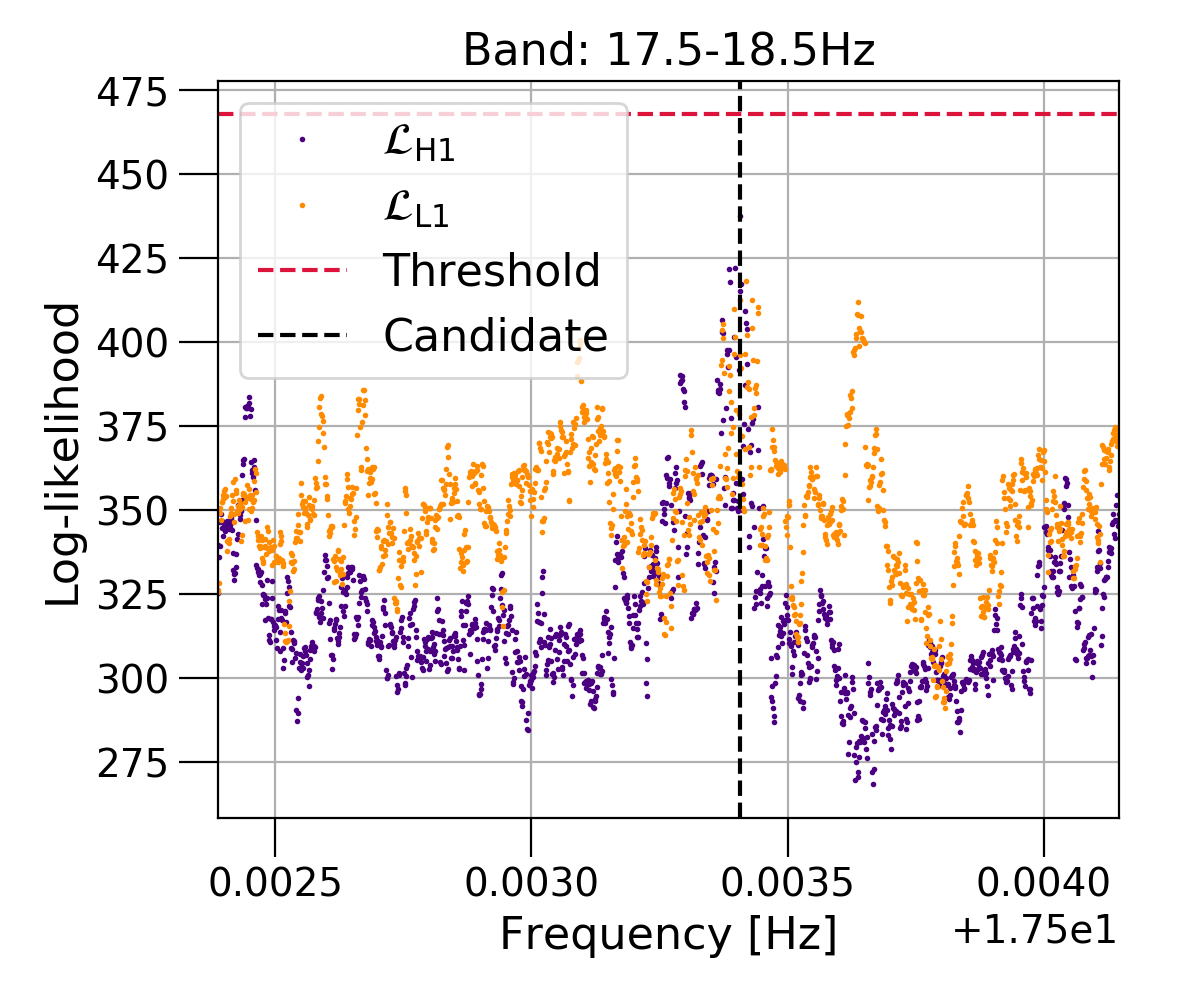}
\end{subfigure}
\begin{subfigure}{0.32\textwidth}
  \centering
  \includegraphics[trim=0.5cm 0.1cm 0.5cm 0.85cm, clip=true,width=1.0\linewidth]{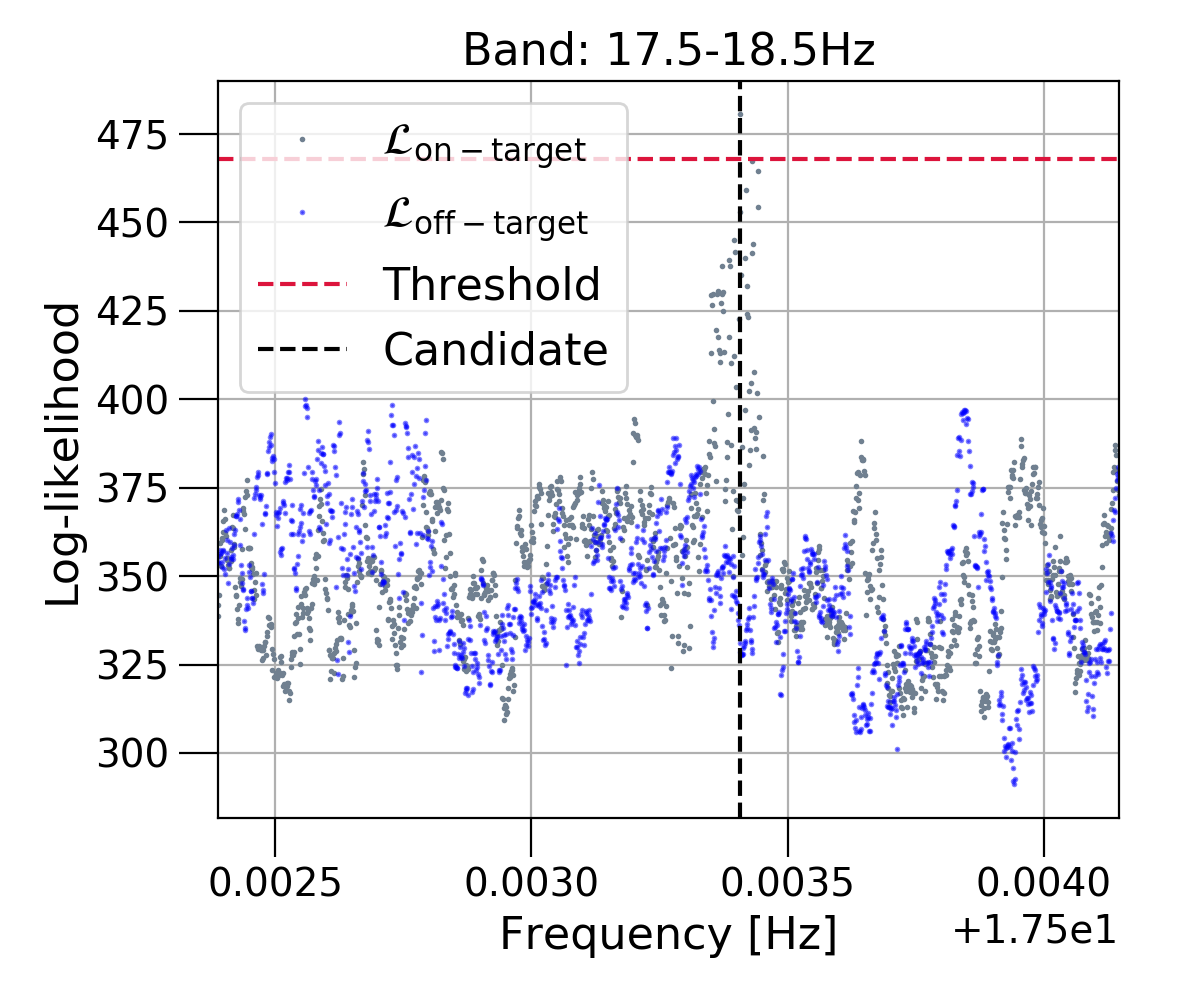}
\end{subfigure}
\caption{\small{Outcome of the single IFO (left) and off$-$target (right) vetoes for the survivor in 17.5$-$18.5Hz sub$-$band of PSR J1718$-$3825, laid out as in Fig.~\ref{fig:PSR_J0532+2200_candidates}.}}
\label{fig:J1718-3825_candidates}
\end{figure}

\begin{figure}[h!]
    \begin{subfigure}{0.32\textwidth}
    \centering
    \includegraphics[trim=0.5cm 0.1cm 0.5cm 0.85cm, clip=true, width=1.0\linewidth]{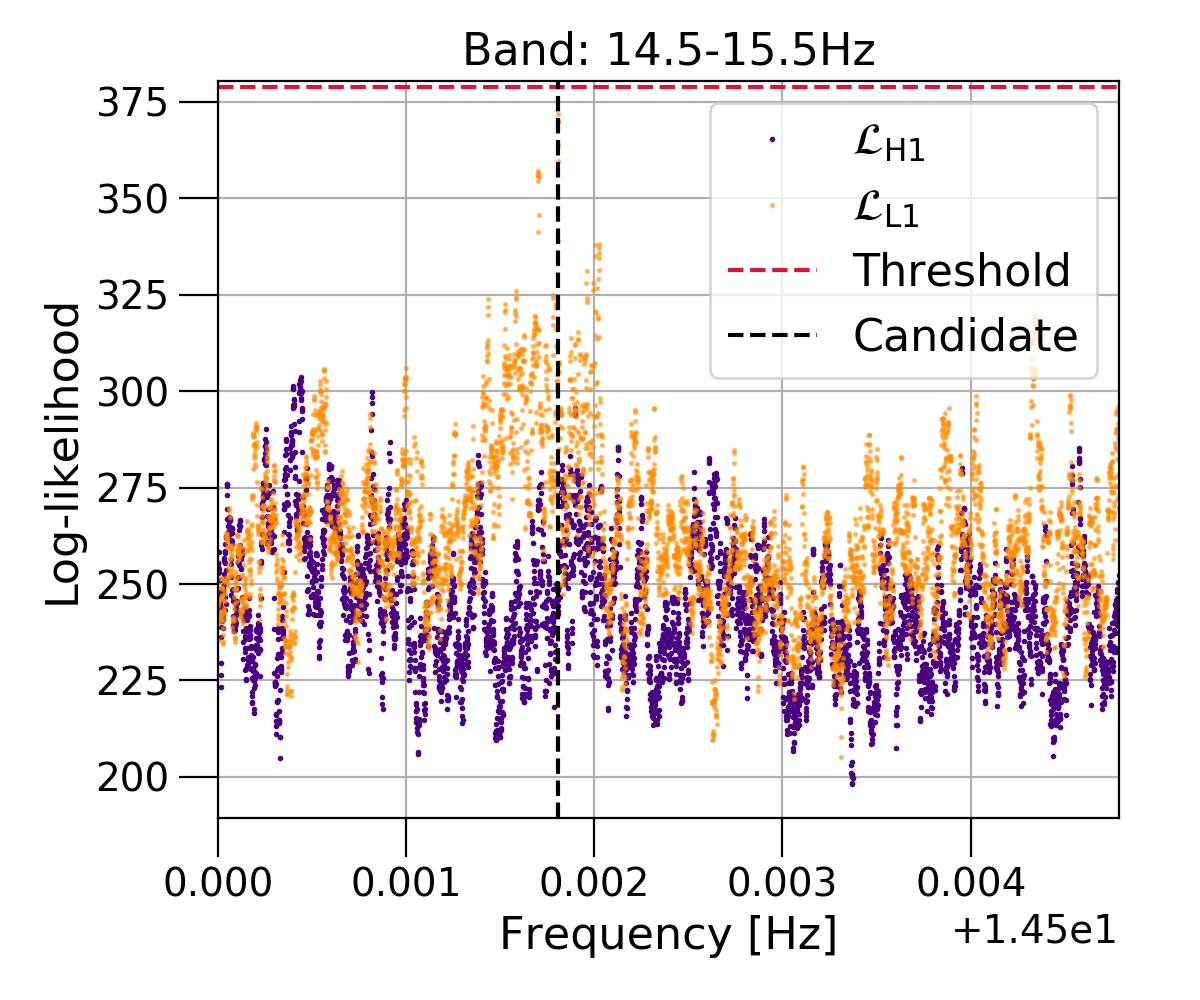}
    \end{subfigure}
    \begin{subfigure}{0.32\textwidth}
    \centering
    \includegraphics[trim=0.5cm 0.1cm 0.5cm 0.85cm, clip=true,width=1.0\linewidth]{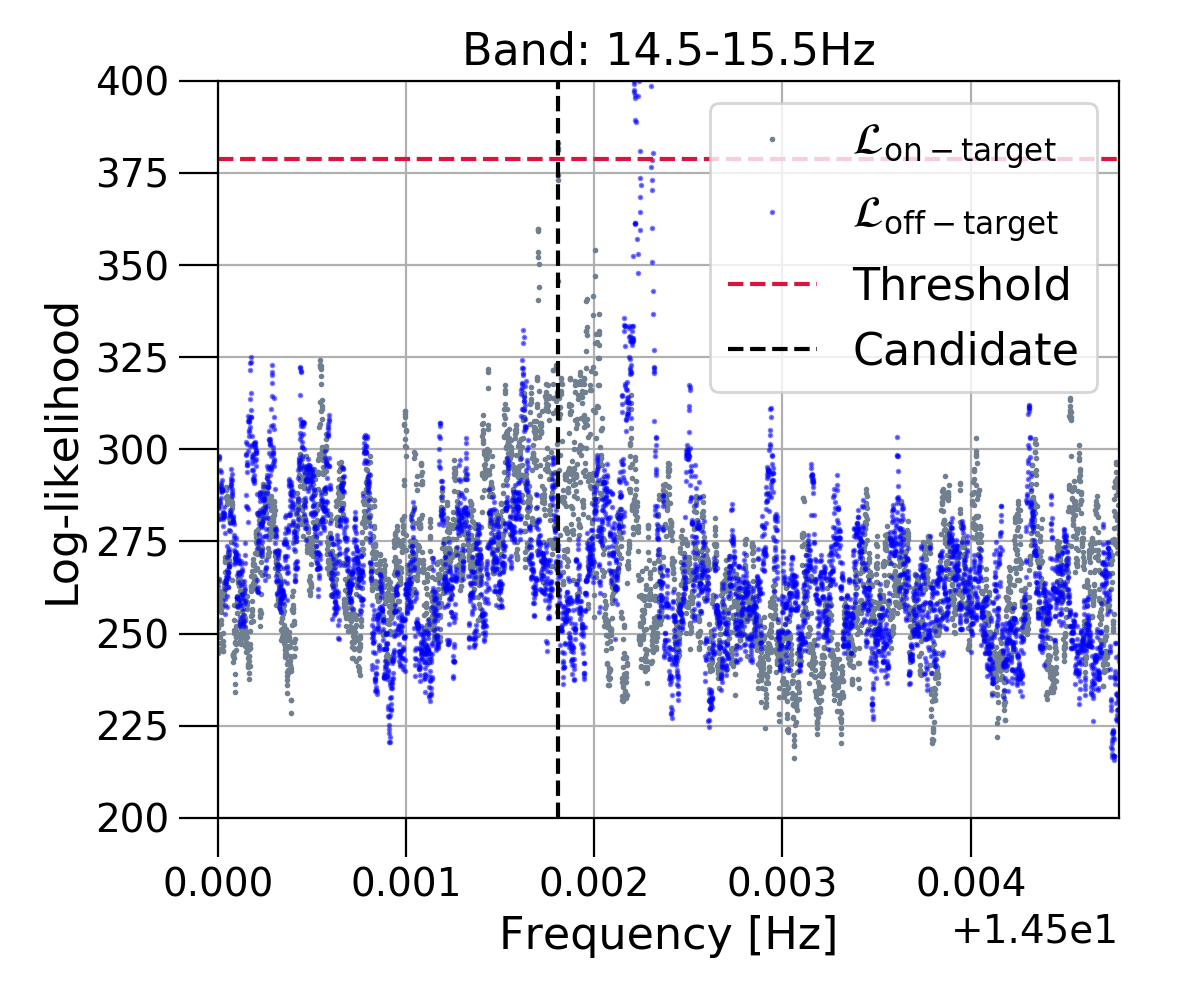}
    \end{subfigure}
    \caption{\small{Outcome of the single IFO (top) and off$-$target (bottom) vetoes for the first survivor in the 14.5$-$15.5Hz sub$-$band of PSR J1831$-$0953, laid out as in Fig.~\ref{fig:PSR_J0532+2200_candidates}.}}
     \label{fig:PSR_J1831-0952_candidates_1}
\end{figure}

\begin{figure}[h!]
    \begin{subfigure}{0.33\textwidth}
    \centering
    \includegraphics[trim=0.5cm 0.1cm 0.5cm 0.85cm, clip=true, width=1.0\linewidth]{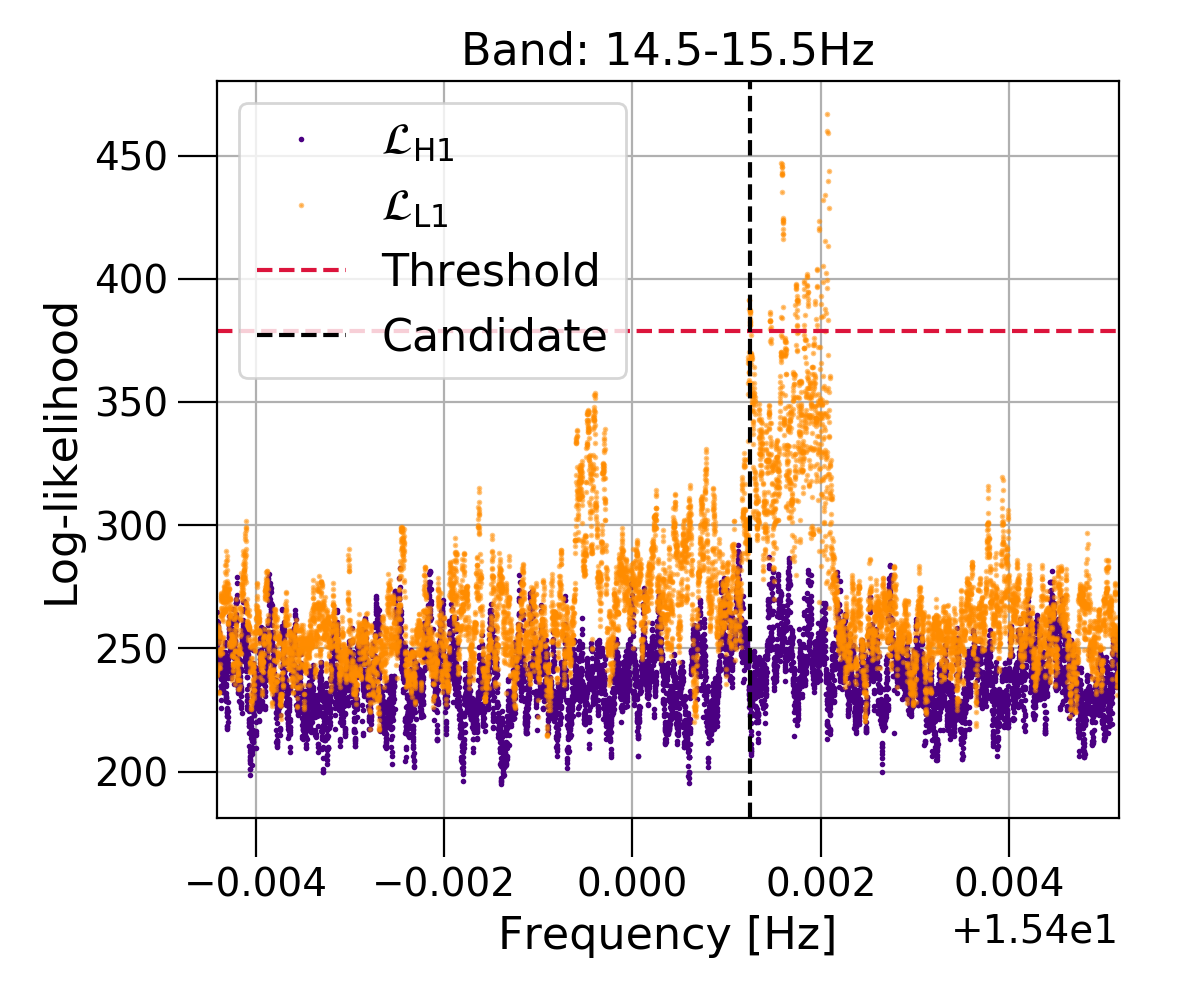}
    \end{subfigure}
    \begin{subfigure}{0.33\textwidth}
    \centering
    \includegraphics[trim=0.5cm 0.1cm 0.5cm 0.85cm, clip=true,width=1.0\linewidth]{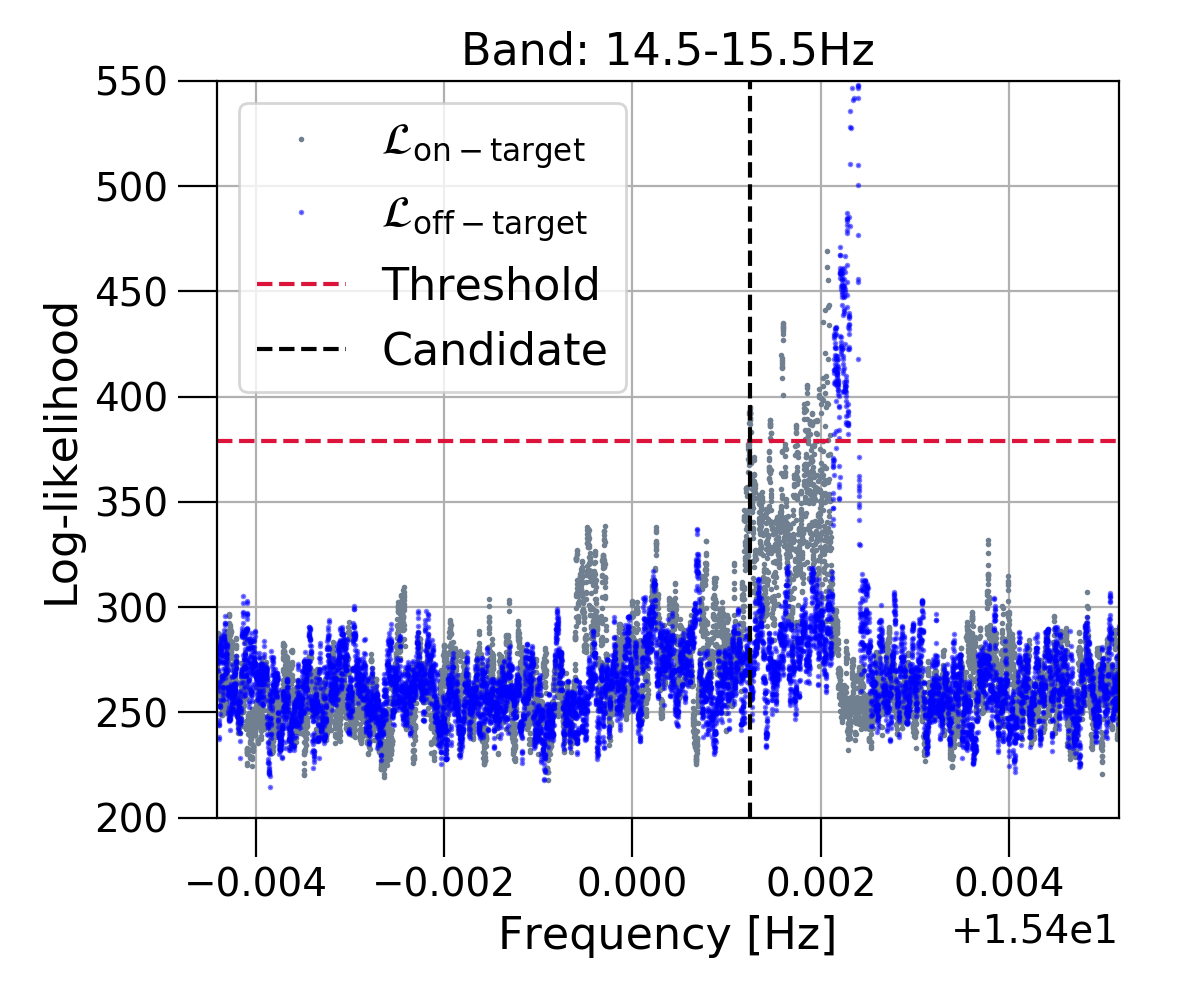}
    \end{subfigure}
    \caption{\small{Outcome of the single IFO (top) and off$-$target (bottom) vetoes for the second survivor in 14.5$-$15.5Hz sub$-$band of PSR J1831$-$0953, laid out as in Fig.~\ref{fig:PSR_J0532+2200_candidates}.}}
     \label{fig:PSR_J1831-0952_candidates_2}
 \end{figure}

\begin{figure}[h!]
    \begin{subfigure}{0.32\textwidth}
    \centering
    \includegraphics[trim=0.5cm 0.1cm 0.5cm 0.85cm, clip=true,width=1.0\linewidth]{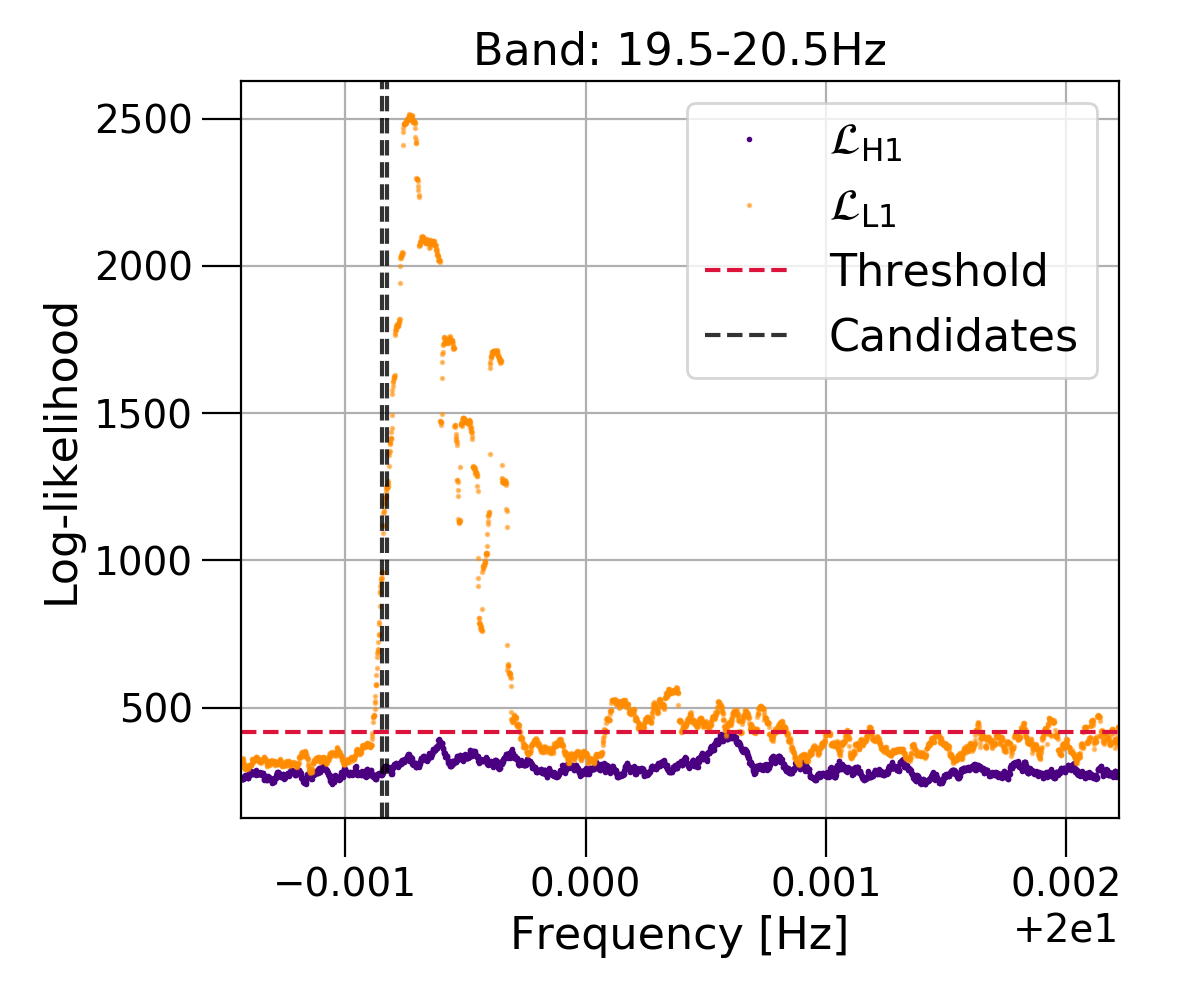}
    \end{subfigure}
    \begin{subfigure}{0.32\textwidth}
    \centering
    \includegraphics[trim=0.5cm 0.1cm 0.5cm 0.85cm, clip=true,width=1.0\linewidth]{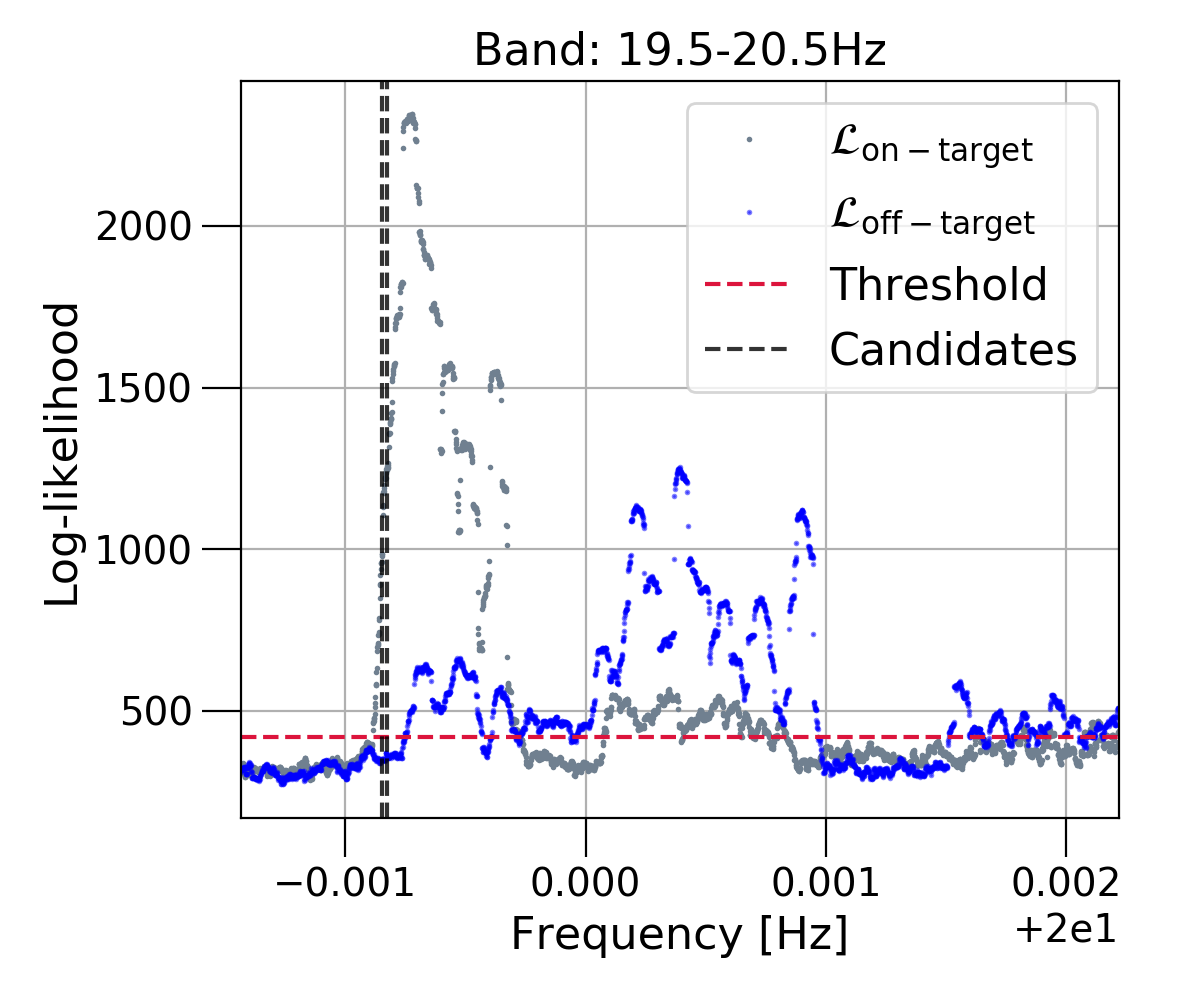}
    \end{subfigure}
    %and 19.5$-$20.5Hz (bottom) sub$-$bands 
    \caption{\small{Outcome of the single IFO (top) and off$-$target (bottom) vetoes for the two survivors in 19.5$-$20.5Hz sub$-$band of PSR J1831$-$1334, laid out as in Fig.~\ref{fig:PSR_J0532+2200_candidates}.}}
    \label{fig:PSR_J1831-0952_candidates_3}
\end{figure}

\noindent \textbf{PSR J1718$-$3825:} We look for a signal from this pulsar in the 13$-$14Hz ($f_*$), 17.5$-$18.5Hz ($4f_*/3$) and 26.5$-$27.5Hz ($2f_*$) sub$-$bands. The search returns 2051 candidates at $f_*$, 165 at $4f_*/3$ and 417 at $2f_*$. Only one candidate survives the data quality vetoes at $4f_*/3$, while none survive at $f_*$ or $2f_*$. 
\begin{enumerate}
    \item \textbf{17.5$-$18.5Hz}: The sole candidate in this sub$-$band is visibly more prominent in the Hanford data than at Livingston, as shown in the top panel of Fig.~\ref{fig:J1718-3825_candidates}. However, it has a comparable $\mathcal{L}$ in both detectors ($\mathcal{L}_{\rm{H1}}=438$, $\mathcal{L}_{\rm{L1}}=412$). The survivor also resurfaces in the off$-$target search with $\mathcal{L}_{\rm{off-target}}\approx~0.8\mathcal{L}_{\rm{on-target}}$. Although we keep this candidate for a follow$-$up study, its characteristics point towards a non$-$astrophysical origin. 
\end{enumerate}
% -----------------------------------------------------------------
\noindent \textbf{PSR J1831-0953:} We search for signals from this pulsar in 14.5$-$15.5Hz ($f_*$), 19.5$-$20.5Hz ($4f_*/3$) and 29$-$30Hz ($2f_*$) sub$-$bands. The search returns 1413 candidates at $f_*$, 263 at $4f_*/3$ and 81 at $2f_*$. Only two candidates survive the data quality vetoes at $f_*$ and $4f_*/3$, while none survive at $2f_*$. 
\begin{enumerate}
    \item \textbf{14.5$-$15.5Hz:} The candidate at 14.50181Hz has $\mathcal{L}_{\rm{L1}}>\mathcal{L}_{\rm{H1}}$ and $\mathcal{L}_{\rm{off-target}}\approx~0.85\mathcal{L}_{\mathrm{on-target}}$, as shown in Fig.~\ref{fig:PSR_J1831-0952_candidates_1}. Although we keep this candidate for a follow$-$up study, the relatively large $\mathcal{L}_{\rm{off-target}}$ points toward a non$-$astrophysical origin. The second survivor at 15.40125Hz is more prominent in the Livingston data ($\mathcal{L}_{\rm{L1}}=384$) than at Hanford ($\mathcal{L}_{\rm{H1}}=229$), as shown in Fig.~\ref{fig:PSR_J1831-0952_candidates_2}. It also resurfaces in the off$-$target search with $\mathcal{L}_{\rm{off-target}}\approx~0.8\mathcal{L}_{\rm{on-target}}$. Although the candidate passes the data quality vetoes, its characteristics point towards a non$-$astrophysical origin.
    \item \textbf{19.5$-$20.5Hz:} The two survivors in this sub$-$band have frequency around 19.999Hz. Both have $\mathcal{L}_{\rm{L1}}\gg\mathcal{L}_{\rm{H1}}$ in the single IFO search, as shown in the top panel of Fig.~\ref{fig:PSR_J1831-0952_candidates_3}. This could be explained by the fact that the Livingston detector is more sensitive than Hanford during O2 \cite{Collaboration_2019}. The candidates also show up in the off$-$target search with $\mathcal{L}_{\rm{off-target}}=~(0.53-0.65)\mathcal{L}_{\rm{on-target}}$. Since these candidates do not strictly satisfy any of the veto criteria, we keep them for a follow$-$up study. Despite this, their proximity to a spectral feature in the Livingston detector points towards a non$-$astrophysical origin. 
\end{enumerate}

\begin{figure}[h!]
    \begin{subfigure}{0.32\textwidth}
    \includegraphics[trim=0.5cm 0.1cm 0.5cm 0.85cm, clip=true, width=1.0\linewidth]{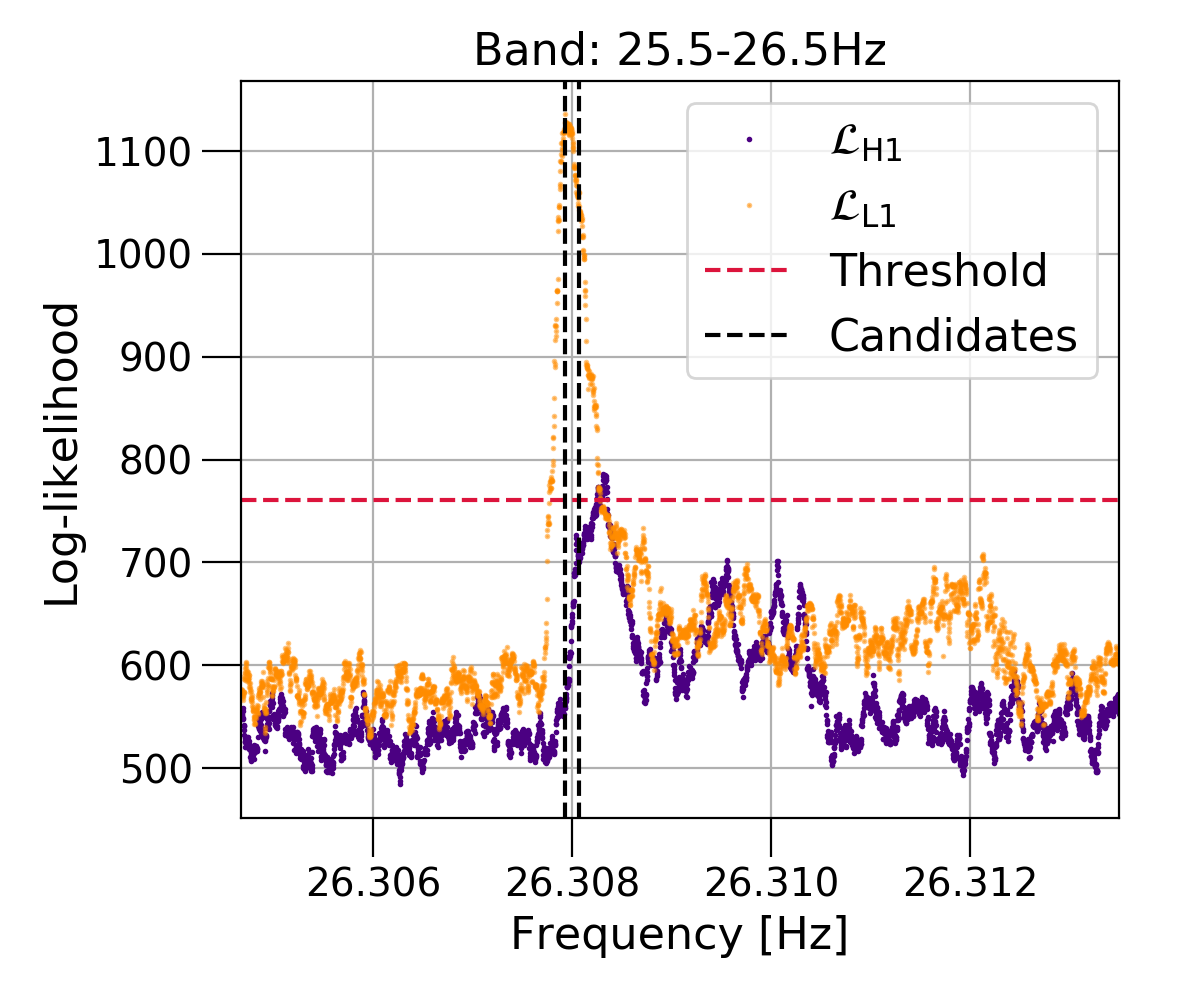}
    \end{subfigure}
    \begin{subfigure}{0.32\textwidth}
    \includegraphics[trim=0.5cm 0.1cm 0.5cm 0.85cm, clip=true,width=1.0\linewidth]{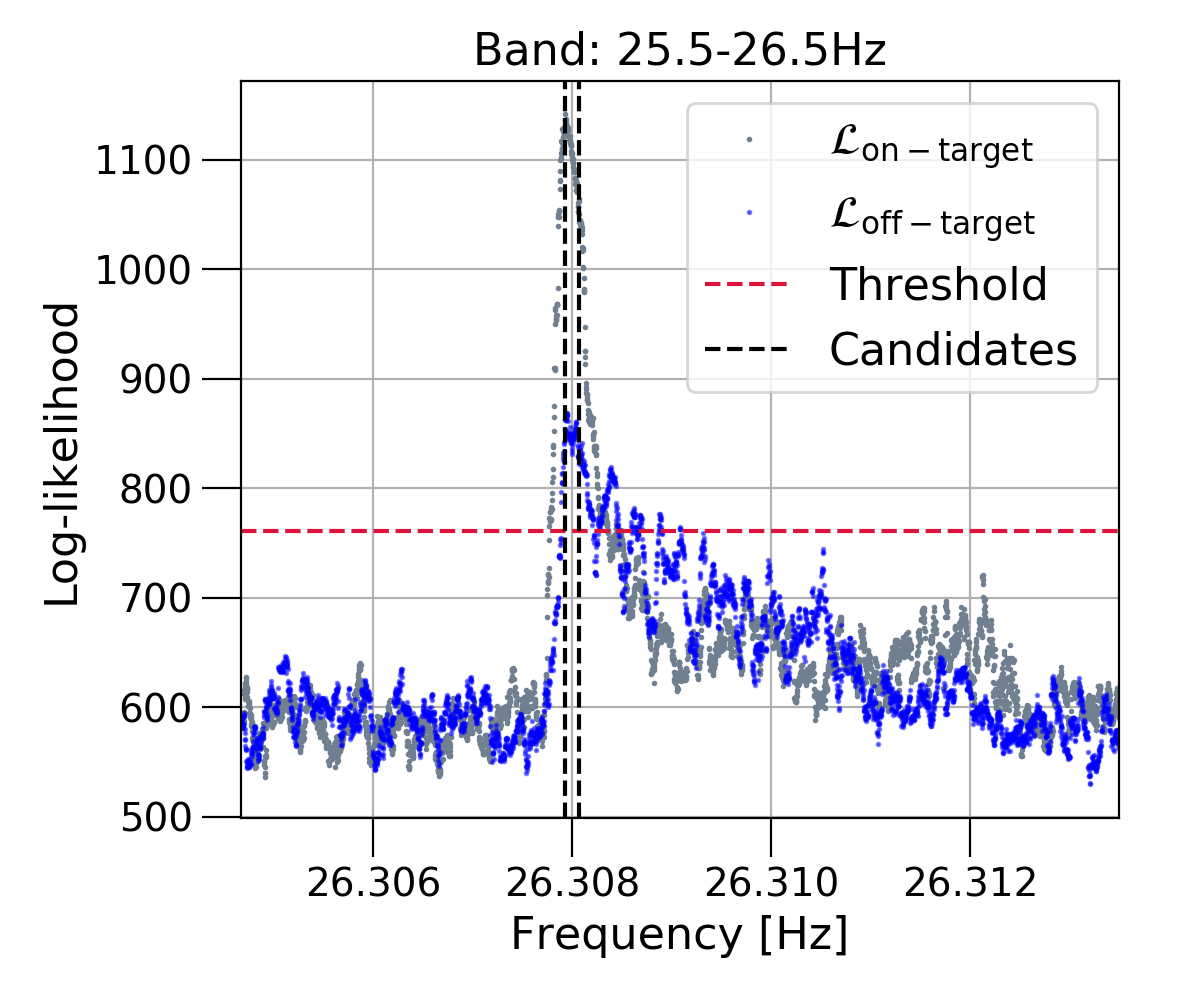}
    \end{subfigure}
\caption{\small{Outcome of the single IFO (top) and off$-$target (bottom) vetoes for the first two survivors in 25.5$-$26.5Hz sub$-$band of PSR J1849$-$0001. Both survivors have frequency around $26.3081\pm0.00014$Hz.}} 
\label{fig:PSR_J1849-0001_candidates_1}
\end{figure}

\begin{figure}[h!]
    \begin{subfigure}{0.32\textwidth}
    \includegraphics[trim=0.5cm 0.1cm 0.5cm 0.85cm,clip=true, width=1.0\linewidth]{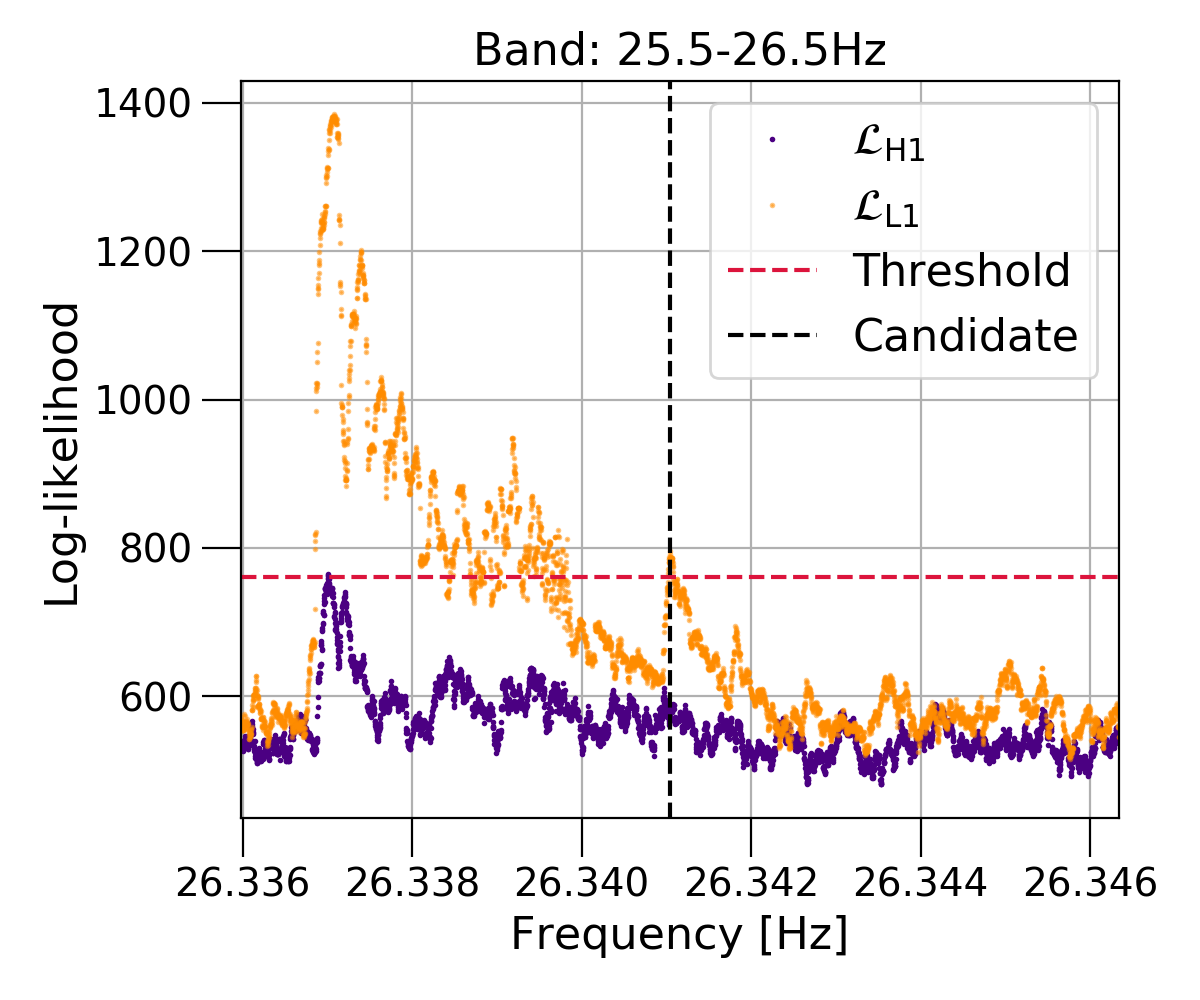}
    \end{subfigure}
    \begin{subfigure}{0.32\textwidth}
    \includegraphics[trim=0.5cm 0.1cm 0.5cm 0.85cm, clip=true,width=1.0\linewidth]{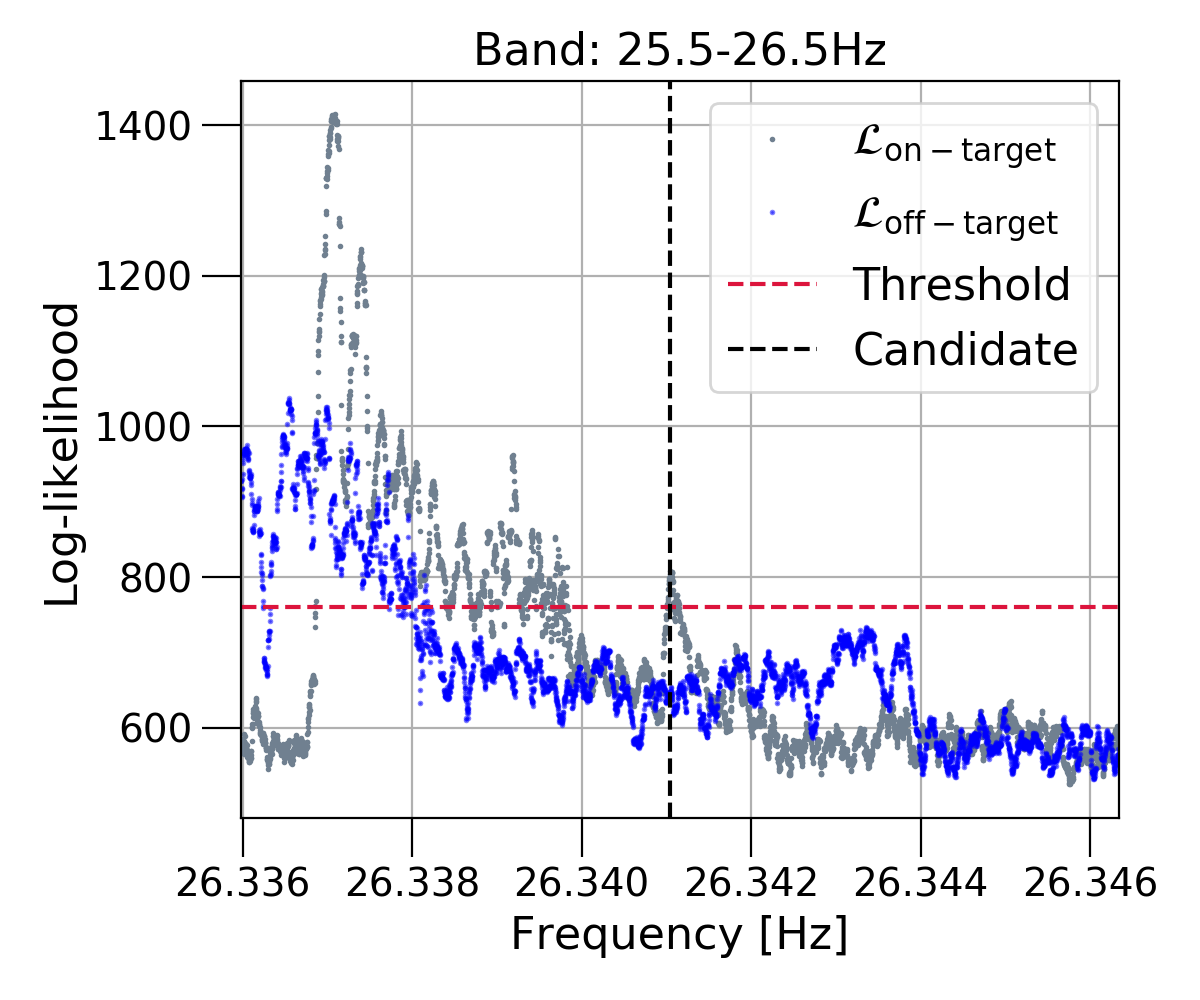}
    \end{subfigure}
\caption{\small{Outcome of the single IFO (top) and off$-$target (bottom) vetoes for the remaining survivor in 25.5$-$26.5Hz sub$-$band of PSR J1849$-$0001, laid out as in Fig.~\ref{fig:PSR_J0532+2200_candidates}.}}
    \label{fig:PSR_J1849-0001_candidates_2}
\end{figure}

% -----------------------------------------------------------------
\noindent \textbf{PSR J1849-0001:} Search for a signal from this pulsar returns 57 candidates in the 25.5$-$26.5Hz ($f_*$) sub$-$band, 61 candidates at 34$-$35Hz ($4f_*/3$) and five in the 51.5$-$52.5Hz ($2f_*$) sub$-$band. Only three candidates survive the data quality vetoes at $f_*$, while none survive at $4f_*/3$ or $2f_*$. 
\begin{enumerate}
    \item \textbf{25.5$-$26.5Hz:} The first two survivors in this sub$-$band have frequency around $26.3081\pm 0.00014$Hz. Both have $\mathcal{L}_{\rm{L1}}>~\mathcal{L}_{\rm{H1}}$ and $\mathcal{L}_{\rm{L1}}\approx~\mathcal{L}_{\rm{2ifo}}$, as shown in the top panel of Fig.~\ref{fig:PSR_J1849-0001_candidates_1}. These candidates also survive the off$-$target veto since they return $\mathcal{L}_{\rm{off-target}}\approx~(0.75-0.85)\mathcal{L}_{\rm{on-target}}$. However, their characteristics point towards a non$-$astrophysical origin. The remaining candidate at 26.3410Hz is more prominent in the Livingston data ($\mathcal{L}_{\rm{L1}}=797$) than at Hanford ($\mathcal{L}_{\rm{H1}}=573$), as shown in Fig.~\ref{fig:PSR_J1849-0001_candidates_2}. It also survives the off$-$target veto as it has $\mathcal{L}_{\rm{off-target}} \approx~0.88\mathcal{L}_{\rm{on-target}}$. Despite this, the characteristics of the survivor as well as its proximity to a spectral feature in the Livingston detector points towards a non$-$astrophysical origin.  
\end{enumerate}

% -----------------------------------------------------------------
\section{Conclusion} \label{sec:Conclusion}
In this paper, we present a search for CW signals from ten pulsars selected on the basis of their high$-$energy emissions and association with TeV sources. The search uses publicly available data from the O2 run of the aLIGO detectors. We use a method which combines the maximum$-$likelihood $\mathcal{F}-$statistic with a HMM to efficiently track the secular spin$-$down and stochastic spin wandering of a CW signal. In addition to being fast, HMM has the ability to search a much wider set of $f_0(t)$ histories than a coherent search with a Taylor series phase model, noting that the electromagnetic pulsations and CW signal are not necessarily locked together.\\

For each target, we scan 1Hz sub$-$bands around $f_*,\ 4f_*/3$ and $2f_*$ and omit bands below 10Hz. Additionally, we choose a coherence timescale $T_{\rm{drift}}={\rm{min}}(T_{\rm{drift}}',T_{\rm{drift}}'')$ to ensure that the signal frequency falls within one frequency bin over a single coherent step. For all targets chosen for this search, the duration of the coherent segments is limited by the secular spin$-$down ($T_{\rm{drift}}'$) and not the stochastic spin wandering ($T_{\rm{drift}}''$). Despite this, it is vital to include stochastic spin wandering in the GW phase model because the CW signal may not be locked to the electromagnetic pulsations; the spin of the gravitational$-$wave$-$emitting quadrupole may wander invisibly a lot more than the electromagnetically timed solid crust.\\
% even when it contributes less than secular spin$-$down to the measured electromagnetic phase residuals, 

The search returns a total of 5,256 candidates across 24 sub$-$bands. Only 12 candidates survive the three data quality vetoes outlined in Section \ref{subsec:Vetoes}. Follow$-$up analysis indicates that seven candidates have $\mathcal{L}_{\rm{off-target}}=~(0.75-0.89)\mathcal{L}_{\rm{on-target}}$ and a single IFO log$-$likelihood (i.e., $\mathcal{L}_{\rm{H1}}$ or $\mathcal{L}_{\rm{L1}}$) which is comparable to $\mathcal{L}_{\rm{on-target}}$. These survivors are marked with an asterisk in Table \ref{tab:survivors}. Although these candidates survive the vetoes, their characteristics point towards a non$-$astrophysical origin. The remaining five candidates only exceed $\mathcal{L}_{\rm{th}}$ in the Livingston data and have $\mathcal{L}_{\rm{off-target}}\leq~0.7\mathcal{L}_{\rm{on-target}}$. This is particularly true for the two survivors in 19.5$-$20.5Hz sub$-$band of PSR J1831$-$0953, both of which have $\mathcal{L}_{\rm{L1}}\approx~\mathcal{L}_{\rm{on-target}}$ and $\mathcal{L}_{\rm{H1}}<\mathcal{L}_{\rm{th}}$. However, we cannot reject these candidates purely on the basis of the outcome of single IFO veto, as Livingston is more sensitive than the Hanford detector during O2~\cite{Collaboration_2019}.\\ 

A dedicated follow$-$up study using the method outlined here, but with data from the O3 run and future observing runs will provide further clarity on the nature of the remaining candidates. While we argue that the 12 candidates identified here are likely non$-$astrophysical in origin, they are potential signals and thus special care should be taken in following up on them. If they do prove to be from instrumental noise, future work should also consider additional vetoes that can be used to effectively rule out such artifacts. \\

A follow$-$up O3 search using the HMM will also provide an opportunity to undertake an injection study to determine upper limits. We do not attempt the exercise here because the phase model used in this paper (a biased random walk defined by the HMM's transition probabilities) differs markedly from the phase models assumed in previous coherent CW searches for these targets (a Taylor series guided by the radio ephemeris). As upper limits are always conditional on the signal model used in the search, it is difficult to make an equivalent comparison. If the true GW signal is exactly locked (or nearly so) to the radio ephemeris, then a coherent search is more sensitive than the semi$-$coherent HMM search by a factor $\sim N_s^{1/4}$ in the characteristic wave strain \cite{Dhurandhar_2008,Aasi_2015_sideband,ScoX1_2017,ScoX1_2019}. However, if the GW signal is displaced from the radio ephemeris and experiences significant spin wandering, a coherent search may miss the signal entirely, and the upper limits from the semi$-$coherent HMM become decisive. In either scenario, the next significant step of placing constraints on the emission mechanism will be achieved once O3 data have been analysed. \\

%it is hard to make general statements about upper limits until a CW signal is actually detected and we have an empirical understanding of the GW emission mechanism as they are conditional on the phase model assumed.

%\section*{Acknowledgments}
\begin{acknowledgments}
The authors would like to express their gratitude towards Ling Sun, Meg Millhouse, Hannah Middleton, Patrick Meyers, Lucy Strang and Julian Carlin for numerous discussions and their ongoing support throughout this project. Additionally, we extend our special thank you to Prof. Gavin Rowell for discussions of the various HESS targets. This research is supported by the Australian Research Council Centre of Excellence for Gravitational Wave Discovery (OzGrav) with project number CE170100004. This work was performed on the OzSTAR national facility at Swinburne University of Technology. The OzSTAR program receives funding in part from the Astronomy National Collaborative Research Infrastructure Strategy (NCRIS) allocation provided by the Australian Government. This research has made use of data, software and/or web tools obtained from the Gravitational Wave Open Science Center (\url{https://www.gw-openscience.org/} ), a service of LIGO Laboratory, the LIGO Scientific Collaboration and the Virgo Collaboration. LIGO Laboratory and Advanced LIGO are funded by the United States National Science Foundation (NSF) as well as the Science and Technology Facilities Council (STFC) of the United Kingdom, the Max-Planck-Society (MPS), and the State of Niedersachsen/Germany for support of the construction of Advanced LIGO and construction and operation of the GEO600 detector. Additional support for Advanced LIGO was provided by the Australian Research Council. Virgo is funded, through the European Gravitational Observatory (EGO), by the French Centre National de Recherche Scientifique (CNRS), the Italian Istituto Nazionale di Fisica Nucleare (INFN) and the Dutch Nikhef, with contributions by institutions from Belgium, Germany, Greece, Hungary, Ireland, Japan, Monaco, Poland, Portugal, Spain.

%This research has made use of data, software and/or web tools obtained from the Gravitational Wave Open Science Center , a service of LIGO Laboratory, the LIGO Scientific Collaboration and the Virgo Collaboration.
\end{acknowledgments}

\appendix
\section{Target parameters\label{Appendix:search_params}}
% -------------------------------------------------------
Tables \ref{Pulsar_J0534}-\ref{Pulsar_J1849} outline the parameters used to search for CW signals from ten pulsars. Pulsars are named in the table titles. For each target, we report the reference time relevant to the measured ephemeris and bounds of the sub$-$bands at $f_*$, $4f_*/3$ and $2f_*$. The coherence timescale $T_{\rm{drift}}$, number of transition steps $N_T$, frequency resolution $\triangle f_{\rm{drift}}$ and the spin$-$down rate measured via electromagnetic observations ($\dot{f_*}$) are also reported. Finally, we quote the average one$-$sided noise PSD $S_h(f_0)^{1/2}$ for each sub$-$band of each target.
% ---------------------------------------------------------

\begin{table}[h]
    \setlength{\tabcolsep}{2pt}
    \centering
    \caption{Search parameters for PSR J0534$+$2200}  \label{Pulsar_J0534}
    \begin{tabular}{l|ccc|l} 
    \hline 
    Parameters  & \multicolumn{3}{c|}{Search} & Units \\
    & $f_*$ & $4f_*/3$ & $2f_*$ &  \\\hline \hline
    Ref time & \multicolumn{3}{c|}{48442.5} & MJD \\
    Band & 29.5$-$30.5 & 39.5$-$40.5 & 59.5$-$60.5 & Hz \\
%    $\mathrm{T_{obs}}$ & 233.6 & 233.6 & 233.6 & days \\
    $T_{\rm{drift}}$ & 10.0 & 8.5 & 7 & hrs  \\
    $N_T$ & 560 & 659 & 801 & - \\ 
    $\triangle f_{\rm{drift}}$ & 1.3889 & 1.6340 & 1.9841 & $\times 10^{-5}$ Hz \\
   $\dot{f_*}$ &  -3.77535 & -5.03380 & -7.55070 & $\times10^{-10}$ Hz$\mathrm{s}^{-1}$ \\
    S$_h(f)^{1/2}$ & 2.4 & 1.3 & 0.86 & $\times 10^{-23}$ Hzs$^{-1}$\\ \hline
    \end{tabular}
\end{table}
% ---------------------------------------------------------
\begin{table}[h]
    \setlength{\tabcolsep}{2pt}
    \centering
    \caption{Search parameters for PSR J0835$-$4510}
    \begin{tabular}{l|ccc|l} 
    \hline 
    Parameters  & \multicolumn{3}{c|}{Search} & Units \\
    & $f_*$ & $4f_*/3$ & $2f_*$ &  \\\hline \hline
    Ref time & \multicolumn{3}{c|}{51559.319} & MJD \\
    Band & 11$-$12 & 14.5$-$15.5 & 22$-$23 & Hz \\
 %   $\mathrm{T_{obs}}$ & 233.6 & 233.6 & 233.6 & days \\
    $T_{\rm{drift}}$ & 49.5 & 43 & 35 & hrs  \\
    $N_T$ & 113 & 130 & 160 & - \\ 
    $\triangle f_{\rm{drift}}$ & 2.8058 & 3.2300 & 3.9683 & $\times 10^{-6}$ Hz \\
    $\dot{f_*}$ & -1.5666 & -2.0888 & -3.1332 & $\times 10^{-11}$ Hzs$^{-1}$ \\
    S$_h(f)^{1/2}$ & 2.5 & 0.48 & 0.060 & $\times 10^{-21}$ Hzs$^{-1}$\\ \hline
    \end{tabular}
\end{table}
% ---------------------------------------------------------
\begin{table}[h!]
    \setlength{\tabcolsep}{2pt}
    \centering
    \caption{Search parameters for PSR J1016$-$5857}
    \begin{tabular}{l|cc|l} 
    \hline 
    Parameters  & \multicolumn{2}{c|}{Search} & Units \\
    &  $4f_*/3$ & $2f_*$ &  \\\hline \hline
    Ref time & \multicolumn{2}{c|}{52717.00000} & MJD\\
    Band & 12$-$13 & 18$-$19 & Hz \\
 %   $\mathrm{T_{obs}}$  & 233.6 & 233.6 & days \\
    $T_{\rm{drift}}$  & 64.0 & 52.0 & hrs  \\
    $N_T$ & 88 & 107 & - \\ 
    $\triangle f_{\rm{drift}}$ & 2.1701 & 2.6455 & $\times 10^{-6}$ Hz \\
    $\dot{f_*}$ & -0.934620 & -1.40193 & $\times 10^{-11}$ Hzs$^{-1}$ \\
    S$_h(f)^{1/2}$  & 1.5 & 0.18 & $\times 10^{-21}$ Hzs$^{-1}$\\  \hline
    \end{tabular}
\end{table}
% ---------------------------------------------------------
\begin{table}[h!]
    \setlength{\tabcolsep}{2pt}
    \centering
    \caption{Search parameters for PSR J1357$-$6429}
    \begin{tabular}{l|c|l} 
    \hline 
    Parameters & Search & Units \\
    &  $2f_*$ &  \\\hline \hline
    Ref time & 52921.00000 & MJD \\
    Band & 11.5$-$12.5 & Hz \\
%    $\mathrm{T_{obs}}$ & 233.6 & days \\
    $T_{\rm{drift}}$ & 38.5 & hrs  \\
    $N_T$ & 146 &  \\ 
    $\triangle f_{\rm{drift}}$ & 3.6075 & $\times 10^{-6}$ Hz \\
    $\dot{f_*}$ & -2.610790 & $\times 10^{-11}$ Hzs$^{-1}$ \\
    S$_h(f)^{1/2}$ & 2.3 & $\times 10^{-21}$ Hzs$^{-1}$\\ \hline
    \end{tabular}
\end{table}
% ---------------------------------------------------------
\begin{table}[h!]
    \setlength{\tabcolsep}{2pt}
    \centering
    \caption{Search parameters for PSR J1420$-$6048}
    \begin{tabular}{l|ccc|l} 
    \hline 
    Parameters  & \multicolumn{3}{c|}{Search} & Units \\
    & $f_*$ & $4f_*/3$ & $2f_*$ &  \\\hline \hline
    Ref time & \multicolumn{3}{c|}{51600.00} & MJD \\
    Band & 14$-$15 & 19$-$20 & 29$-$30 & Hz \\
%    $\mathrm{T_{obs}}$ & 233.6 & 233.6 & 233.6 & days \\
    $T_{\rm{drift}}$ & 46.5 & 40 & 33 & hrs  \\
    $N_T$ & 121 & 140 & 170 &  \\ 
    $\triangle f_{\rm{drift}}$ & 2.9869 & 3.4722 & 4.2088 & $\times 10^{-6}$ Hz \\
    $\dot{f_*}$ & -1.78912 & -2.38549 & -3.57824 & $\times 10^{-11}$ Hzs$^{-1}$ \\
    S$_h(f)^{1/2}$ & 5.1 & 1.2 & 0.25 & $\times 10^{-22}$ Hzs$^{-1}$\\  \hline
    \end{tabular}
\end{table}
% ---------------------------------------------------------
\begin{table}[h!]
    \setlength{\tabcolsep}{2pt}
    \centering
    \caption{Search parameters for PSR J1513$-$5908 }
    \begin{tabular}{l|c|l} 
    \hline 
    Parameters  & Search & Units \\
     & $2f_*$ &  \\\hline \hline
    Ref time & 55336 & MJD \\
    Band & 12.5$-$13.5 & Hz \\
%    $\mathrm{T_{obs}}$  & 233.6 & days \\
    $T_{\rm{drift}}$ & 17 & hrs  \\
    $N_T$  & 330 &  \\ 
    $\triangle f_{\rm{drift}}$ & 8.1699 & $\times 10^{-6}$ Hz \\
    $\dot{f_*}$ & -1.33062112& $\times 10^{-10}$ Hzs$^{-1}$ \\
    S$_h(f)^{1/2}$ & 1.2 & $\times 10^{-21}$ Hzs$^{-1}$\\  \hline
    \end{tabular}
\end{table}
% ---------------------------------------------------------
\begin{table}[h!]
    \setlength{\tabcolsep}{2pt}
    \centering
    \caption{Search parameters for PSR J1718$-$3825}
    \begin{tabular}{l|ccc|l} 
    \hline 
    Parameters  & \multicolumn{3}{c|}{Search} & Units \\
    & $f_*$ & $4f_*/3$ & $2f_*$ &  \\\hline \hline
    Ref time & \multicolumn{3}{c|}{51184.000} & MJD \\
    Band & 13$-$14 & 17.5$-$18.5 & 26.5$-$27.5 & Hz \\
%    $\mathrm{T_{obs}}$ & 233.6 & 233.6 & 233.6 & days \\
    $T_{\rm{drift}}$ & 127.5 & 110.5 & 90 & hrs  \\
    $N_T$ & 44 & 51 & 62 &  \\ 
    $\triangle f_{\rm{drift}}$ & 1.0893 & 1.2569 & 1.5432 & $\times 10^{-6}$ Hz \\
    $\dot{f_*}$ & -2.371346 & -3.16179 & -4.742692 & $\times 10^{-12}$ Hzs$^{-1}$ \\
    S$_h(f)^{1/2}$ & 8.0 & 2.5 & 0.3 & $\times 10^{-22}$ Hzs$^{-1}$\\ \hline
    \end{tabular}
\end{table}
% ---------------------------------------------------------
\begin{table}[h!]
    \setlength{\tabcolsep}{2pt}
    \centering
    \caption{Search parameters for PSR J1826$-$1334}
    \begin{tabular}{l|cc|l} 
    \hline 
    Parameters  & \multicolumn{2}{c|}{Search} & Units \\
     & $4f_*/3$ & $2f_*$ &  \\\hline \hline
    Ref time & \multicolumn{2}{c|}{54200} & MJD \\
    Band     & 12.5$-$13.5 & 19$-$20        & Hz \\
%    $\mathrm{T_{obs}}$       & 233.6      & 233.6       & days \\
    $T_{\rm{drift}}$  & 63         & 51.5        & hrs  \\
    $N_T$           & 89         & 109         &  \\ 
    $\triangle f_{\rm{drift}}$ & 2.2046     & 2.6969      & $\times 10^{-6}$ Hz \\
    $\dot{f_*}$     & -0.97417325 & -1.4612599 & $\times 10^{-11}$ Hzs$^{-1}$ \\
    S$_h(f)^{1/2}$  & 1.2        & 0.12        & $\times 10^{-21}$ Hzs$^{-1}$\\ \hline
    \end{tabular}
%\hspace{5cm}  
\end{table}
% ---------------------------------------------------------
\begin{table}[h!]
    \setlength{\tabcolsep}{2pt}
    \centering
    \caption{Search parameters for PSR J1831$-$0952}
    \begin{tabular}{l|ccc|l} 
    \hline 
    Parameters  & \multicolumn{3}{c|}{Search} & Units \\
    & $f_*$ & $4f_*/3$ & $2f_*$ &  \\\hline \hline
    Ref time & \multicolumn{3}{c|}{52412.0000} & MJD \\
    Band     & 14.5$-$15.5 & 19.5$-$20.5 & 29$-$30 & Hz \\
%    $\mathrm{T_{obs}}$  & 233.6     & 233.6      & 233.6     & days \\
    $T_{\rm{drift}}$ & 145       & 125.5      & 102.5     & hrs  \\
    $N_T$            & 39        & 45         & 55        &  \\ 
    $\triangle f_{\rm{drift}}$  & 0.95785   & 1.1067     & 1.3550    & $\times 10^{-6}$ Hz \\
    $\dot{f_*}$      & -1.839595 & -2.452793  & -3.679190 & $\times 10^{-12}$ Hzs$^{-1}$ \\
    S$_h(f)^{1/2}$   & 4.8       & 1.0        & 0.2       & $\times 10^{-22}$ Hzs$^{-1}$\\   \hline
    \end{tabular}
\end{table}
% ---------------------------------------------------------
\begin{table}[h!]
    \setlength{\tabcolsep}{2pt}
    \centering
    \caption{Search parameters for PSR J1849$-$0001} \label{Pulsar_J1849}
    \begin{tabular}{l|ccc|l} 
    \hline 
    Parameters  & \multicolumn{3}{c|}{Search} & Units \\
    & $f_*$ & $4f_*/3$ & $2f_*$ &  \\\hline \hline
    Ref time & \multicolumn{3}{c|}{55535.285052933} & MJD \\
    Band     & 25.5$-$26.5 & 34$-$35  & 51.5$-$52.5 & Hz \\
%    $\mathrm{T_{obs}}$  & 233.6     & 233.6  & 233.6   & days \\
    $T_{\rm{drift}}$ & 63        & 55     & 44.5    & hrs  \\
    $N_T$            & 89        & 102    & 126     &  \\ 
    $\triangle f_{\rm{drift}}$  & 2.2046    & 2.5253 & 3.1210  & $\times 10^{-6}$ Hz \\
    $\dot{f_*}$      & -9.59      & -12.79  & -19.18   & $\times 10^{-12}$ Hzs$^{-1}$ \\
    S$_h(f)^{1/2}$   & 3.4       & 1.5    &  0.90   & $\times 10^{-23}$ Hzs$^{-1}$\\  \hline
    \end{tabular}
\end{table}
\newpage
% =============================================================================
\section{Spin$-$wandering timescale} \label{Appendix:SW_derivation}
 The strength of timing noise, inferred from timing the radio pulsation of a pulsar, is quantified by $\mathrm{\langle\triangle\Phi^2_{EM}\rangle^{1/2}}$, the rms phase residual separating the actual signal and the best fit Taylor series which typically contains terms up to and including first frequency derivatives \cite{Jones_2004}. For random walks in rotation phase, frequency and spin$-$down rate, the rms phase residual scales with the duration of the observation ($\tau$) as
\begin{equation}
    \mathrm{\langle\triangle\Phi^2_{EM}\rangle^{1/2}} = k\tau^{(\beta-1)/2}, \label{residual_PN}
\end{equation}
where $k$ is a normalisation parameter and $\beta$ is the exponent of a power$-$law PSD used to model the observed timing noise \cite{Shannon_2016,Parthasarathy_2019}. This power$-$law PSD is defined by
\begin{equation}
    P_{\rm{red}}(f) = \frac{A^2_{\rm{red}}}{12\pi^2}\left(\frac{f}{f_{\rm{yr}}}\right)^{-\beta}, \label{eq:power_law_spectrum}
\end{equation}

where, $A_{\rm{red}}$ is the red$-$noise amplitude in units of $\rm{yr}^{3/2}$ and $f_{\rm{yr}}$ is the reference frequency of 1 cycle per year. \\

In this HMM search, the stochastic spin wandering timescale ($T_{\rm{drift}}''$) is chosen such that the signal frequency falls in one frequency bin during a single coherent step. This sets the frequency resolution of the search $\left(\triangle f_{\rm{drift}}\right)$ to be 
\begin{equation}
\triangle f_{\rm{drift}} = \frac{1}{2(T''_{\rm{drift}})}. \label{TN_drift}
\end{equation}
We also require the change in the signal frequency due to the stochastic spin wandering over [$t,t+T''_{\rm{drift}}$] to satisfy 
\begin{equation}
    \triangle f_{\rm{drift}}\geq\langle\triangle f_{\rm{EM}}^2\rangle^{1/2} ,
\end{equation}
where $\langle\triangle f_{\rm{EM}}^2 \rangle^{1/2}$ denotes the rms frequency residual used to estimate the scale of frequency wandering. We obtain an estimate for $\langle\triangle f_{\rm{EM}}^2 \rangle^{1/2}$ by differentiating Eq. \ref{residual_PN} with respect to $\tau$ \cite{Cordes_1980};
\begin{equation}
\langle\triangle f^2_{\rm{EM}}\rangle^{1/2} = \frac{k(\beta-1)}{2}\ \tau^{(\beta-3)/2} \label{rms_frequency}.
\end{equation}

By combining Eqs. \ref{TN_drift} and \ref{rms_frequency}, we arrive at the following expression for $T''_{\rm{drift}}$, which depends on the parameters $k$ and $\beta$.
\begin{align}
     T''_{\rm{drift}} &\propto \left[\frac{1}{k(\beta-1)}\right]^{2/(\beta-1)} \label{TN_drift2}
\end{align}
 We estimate $T''_{\rm{drift}}$ for nine out of the ten pulsars in this study by first solving for parameter $k$ in Eq. \ref{residual_PN}. To do this, we estimate $\mathrm{\langle\triangle\Phi^2_{EM}\rangle^{1/2}}$ by
\begin{equation}
    \langle\triangle\Phi^2_{\mathrm{EM}}\rangle^{1/2} = \mathrm{Res}\ f_{\rm{EPOCH}}, \label{eq:PR_timing_res}
\end{equation}
where $\rm{Res}$ denotes the rms timing residuals (in seconds) accumulated over a reference epoch used in the timing solution and $f_{\rm{EPOCH}}$ denotes the star's rotation frequency at the mid$-$point of this epoch. This combined with an appropriate choice of value for the $\beta$ parameter is then used to obtain an estimate for the normalisation parameter $k$ and thus $T''_{\rm{drift}}$.\\
  
  We cannot find timing residuals for PSR J1513$-$5908 in the literature and thus estimate it using the fit parameters from its power$-$law PSD \cite{Parthasarathy_2019} 
\begin{equation}
    \mathrm{Res} = \sqrt{\frac{A_{\rm{red}}^2}{12\pi^2}\frac{1}{(\beta-1)}\left(\frac{\tau^{\beta-1}}{{\tau_{\rm{yr}}}^{\beta}}\right)}, \label{eq:PR_power_law}
\end{equation}
where, $A_{\rm{red}}$ and $\beta$ are parameters as defined above, $\tau$ is the total observation time (in years) used to obtain a timing solution and $\tau_{\rm{yr}}$ is the reference timescale of 1 year. 
% $\Gamma(\beta)$ is the gamma function\footnote{\url{https://mathworld.wolfram.com/GammaFunction.html}}

The parameters used to estimate the stochastic spin wandering timescales for all pulsars are summarised in Table \ref{tab:Timing_noise_2}. 

\begin{table*}
    \setlength{\tabcolsep}{7pt}
    \renewcommand{\arraystretch}{1.2}
    \centering
    \caption{\small{Stochastic spin wandering timescale for 10 pulsars. Names of the pulsars and the reference epoch used for the timing solution are reported in columns 1 and 2, respectively. In column~3, we report the star's rotation frequency at the mid$-$point of the reference epoch. The total observation period $\tau$ (in years) and the accumulated timing residuals (in milliseconds) over this time period are reported in columns 4 and 5, respectively. The rms phase residuals $\mathrm{\langle\triangle\Phi^2_{EM}\rangle^{1/2}}$ are reported in column~6. The stochastic spin wandering timescale estimated using Eq. \ref{TN_drift2} is given in column~7.}}
    \begin{tabular}{c|c|c|c|c|c|c|c}
        \hline 
     Pulsar & Date Range        & $f_{\rm{EPOCH}}$   &  $\tau$  & Res & $\mathrm{\langle\triangle\Phi^2_{EM}\rangle^{1/2}}$ &  $\mathrm{T''_{\rm{drift}}}$ & Refs  \\ 
     $[$J2000]       & [MJD]     & [Hz]           &  [yrs] & [ms]  & [Cycles]  & [yrs]           &       \\ \hline \hline
    J0534$+$2200 & 48422$-$48452 & 29.9466088403 & 0.082 & 0.1 & 0.003 & 1.902  & \cite{Lyne_1993}\footnote{\url{http://www.jb.man.ac.uk/pulsar/crab/crab2.txt}}      \\
     
    J0835$-$4510 & 54200$-$55000   & 11.1856              & 2.19   & 440      & 4.922 &  0.609   &   \cite{Weltevrede_2010}   \\
     J1016$-$5857 & 51451$-$52002   & 9.31274229245(4)     & 1.51    & 4.3      & 0.0401 & 2.87        &  \cite{Camilo_2001}    \\
     J1357$-$6429 & 53487$-$53714   & 6.019298216(5)      & 0.62   & 3.6      & 0.0217 & 1.51      &  \cite{Zavlin_2007}     \\
     J1420$-$6048 & 54200$-$54600    & 14.66275659824       & 1.09    & 70     & 1.02 & 0.57   & \cite{Weltevrede_2010}     \\
     J1513$-$5908 & 54220$-$58469   & 6.59709182778(19)   & 11.6   & 298.2   & 1.97 & 4.52   &  \cite{Parthasarathy_2019}    \\
    J1718$-$3825 & 54200$-$55000    & 13.386880856760     & 2.19   & 27       & 0.361 &  1.73     & \cite{Weltevrede_2010}       \\
  J1826$-$1334 & 48486.4-52793.4 & 9.856   & 11.8    & 2267.62   & 22.35 & 1.79 &  \cite{Hobbs_2010}    \\
     J1831$-$0952 & 51302$-$53523  & 14.8661645038(14)    & 6      & 2.0 & 0.0297 & 13.04     &  \cite{Lorimer_2006}    \\
     J1849$-$0001 & 58166.4$-$58355.9  & 25.9590178660(19) & 0.52  & 0.525     & 0.0136 & 1.52      & \cite{Bogdanov_2019}    \\ \hline 
    \end{tabular} \label{tab:Timing_noise_2}
\end{table*}

\section{Off$-$target veto} \label{Appendix:Off-target_veto}
Here, we outline the procedure used to select an appropriate offset for the off$-$target veto. It is based on injecting signals into the aLIGO O2 data. To simplify the study, we use the same search parameters as the ones used in the 59.5$-$60.5Hz sub$-$band of PSR J0534$+$2200. The source and search parameters are outlined in Tables \ref{off-inject} and \ref{off-search}. We conservatively use a coherence time of 7hrs for this study since shorter coherence times exhibit less sensitivity to sky position, as the Earth experiences less diurnal motion.\\

\begin{table}[h!]
    \centering
    \setlength{\tabcolsep}{7pt}
    \caption{\small{Injection parameters for the off$-$target veto study.}}  \label{off-inject}
    \begin{tabular}{l|c|l} 
    \hline 
    Parameters  & Value & Units \\ \hline \hline
    $\cos{\iota}$ & 1 & - \\
    $\alpha_{\rm{source}}$ & [0, $2\pi$) & rad \\
    $\delta_{\rm{source}}$ & 0.3842248 & rad \\
    $h_0$ & 2.5$\times10^{-23}$ & - \\
    $T_{\rm{drift}}$ & 7 & hrs  \\
    $T_{\rm{obs}}$ & 233.6 & days \\ \hline
%    $f_{\rm{min}}$ & .... &  \\
    \end{tabular}
\end{table}

\begin{table}[h!]
    \centering
    \setlength{\tabcolsep}{7pt}
    \caption{\small{Search parameters for the off$-$target veto study. Both $\alpha_{\rm{offset}}$ and $\delta_{\rm{offset}}$ are set to 0 for an on$-$target search. For an off$-$target search, we experiment with three different offsets for $\alpha_{\rm{search}}$ while fixing $\delta_{\rm{offset}}$ to 0.003 rad (i.e., 10min).}}  \label{off-search}
    \begin{tabular}{l|c|l} 
    \hline 
    Parameters  & Value & Units \\ \hline \hline
    Band & 59.5$-$60.5 & Hz \\
    $\alpha_{\rm{search}}$ & $\alpha_{\rm{source}}$ + $\alpha_{\rm{offset}}$ & rad \\
    $\delta_{\rm{search}}$ & $\delta_{\rm{source}}$ + $\delta_{\rm{offset}}$ & rad \\
    $T_{\rm{drift}}$ & 7 & hrs  \\
    $N_T$ & 801 & - \\
    $\triangle f_{\rm{drift}}$ & 1.9841$\times 10^{-5}$ & Hz \\
    $\dot{f_*}$ & -7.55070$\times10^{-10}$ & Hz$\mathrm{s}^{-1}$ \\ \hline
   \end{tabular}
\end{table}

\begin{figure}
    \centering
    \includegraphics[width=0.47\textwidth]{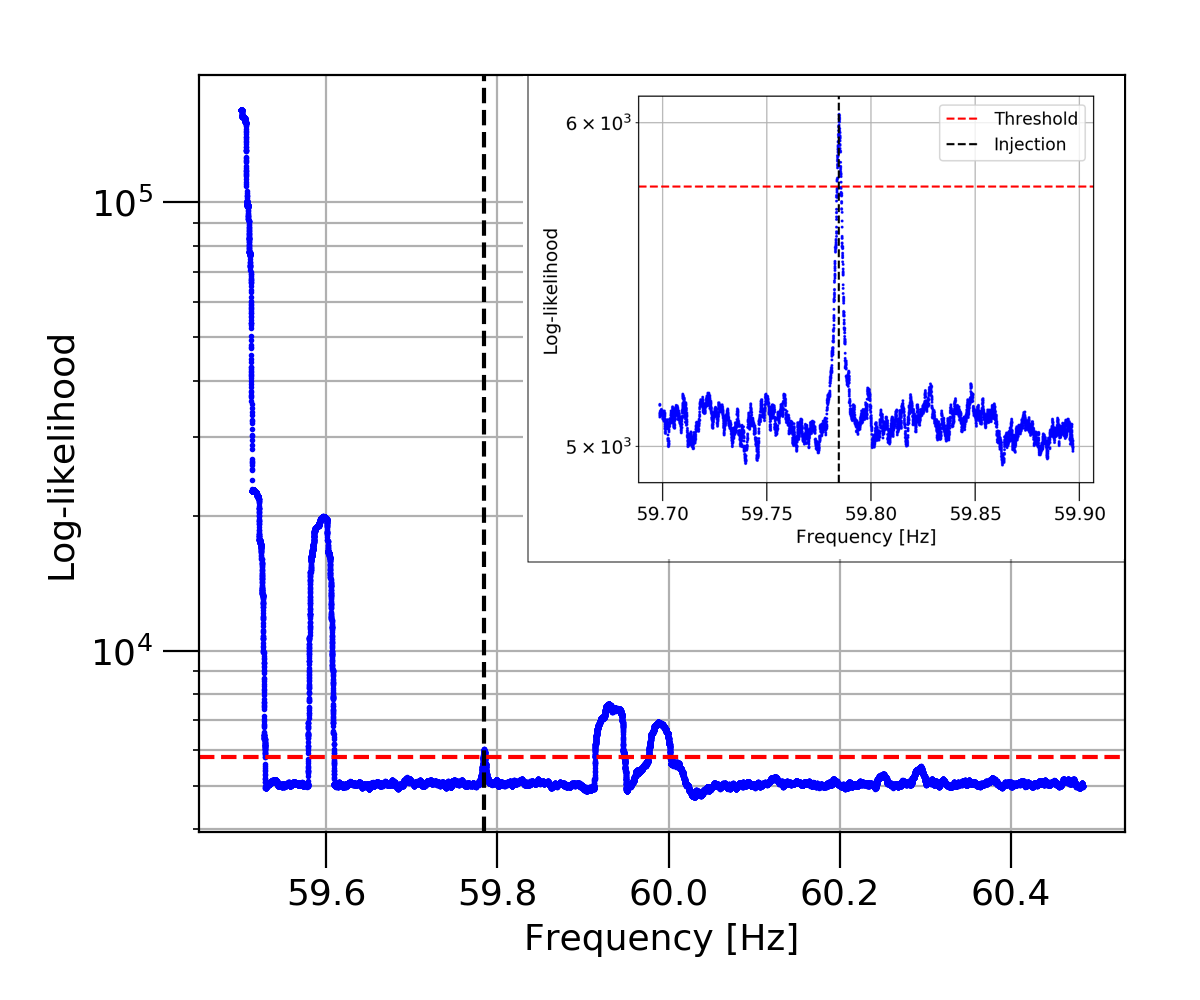}
    \caption{\small{Signal injected into the data from O2 run of the aLIGO detectors. The log$-$likelihood ($\mathcal{L}$) is plotted against the terminating frequency of associated Viterbi paths (blue dots). The location of the injected signal is indicated by the black vertical line. The red horizontal dashed line indicates the Gaussian threshold for this sub$-$band (i.e., $\mathcal{L}_{\rm{th}}$). Insert: A magnified view of the injected signal, laid out as in the main figure.}}
    \label{fig:off-target_inject}
\end{figure}

Here, we experiment with three different values of $\alpha_{\rm{offset}}$, the offset in right ascension of the search from source's location. These include $\alpha_{\rm{offset}}$ of 0.262 rad (1hr), 0.524 rad (2hrs) and 0.785 rad (3hrs). In each case, we set $\delta_{\rm{offset}}$ (i.e., the offset in search declination) to 0.003 rad (10min). Note that the $\delta_{\rm{offset}}\ll \alpha_{\rm{offset}}$ because the response of the two detectors changes with $\delta_{\rm{search}}$, which would artificially adjust $\mathcal{L}$ of the recovered signal \cite{Jaranowski_1998}. For each $\alpha_{\rm{offset}}$, we repeat the following procedure for 100 realisations. 
\begin{enumerate}
    \item \textbf{Generate a fake signal:} The parameters of the injected signal are outlined in Table \ref{off-inject}. We use $h_0 = 2.5\times10^{-23}$, which corresponds to the $95\%$ upper limit on the detectable signal ($h_0$) for this sub$-$band. Additionally, we generate a Viterbi path where the signal frequency follows a biased random walk as described by the transition matrix in Eq. $\ref{eq:transition_matrix}$ of Section \ref{subsec:HMM_framework}. 
    
    \item \textbf{Inject signal in LIGO data:} We use the \texttt{lalapps$\_$Makefakedata$\_$v4} tool in the LAL suite \cite{LALapps_2018} to inject fake signals into the Hanford and Livingston detector data. To randomise the search between each run, we sample $\alpha_{\rm{source}}$ (i.e., right ascension of the source) from a uniform distribution of values between [0, 2$\pi$) while fixing $\delta_{\rm{source}} = 0.3842248$ rad from EM measurements (i.e., $\rm{DEC}_{\rm{source}} =+20^\circ00'52.06''$).  % because the sensitivity of the depends on delta not alpha. 
    
    \item \textbf{Search for the injected signal:} We search for the injected signal in O2 data using a combination of the $\mathcal{F}-$statistic and Viterbi algorithm. As shown in Fig.~\ref{fig:off-target_inject}, this LIGO sub$-$band is contaminated by noise lines. Therefore, we choose to restrict the analysis to a 0.2Hz wide region around the injected signal frequency. The search is then performed at an on$-$target (i.e., at the same location as the injected signal) and off$-$target sky position. 
    
    \item \textbf{Compare results:} We compare log$-$likelihood and Viterbi path of the candidate from the on$-$ and off$-$target searches. The two Viterbi paths are classified as overlapping if the off$-$target path lies within the extended on$-$target path, which accounts for the frequency shift expected due to the Doppler modulation of the Earth.
  
\end{enumerate}

\begin{figure}[h!]
\begin{subfigure}{0.40\textwidth}
  \centering
  % include second image
  \includegraphics[width=1\linewidth]{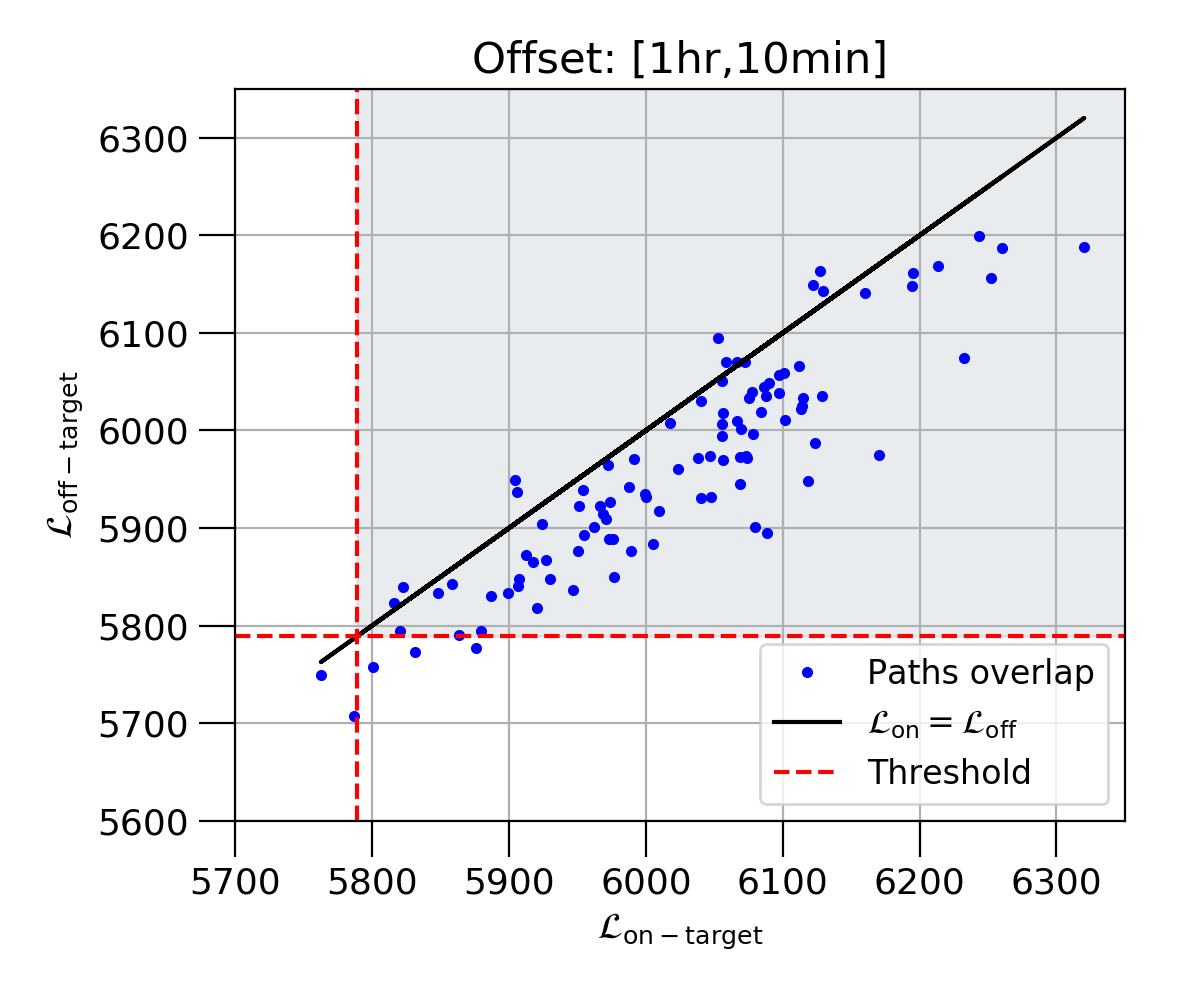}
\end{subfigure}
\begin{subfigure}{0.40\textwidth}
  \centering
  % include second image
  \includegraphics[width=1\linewidth]{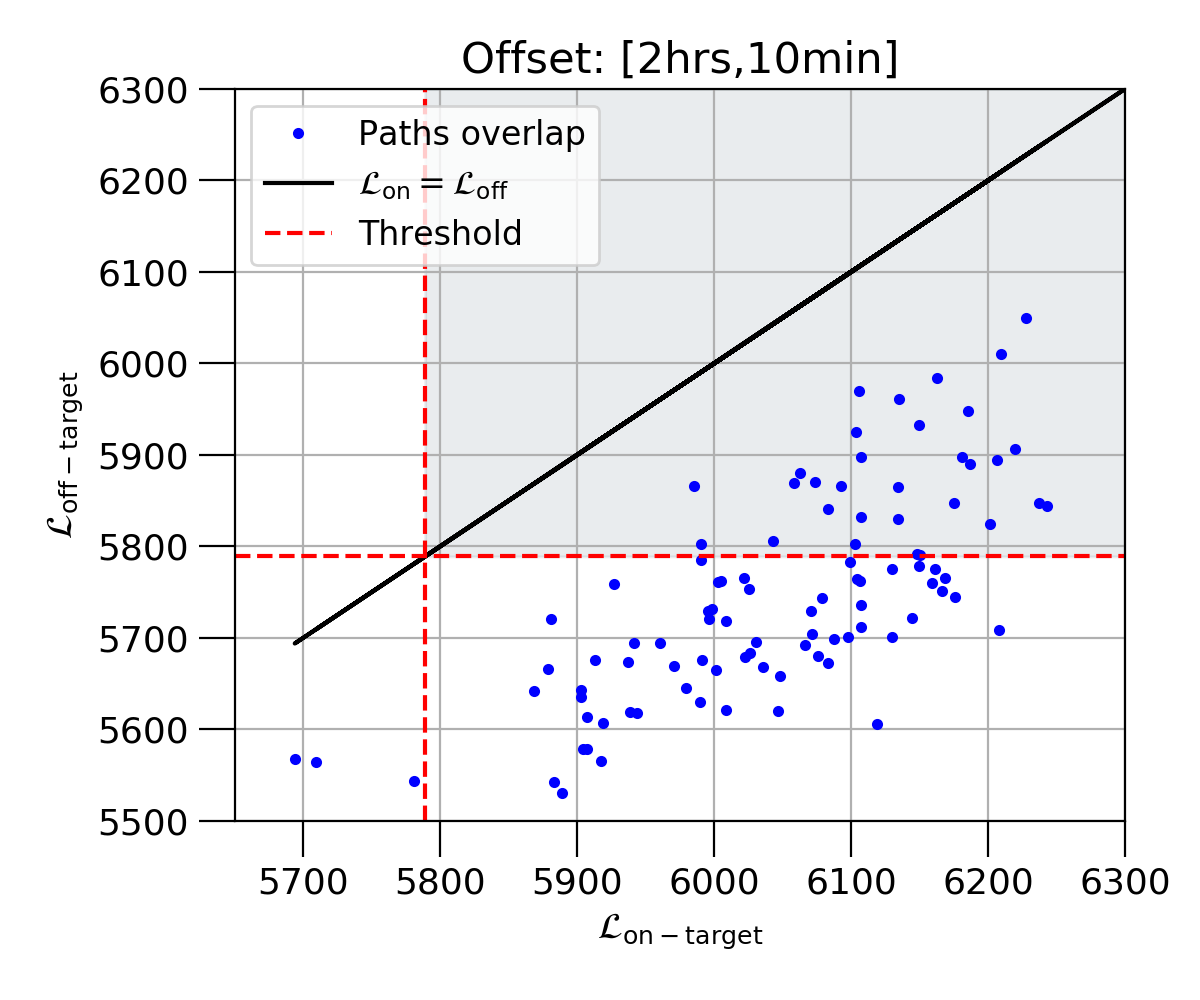}
\end{subfigure}
\begin{subfigure}{0.40\textwidth}
  \centering
  % include second image
  \includegraphics[width=1\linewidth]{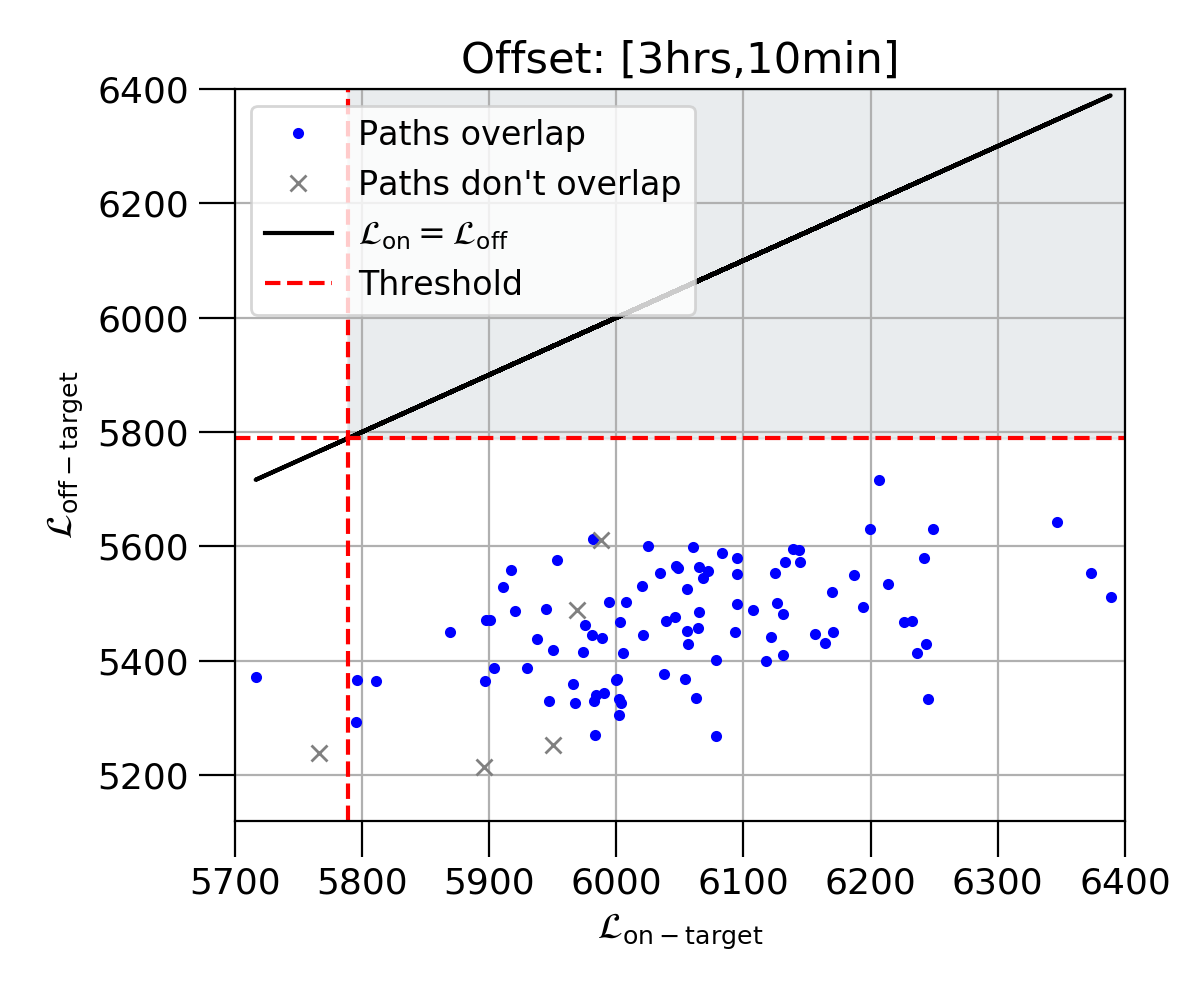}
\end{subfigure}
\caption{\small{Simulation results for the off$-$target veto study. In each case, we plot the off$-$target $\mathcal{L}$ against the on$-$target $\mathcal{L}$ for 100 realisations. The search (RA,DEC) is offset from the source location by [$1\rm{hr}, 10\rm{min}$] (top), [$2\rm{hr}, 10\rm{min}$] (middle) and [$3\rm{hr}, 10\rm{min}$] (bottom). The red vertical and horizontal dashed lines indicate the Gaussian log$-$likelihood ($\mathcal{L}_{\rm{th}}$) for this sub$-$band. The black diagonal line indicates the case where $\mathcal{L}_{\rm{on-target}}=\mathcal{L}_{\rm{off-target}}$. The crosses in the bottom plot indicate the Viterbi paths which do not overlap. The shaded gray region indicates the area where both on$-$ and off$-$target searches return candidates above the Gaussian threshold.}} \label{fig:Off-target_sim}
\end{figure}

The outcomes of the injection studies are presented in Fig.~\ref{fig:Off-target_sim}. We plot $\mathcal{L}_{\rm{off-target}}$ against $\mathcal{L}_{\rm{on-target}}$ for three cases. The top panel shows the results for the case where the search is offset from the source's location by ($1\rm{hr}, 10\rm{min}$) in right ascension and declination, respectively. The search returns candidates with $\mathcal{L}_{\rm{off-target}}>~\mathcal{L}_{\rm{th}}$ 95$\%$ of the time. This is problematic if the off$-$target veto criterion is based on comparing $\mathcal{L}_{\rm{off-target}}$ with $\mathcal{L}_{\rm{th}}$, as is the case in Ref.~\cite{Middleton_2020}. This offset and veto criterion would incorrectly reject astrophysical signals 95$\%$ of the time. Additionally, both on$-$ and off$-$target searches return comparable $\mathcal{L}$, as most of the data points lie along to the $\mathcal{L}_{\rm{on-target}}=\mathcal{L}_{\rm{off-target}}$ line. Hence we cannot safely discern between an astrophysical signal and a noise artifact in the detector, which would be expected to produce comparable $\mathcal{L}$ in on$-$ and off$-$target searches. \\ 
% which uses Viterbi score as a detection statistic

 The middle panel shows the results for the case where the off$-$target search is offset from the source's location by ($2\rm{hr}, 10\rm{min}$) in right ascension and declination. Approximately 67$\%$ of the above$-$threshold, on$-$target candidates are sub$-$threshold in the off$-$target search. Although this is significantly better than the above case, the off$-$target veto would incorrectly reject approximately 33$\%$ of the true astrophysical signals, provided the criterion is set to reject candidates with $\mathcal{L}_{\rm{off-target}}\geq~\mathcal{L}_{\rm{th}}$. None of the candidates in this search have $\mathcal{L}_{\rm{off-target}}>\mathcal{L}_{\rm{on-target}}$. This implies that this off$-$target offset would produce a noticeable (but not significant) change in $\mathcal{L}$ of an astrophysical candidate. Therefore, it may still be difficult to discern a true astrophysical signal from a noise artifact.\\ 
    
We show the results for a search offset of ($3\rm{hr}, 10\rm{min}$) in the bottom panel of Fig.~\ref{fig:Off-target_sim}. The crosses indicate the Viterbi paths which do not overlap since the recovered off$-$target path doesn't lie within the extended on$-$target path, which accounts for the Doppler modulation of the Earth. In three out of five cases, this occurs because $\mathcal{L}_{\rm{off-target}}$ of the injected signals is comparable to the background noise. Since the algorithm backtracks the Viterbi path with the highest $\mathcal{L}$ in a 0.2Hz wide region around the injected signal frequency, a non$-$overlapping Viterbi path is recovered in the off$-$target search. In the remaining two cases, the recovered path lies just outside the extended on$-$target path and thus deemed to be non$-$overlapping. The results here clearly indicate that this offset should reduce $\mathcal{L}_{\rm{off-target}}\approx 0.9\mathcal{L}_{\rm{on-target}}$ for an astrophysical signal. This should be sufficient to easily distinguish between an astrophysical signal and a noise artifact, which is expected to produce comparable $\mathcal{L}$ in both on$-$ and off$-$target searches. Additionally, we can also compare $\mathcal{L}_{\rm{off-target}}$ with $\mathcal{L}_{\rm{th}}$ for candidates that are marginally above$-$threshold in the on$-$target search. This is because the offset should yield $\mathcal{L}_{\rm{off-target}}<\mathcal{L}_{\rm{th}}$ for signals close to the Gaussian threshold. \\

Based on these results, we conclude that an offset of $3\rm{hrs}$ in right ascension and $10\rm{min}$ in declination for the off$-$target search should produce a noticeable change in $\mathcal{L}$ of an astrophysical signal. Our most conservative estimates suggest that with this offset, a real signal would produce $\mathcal{L}_{\rm{off-target}}\leq0.9\mathcal{L}_{\rm{on-target}}$. Additionally, searches with longer coherence times would be expected to produce significantly lower $\mathcal{L}_{\rm{off-target}}$. Therefore, we choose to reject candidates with $\mathcal{L}_{\rm{off-target}}\geq0.9\mathcal{L}_{\rm{on-target}}$ in this study. This should reject noise lines without unduly affecting the false alarm or false dismissal probabilities. 

% --------------------------------------------------
\section{Candidates} \label{Appendix:Results_full}
\begin{figure}[h]
\begin{subfigure}{0.47\textwidth}
  \centering
  % include second image
  \includegraphics[width=0.9\linewidth]{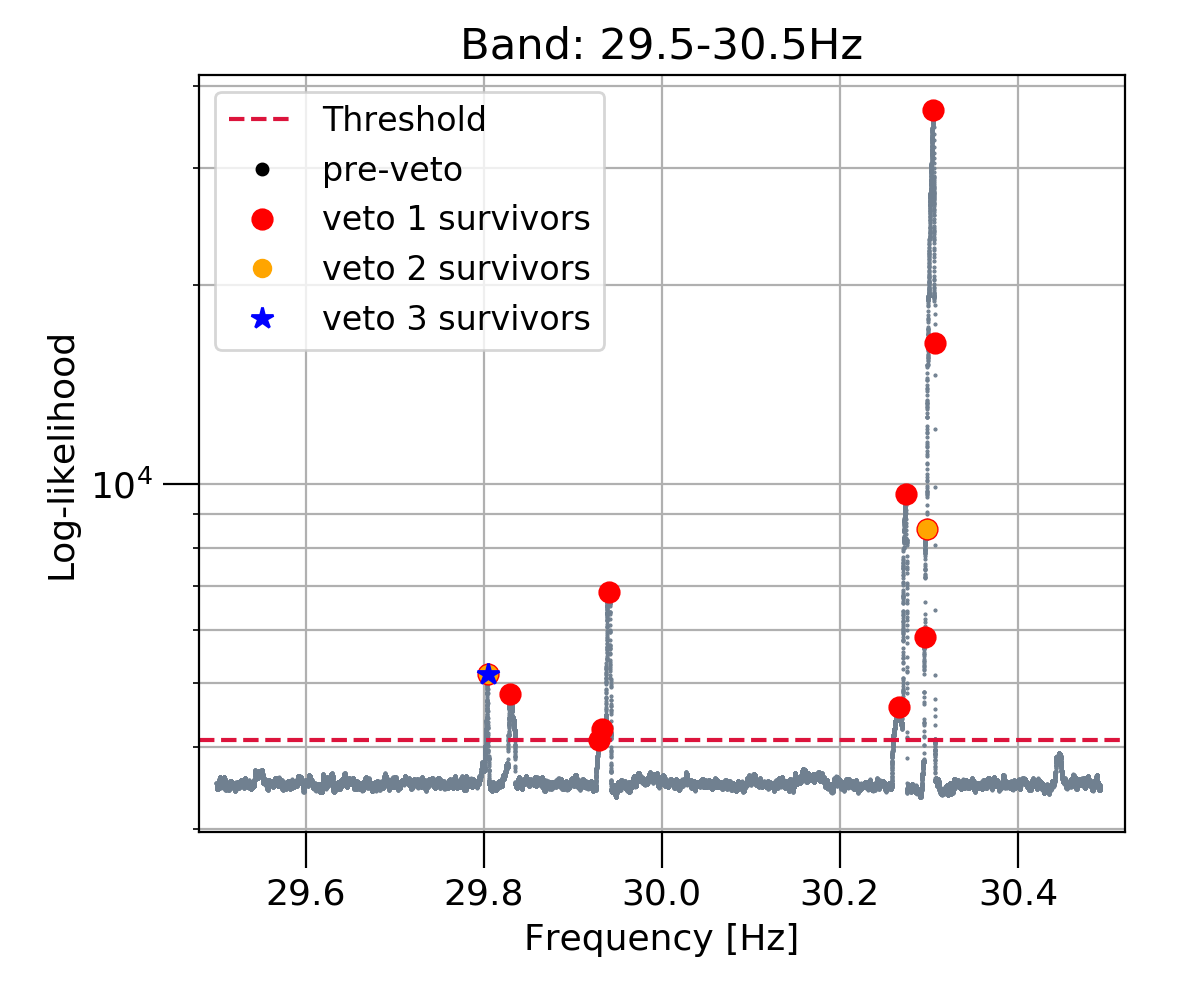}
\end{subfigure}
\begin{subfigure}{0.47\textwidth}
  \centering
  % include first image
  \includegraphics[width=0.9\linewidth]{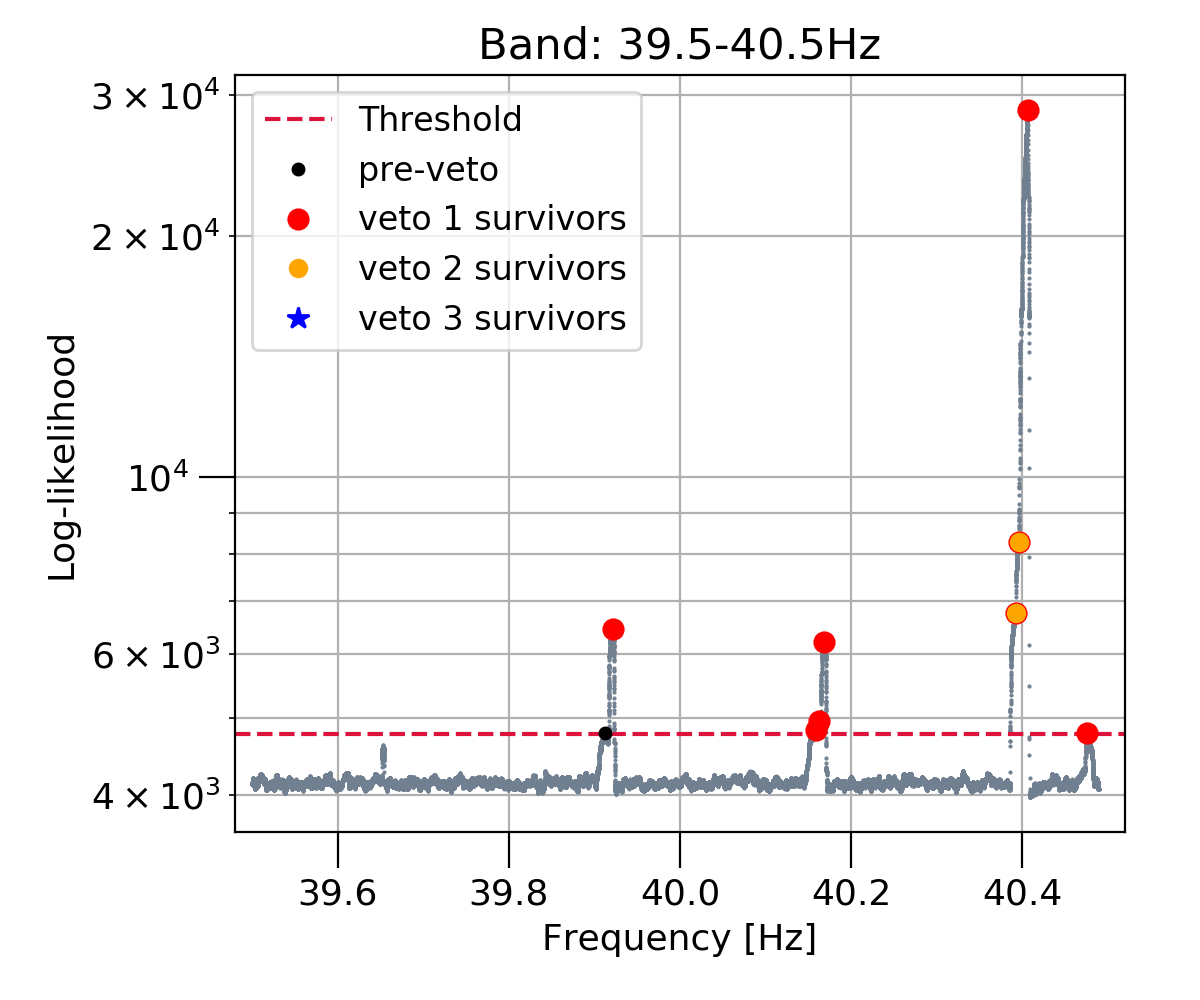}
\end{subfigure}
\begin{subfigure}{0.47\textwidth}
  \centering
  % include second image
  \includegraphics[width=0.9\linewidth]{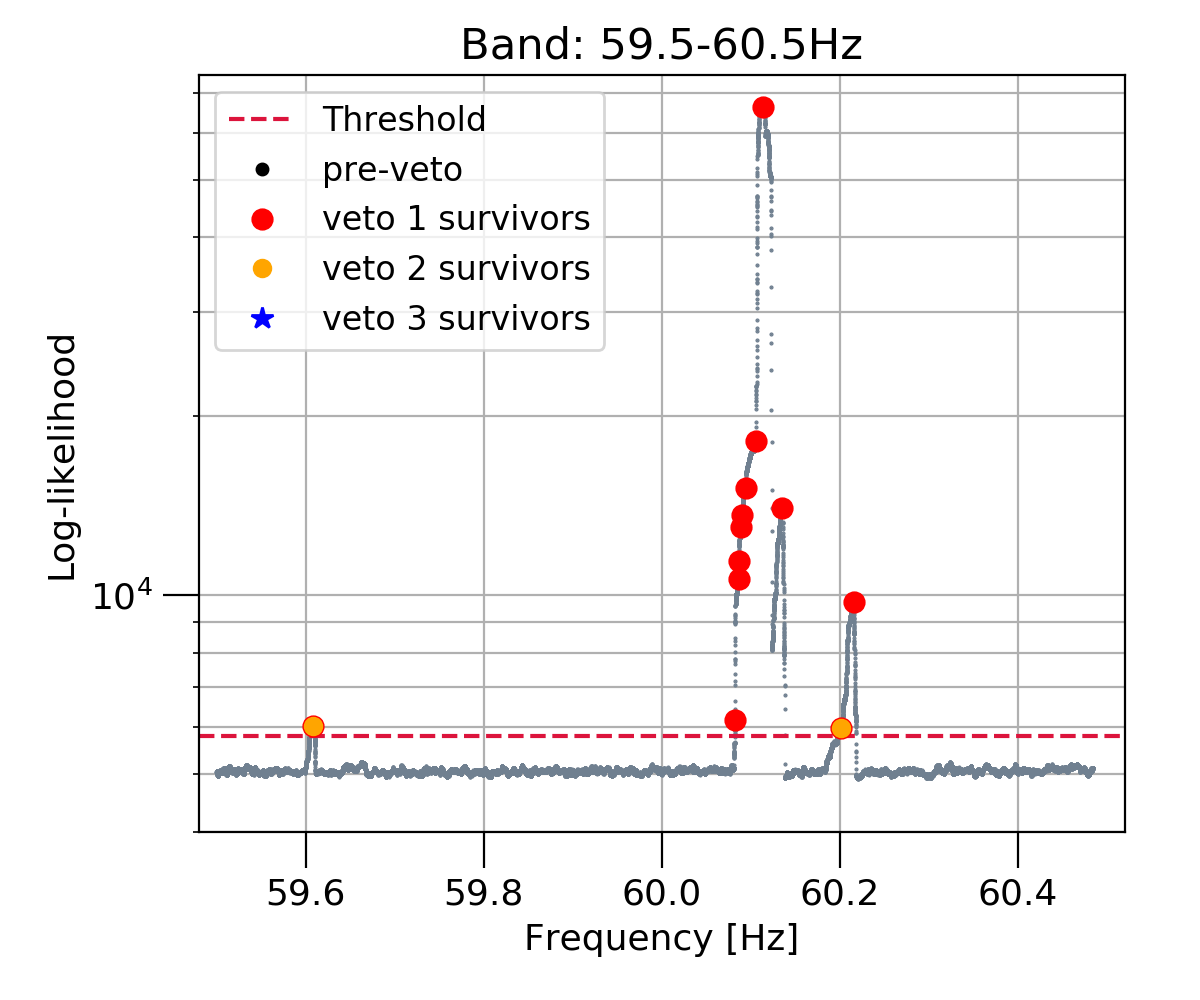}
\end{subfigure}
\caption{\small{Search results for three sub$-$bands of PSR J0534+2200. The log$-$likelihood ($\mathcal{L}$) is plotted against the terminating frequency of associated Viterbi paths (gray dots) at $f_*$ (top), $4f_*/3$ (middle) and $2f_*$ (bottom). The Gaussian threshold (red dashed line) is calculated using 100 noise realisations and indicated for each sub$-$band. Candidates before the vetoing procedures (black dots), after the known lines veto (red circles), single IFO veto (yellow circle) and the off$-$target veto (blue star) are overlaid in each case. The search returns 11, nine and 12 candidates in bands corresponding to $f_*$, $4f_*/3$ and $2f_*$. Only one candidate survives the three vetoes at $f_*$ while none survive at $4f_*/3$ or $2f_*$.}}
\label{Res:PSR_J0534_2200}
\end{figure}

\begin{figure}[h]
\begin{subfigure}{0.47\textwidth}
  \centering
  % include second image
  \includegraphics[width=0.9\linewidth]{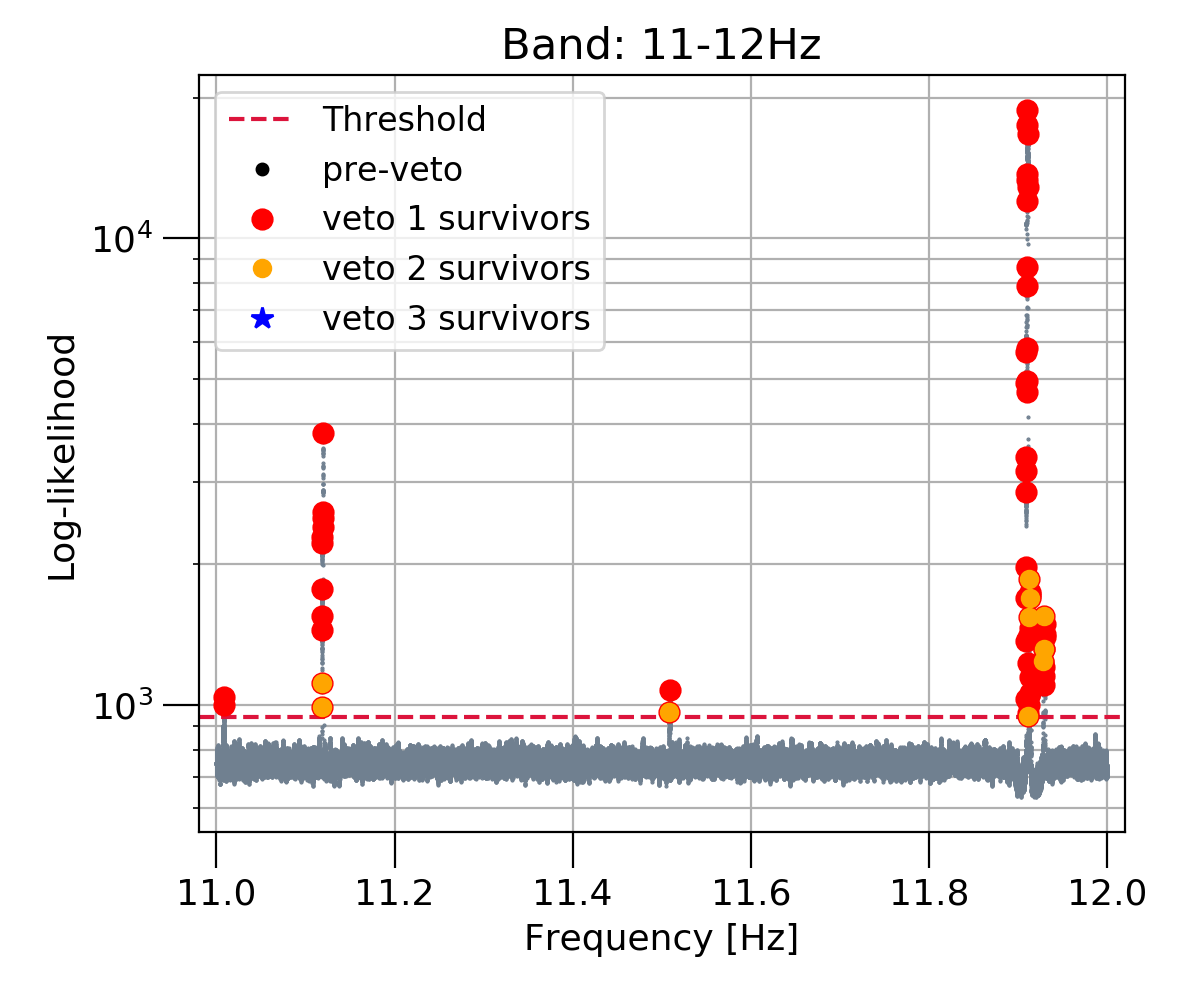}
\end{subfigure}
\begin{subfigure}{0.47\textwidth}
  \centering
  % include first image
  \includegraphics[width=0.9\linewidth]{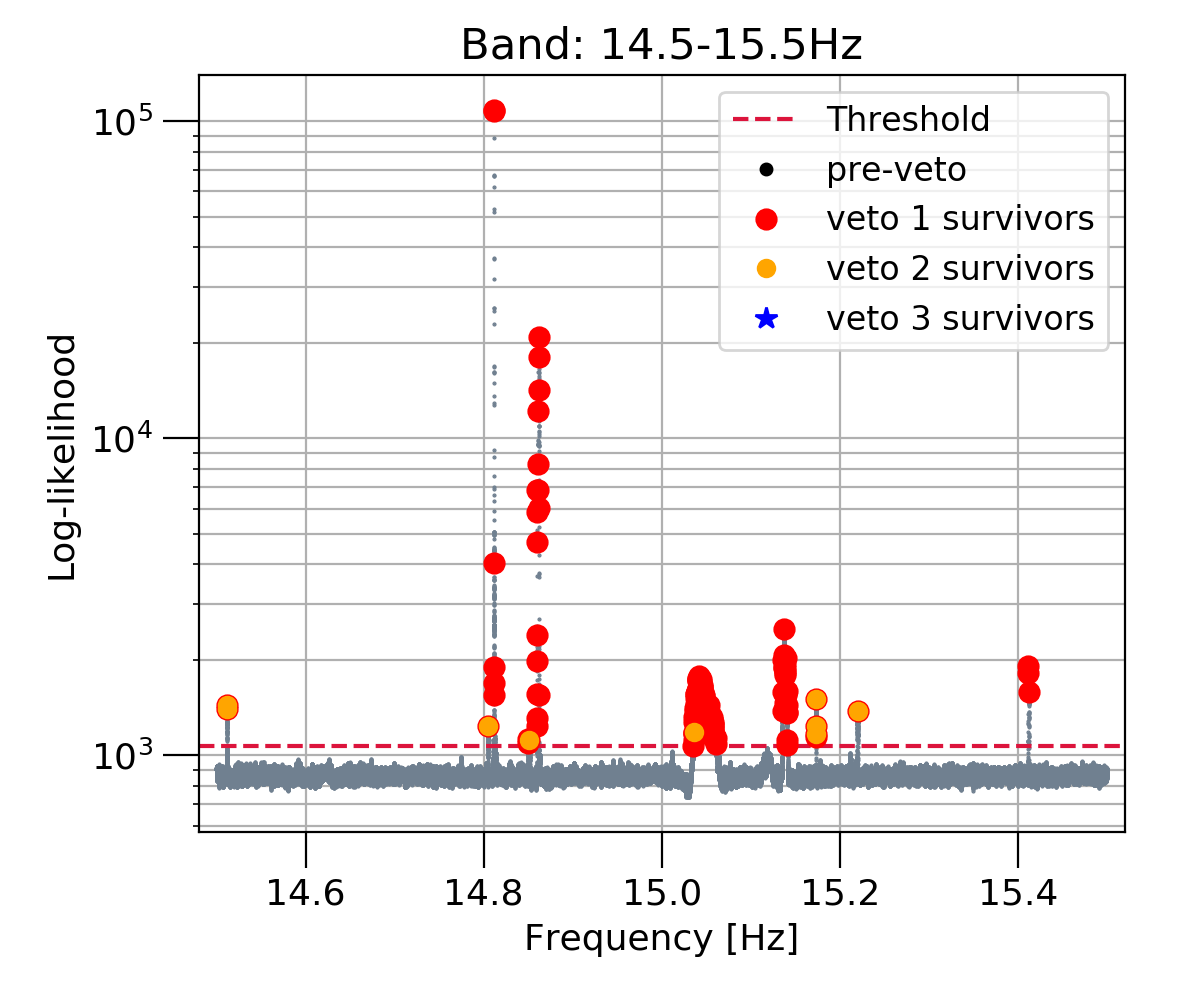}
\end{subfigure}
\begin{subfigure}{0.47\textwidth}
  \centering
  % include second image
  \includegraphics[width=0.9\linewidth]{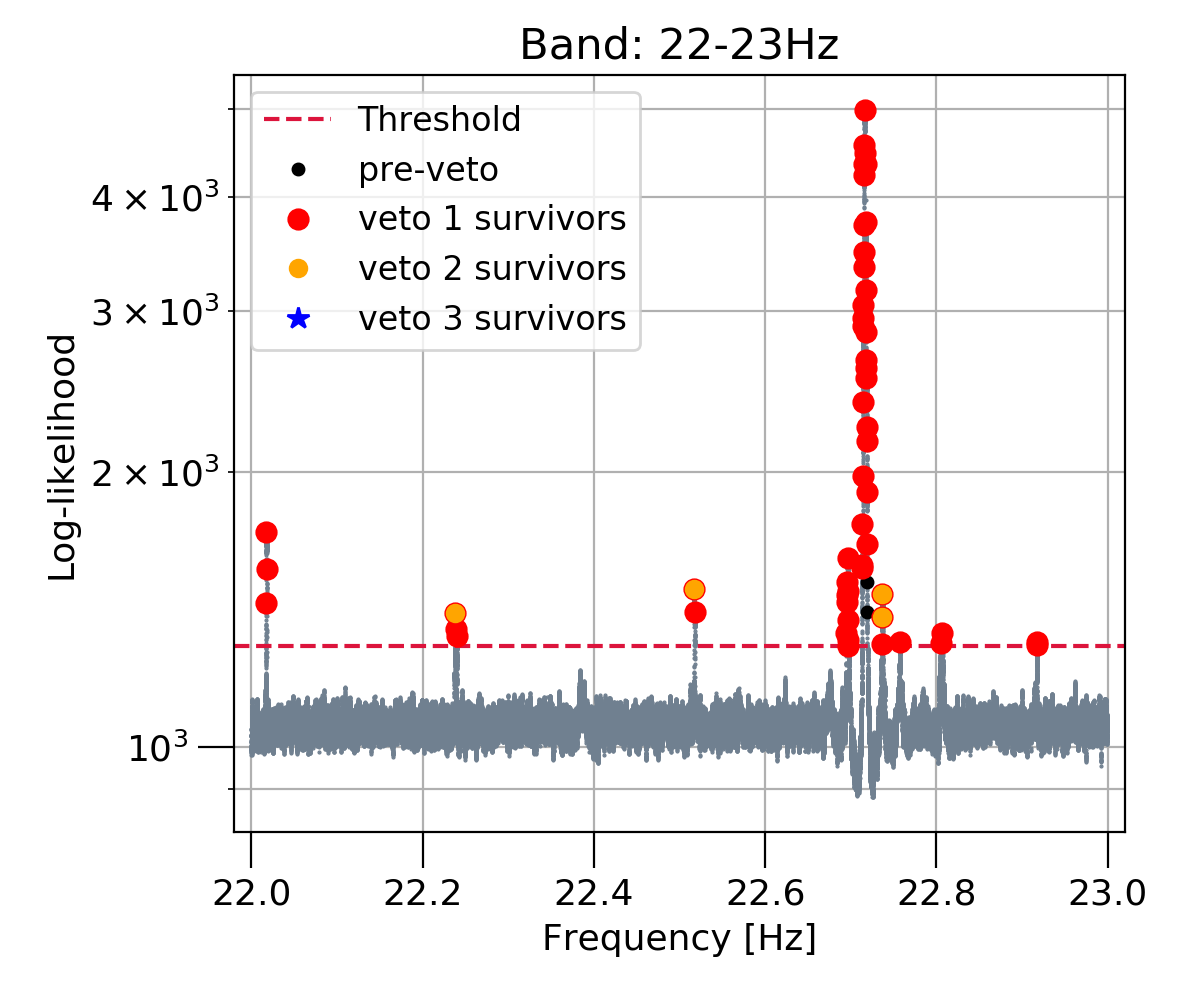}
\end{subfigure}
\caption{\small{Search results for PSR J0835$-$4510, laid out as in Fig.~\ref{Res:PSR_J0534_2200}. There are 64, 196 and 54 candidates in sub$-$bands corresponding to signal at $f_*$, $4f_*/3$ and $2f_*$. None of the candidates survive the three data quality vetoes.}}
\label{Res:PSR_J0835-4510}
\end{figure}

\begin{figure}[h]
\begin{subfigure}{0.47\textwidth}
  \centering
  % include second image
  \includegraphics[width=0.9\linewidth]{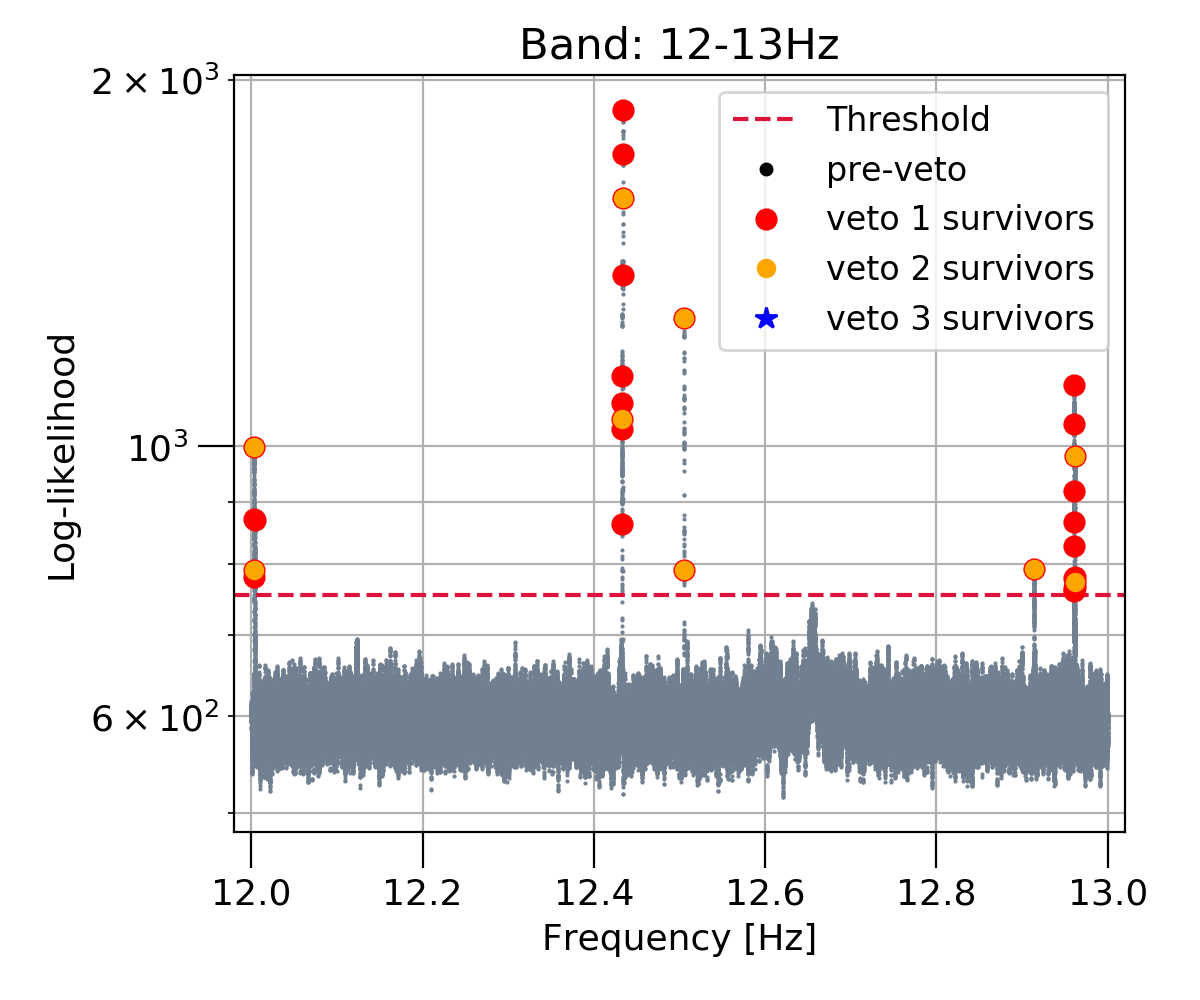}
\end{subfigure}
\begin{subfigure}{0.47\textwidth}
  \centering
  % include first image
  \includegraphics[width=0.9\linewidth]{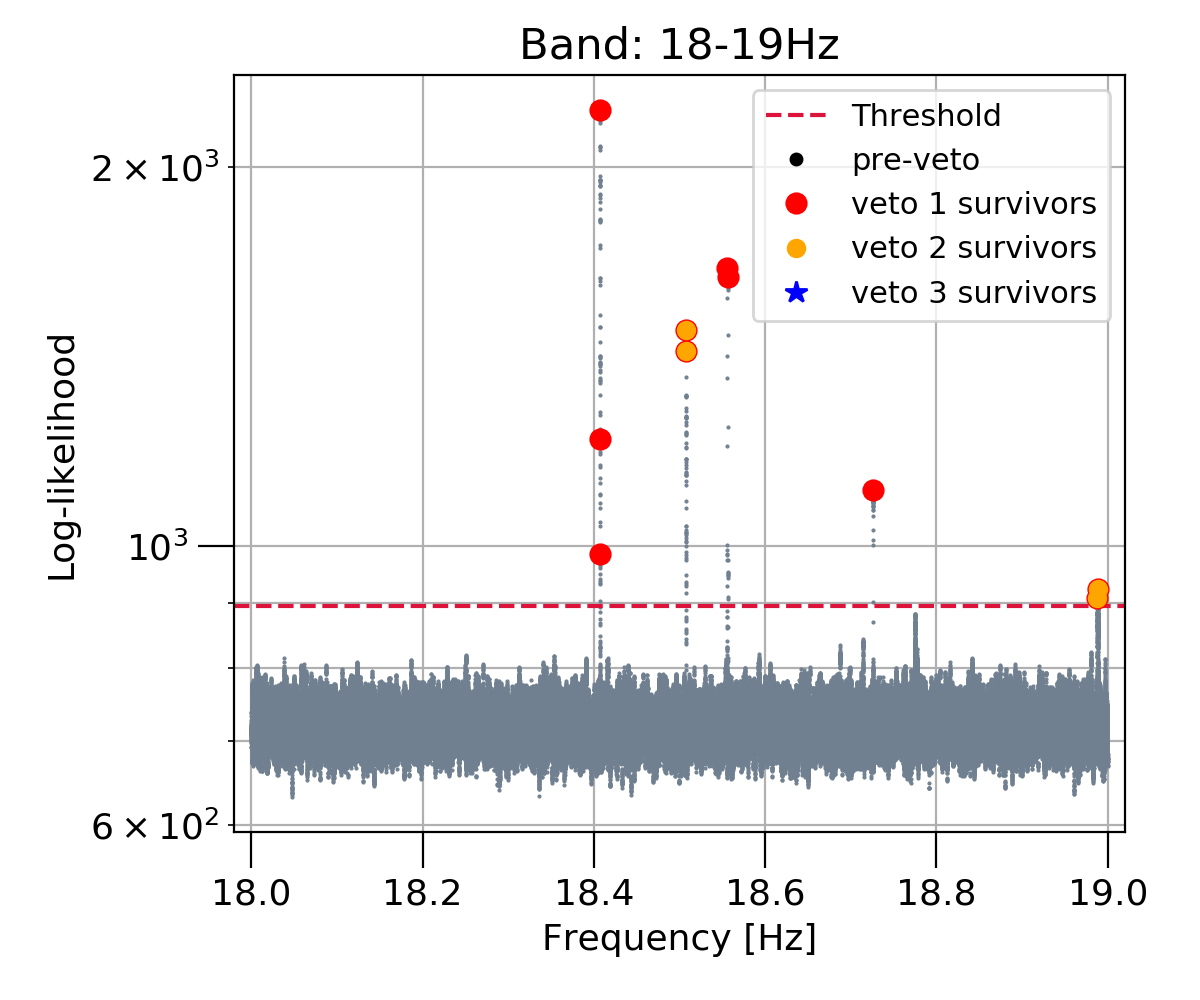}
\end{subfigure}
\caption{\small{Search results for $4f_*/3$ (top) and $2f_*$ (bottom) sub$-$band of PSR J1016$-$5857, laid out as in Fig.~\ref{Res:PSR_J0534_2200}. The search returns 29 candidates at $4f_*/3$ and 10 at $2f_*$. None of these candidate survives the three data quality vetoes.}}
\label{Res:PSR_J1016-5857}
\end{figure}

\begin{figure}[h]
\begin{subfigure}{0.47\textwidth}
  \centering
  % include second image
  \includegraphics[width=0.9\linewidth]{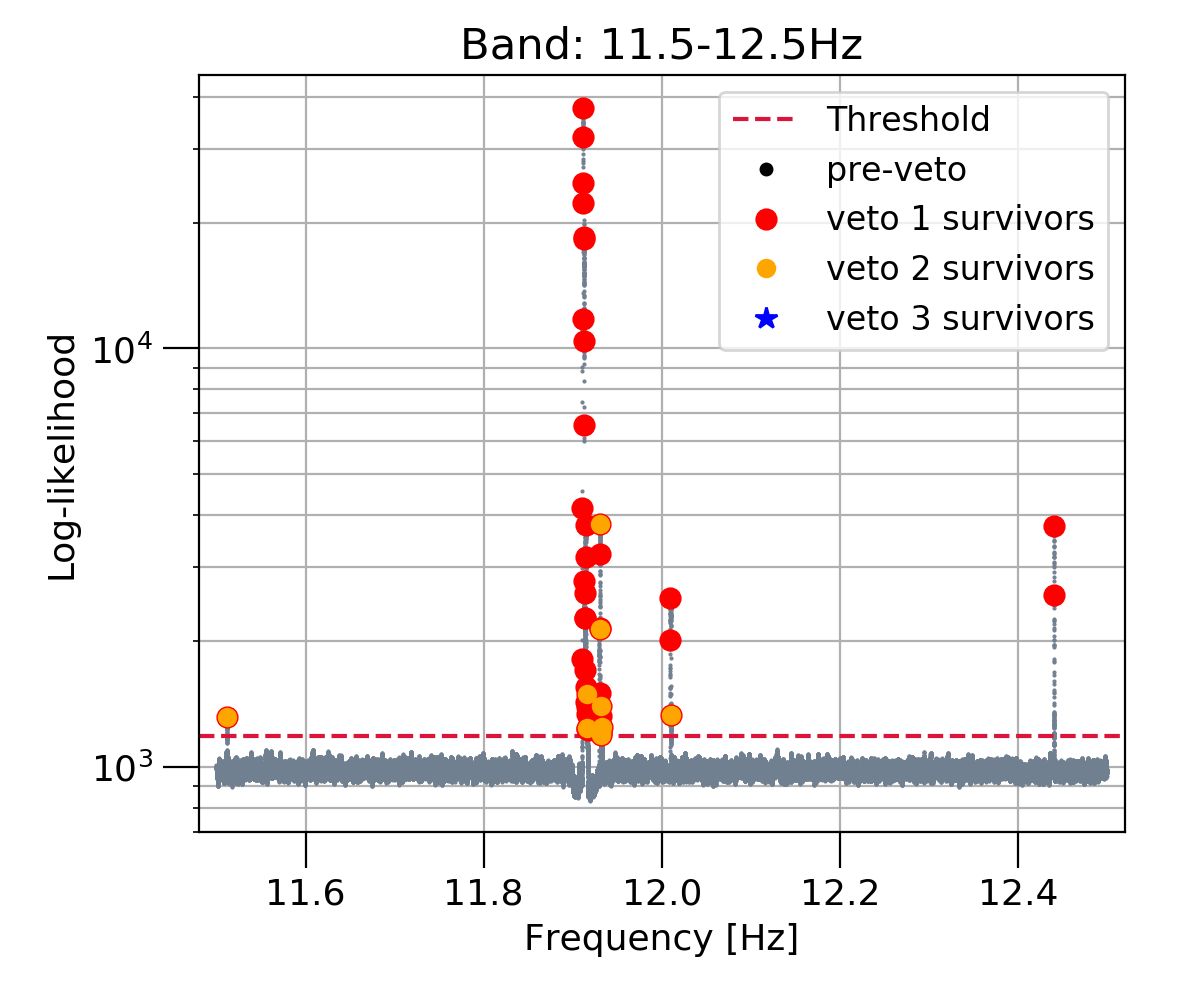}
\end{subfigure}
\caption{\small{Search results for $2f_*$ sub$-$band of PSR J1357$-$6429, laid out as in Fig.~\ref{Res:PSR_J0534_2200}. There are 44 candidates in this sub$-$band, none of which survives the three data quality vetoes.}}
\label{Res:PSR_J1357-6429}
\end{figure}

\begin{figure}[h]
\begin{subfigure}{0.47\textwidth}
  \centering
  % include second image
  \includegraphics[width=0.9\linewidth]{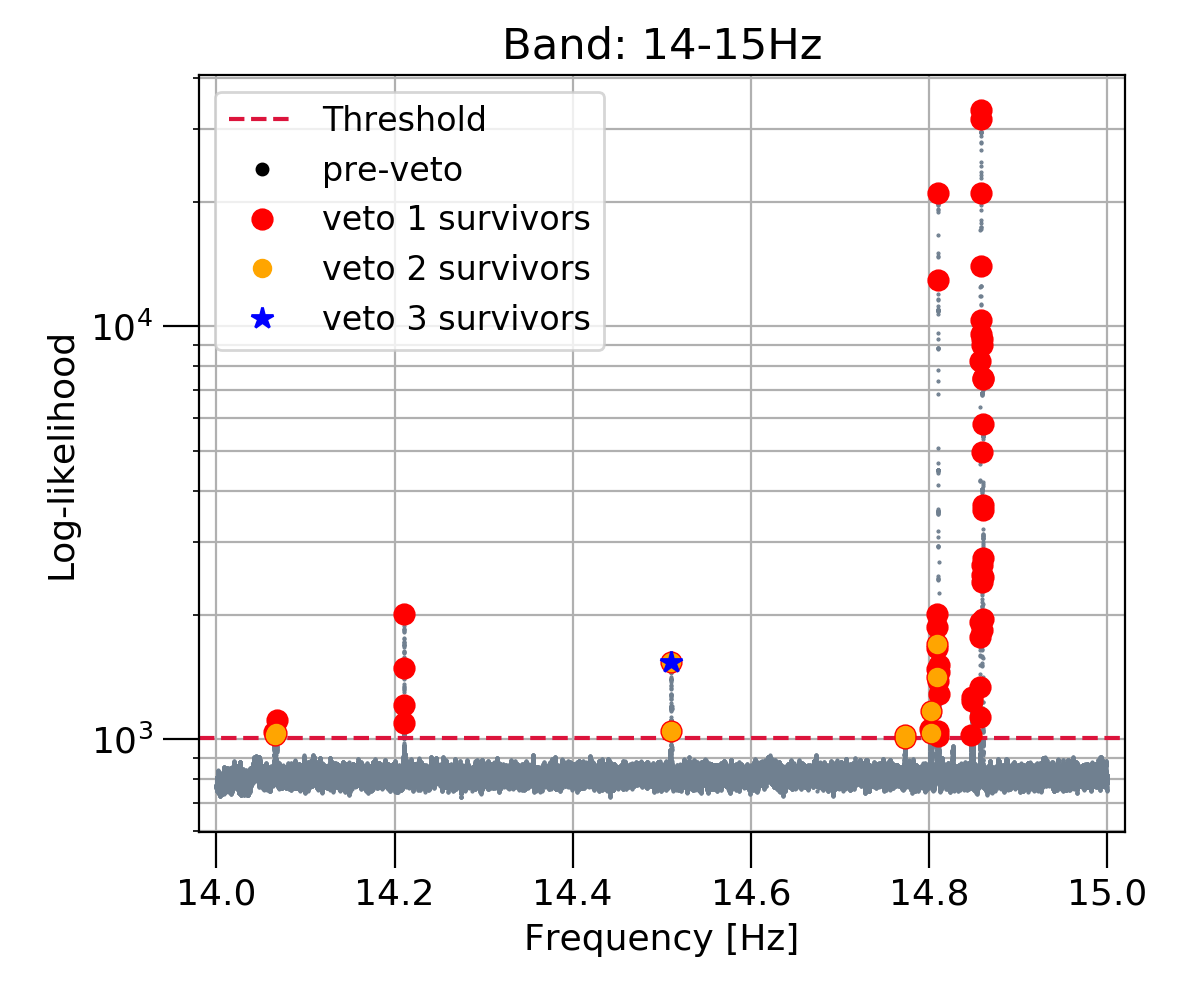}
\end{subfigure}
\begin{subfigure}{0.47\textwidth}
  \centering
  % include first image
  \includegraphics[width=0.9\linewidth]{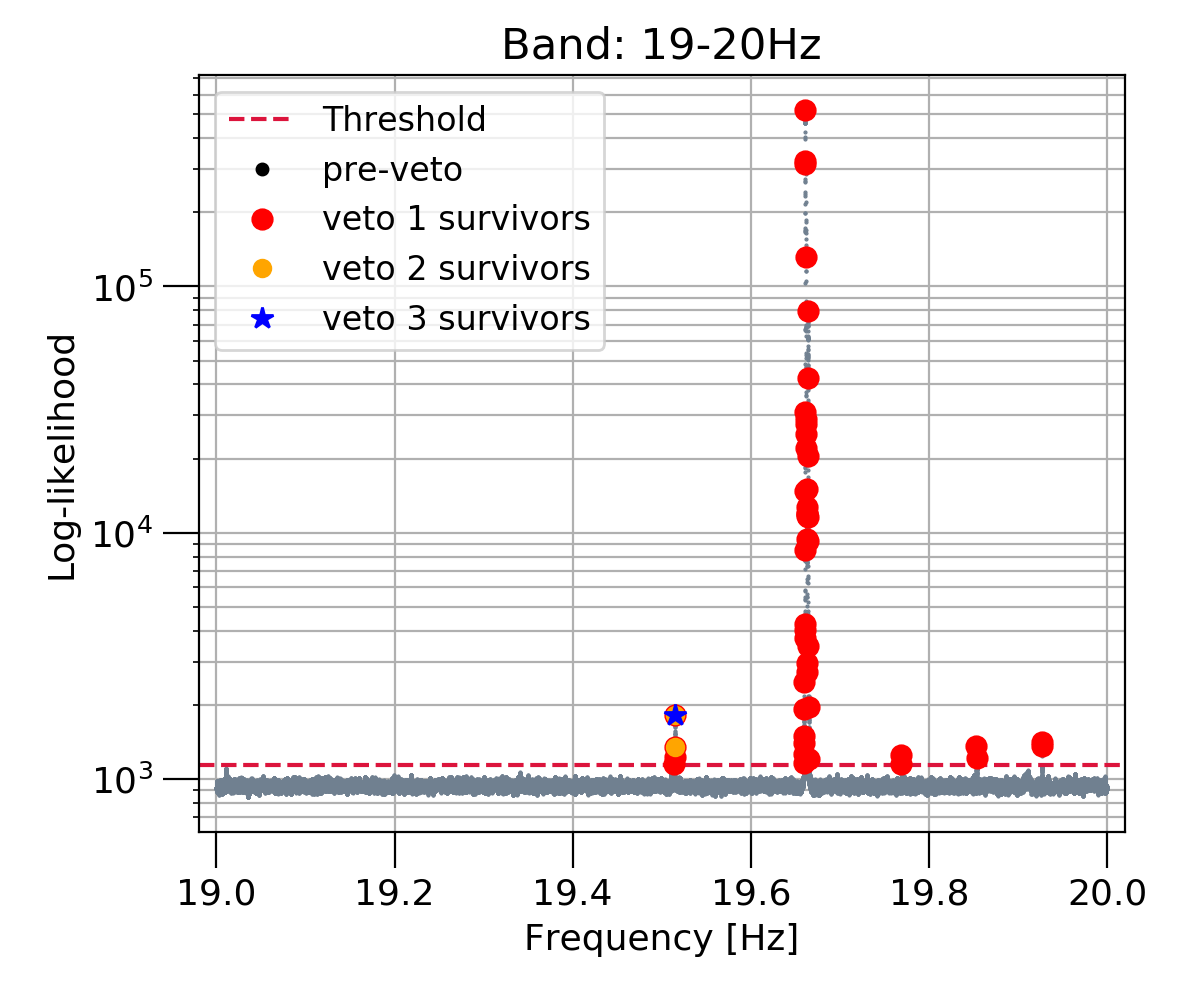}
\end{subfigure}
\begin{subfigure}{0.47\textwidth}
  \centering
  % include second image
  \includegraphics[width=0.9\linewidth]{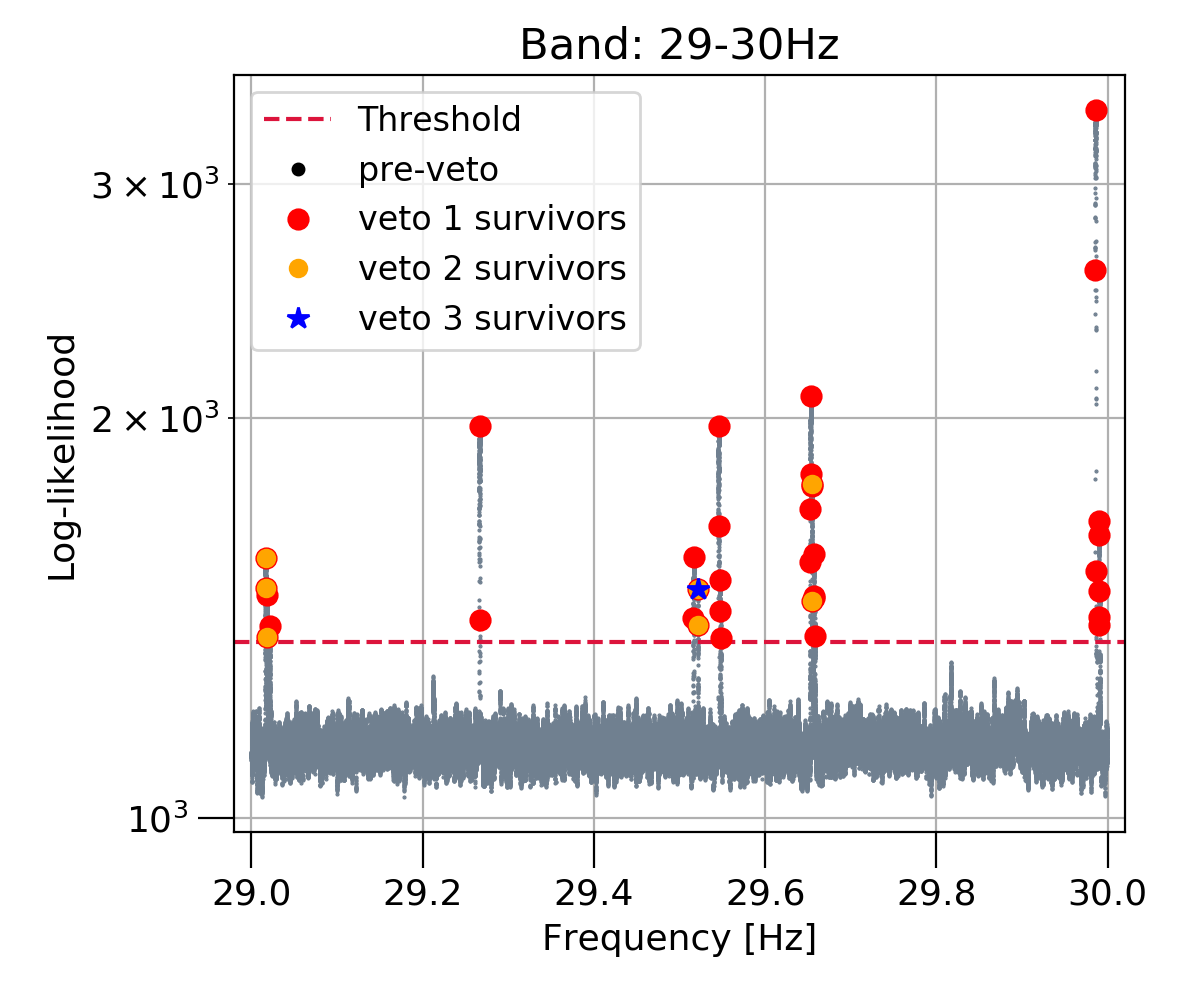}
\end{subfigure}
\caption{\small{Search results for three sub$-$bands of PSR J1420$-$6048, laid out as in Fig.~\ref{Res:PSR_J0534_2200}. The search reveals 58, 46 and 35 candidates in bands corresponding to $f_*$, $4f_*/3$ and $2f_*$. After the three data quality vetoes, only one candidate remains at $f_*$, $4f_*/3$ and $2f_*$.}}
\label{Res:PSR_J1420-6048}
\end{figure}

\begin{figure}[h]
\begin{subfigure}{0.47\textwidth}
  \centering
  % include second image
  \includegraphics[width=0.9\linewidth]{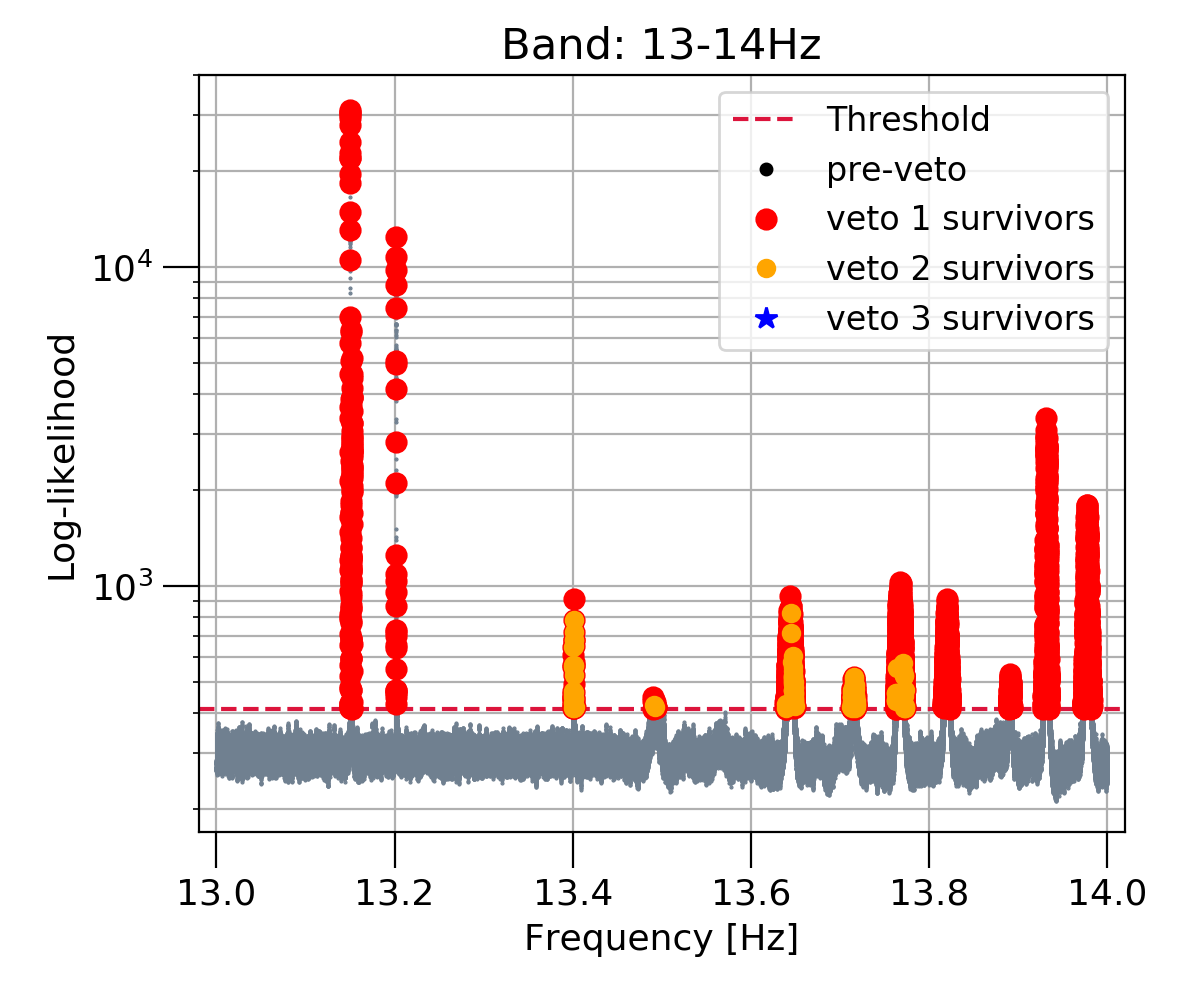}
\end{subfigure}
\begin{subfigure}{0.47\textwidth}
  \centering
  % include first image
  \includegraphics[width=0.9\linewidth]{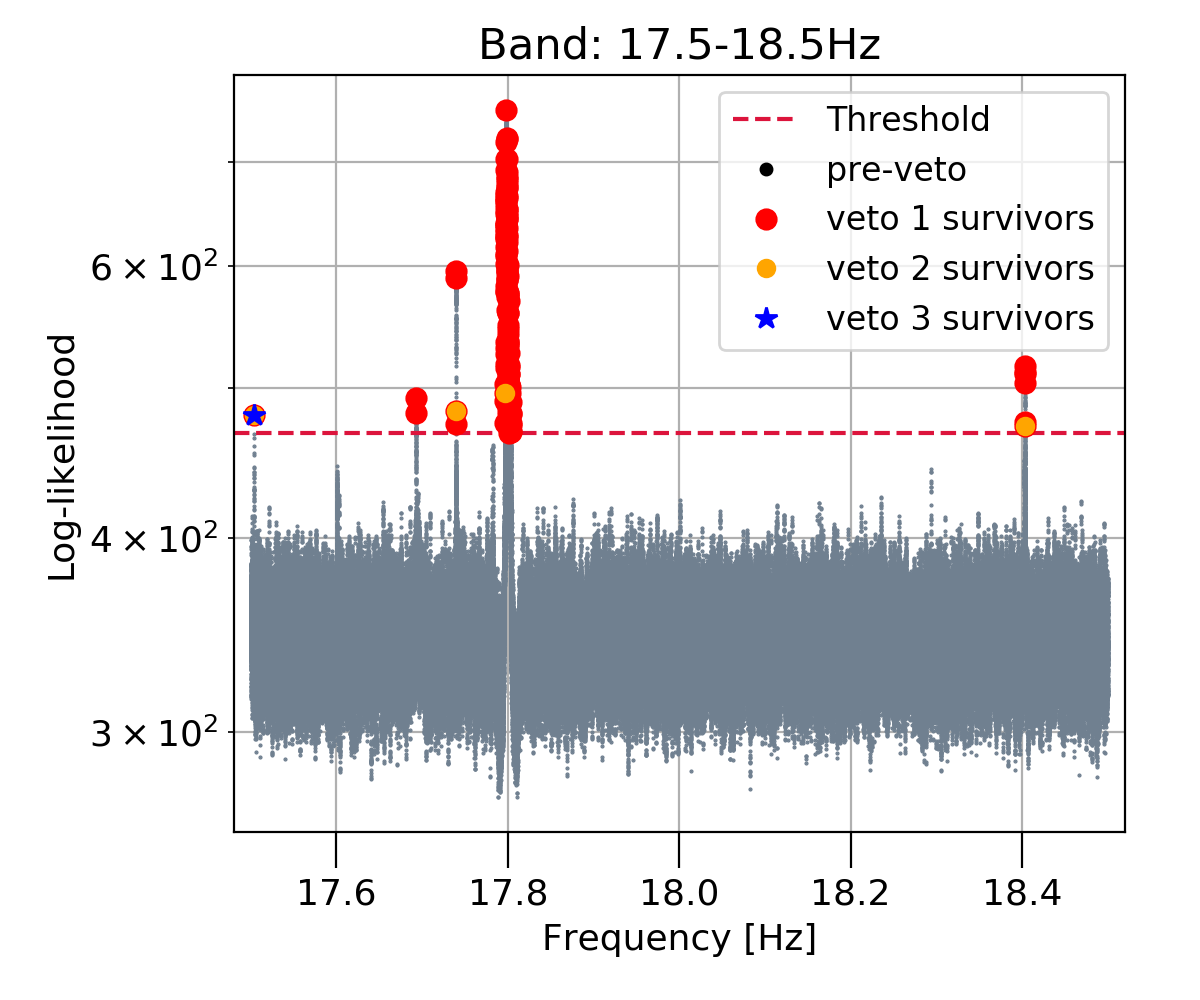}
\end{subfigure}
\begin{subfigure}{0.47\textwidth}
  \centering
  % include second image
  \includegraphics[width=0.9\linewidth]{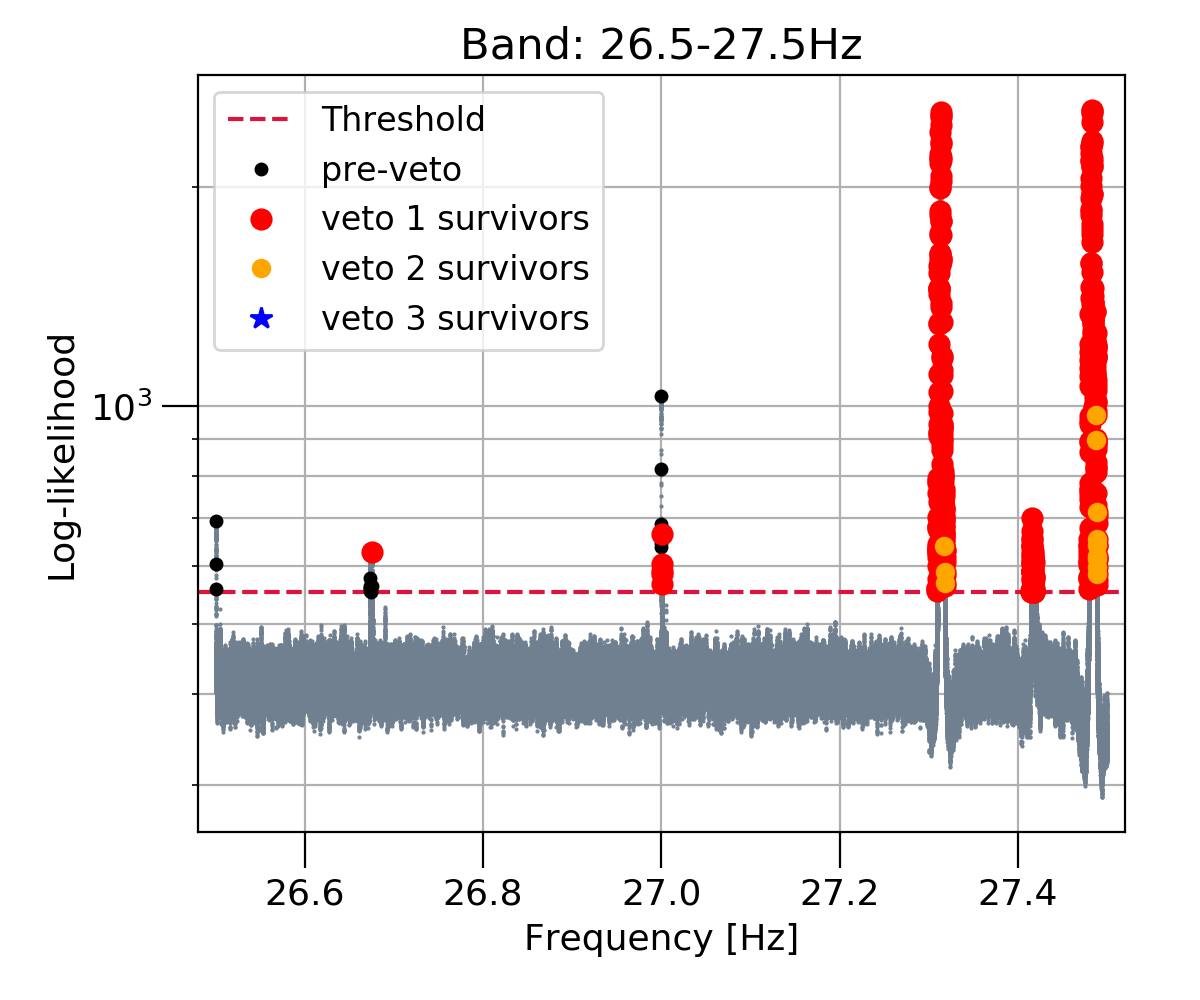}
\end{subfigure}
\caption{\small{Search results for the three sub$-$bands of PSR J1718$-$3825, laid out as in Fig.~\ref{Res:PSR_J0534_2200}. There are 2051, 165 and 417 candidates in sub$-$bands corresponding to a signal at $f_*$, $4f_*/3$ and $2f_*$. After the three data quality vetoes, only one candidate survives at $4f_*/3$ while none survive at $f_*$ or $2f_*$.}}
\label{Res:PSR_J1718-3825}
\end{figure}

\begin{figure}[h]
\begin{subfigure}{0.47\textwidth}
  \centering
  % include second image
  \includegraphics[width=0.9\linewidth]{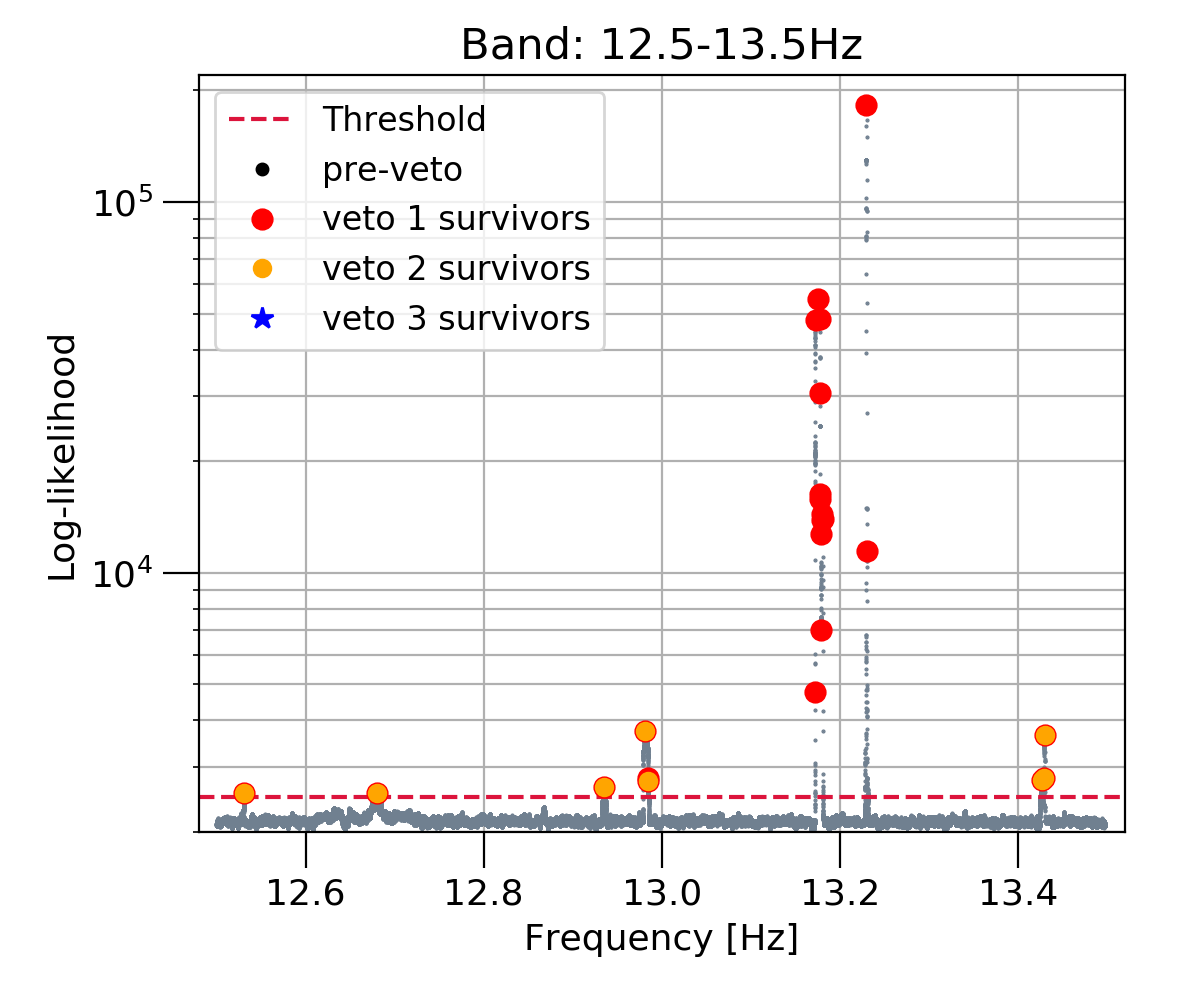}
\end{subfigure}
\caption{\small{Search results for the 2$f_*$ sub-band of PSR J1513$-$5908, laid out as in Fig.~\ref{Res:PSR_J0534_2200}. The search returns 23 candidates in this sub$-$band, none of which survive the three data quality vetoes.}}
\label{Res:PSR_J1513-5908}
\end{figure}

\begin{figure}[h]
\begin{subfigure}{0.47\textwidth}
  \centering
  % include second image
  \includegraphics[width=0.9\linewidth]{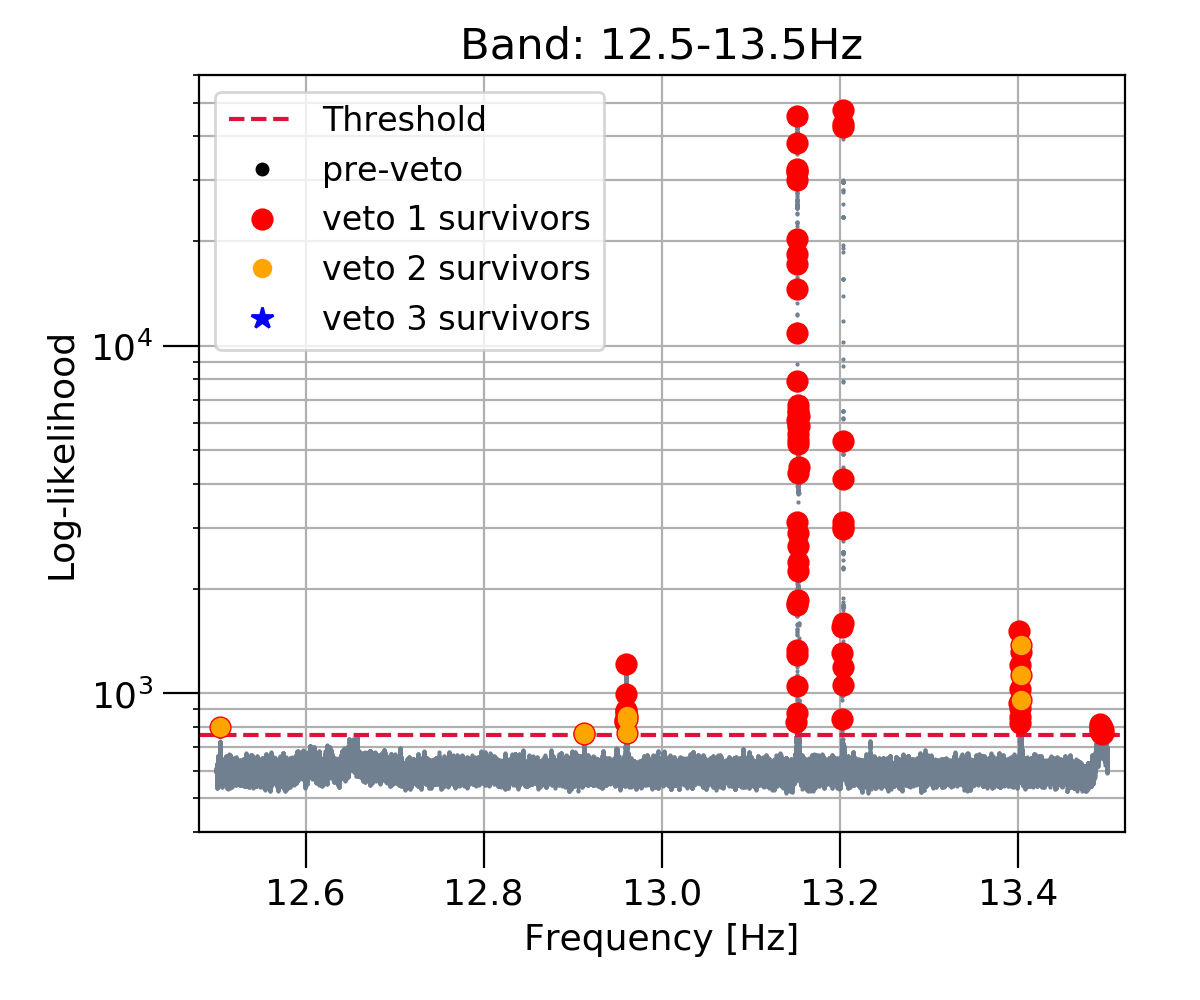}
\end{subfigure}
\begin{subfigure}{0.47\textwidth}
  \centering
  % include first image
  \includegraphics[width=0.9\linewidth]{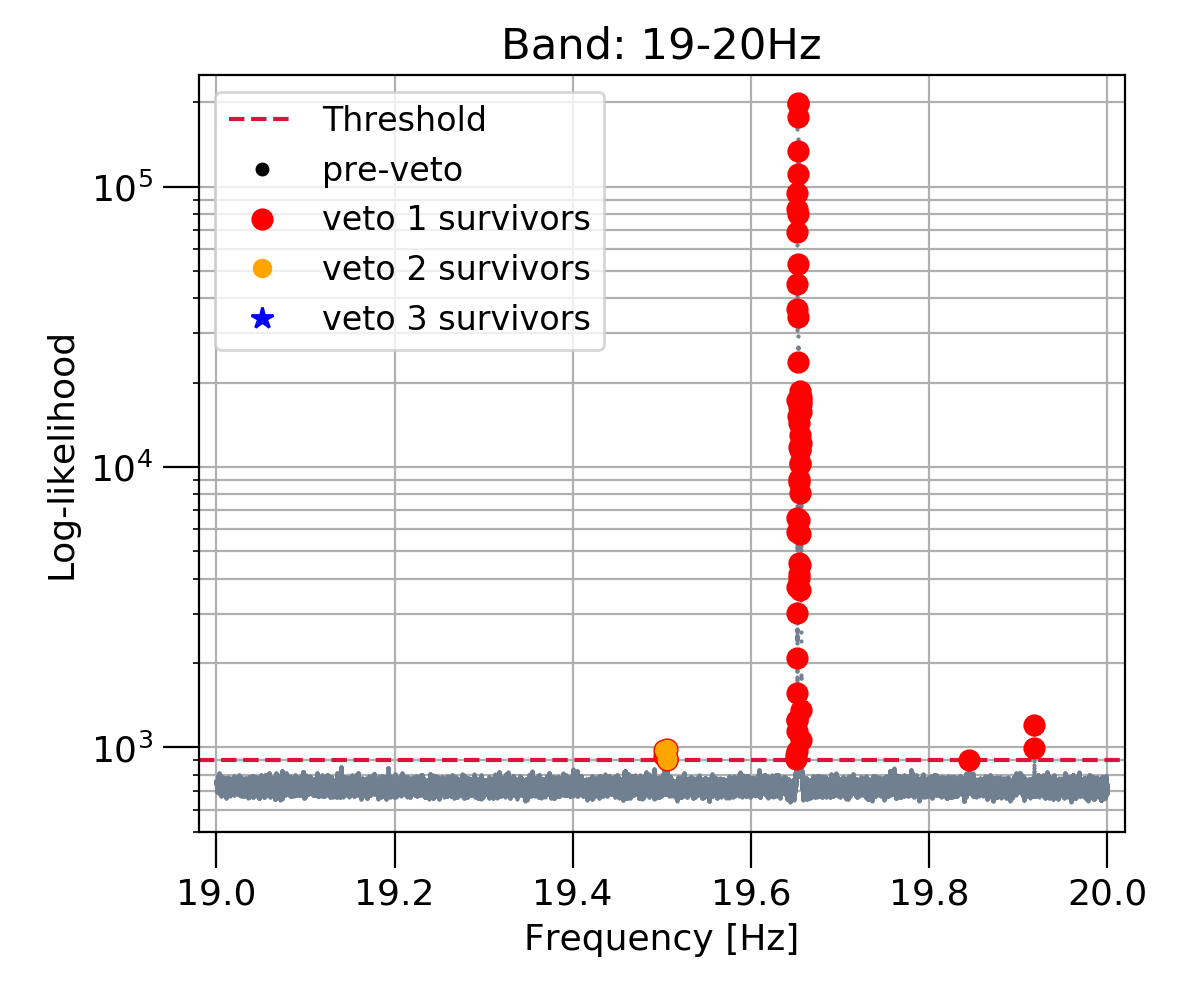}
\end{subfigure}
\caption{\small{Search results for two sub$-$bands of PSR J1826$-$1334, laid out as in Fig.~\ref{Res:PSR_J0534_2200}. There are 89 and 63 candidates in sub$-$bands corresponding to a signal at $4f_*/3$ and $2f_*$, respectively. None of the candidates survive the three data quality vetoes at $4f_*/3$ or $2f_*$.}}
\label{Res:PSR_J1826-1334}
\end{figure}

\begin{figure}[h!]
\begin{subfigure}{0.47\textwidth}
  \centering
  % include second image
  \includegraphics[width=0.9\linewidth]{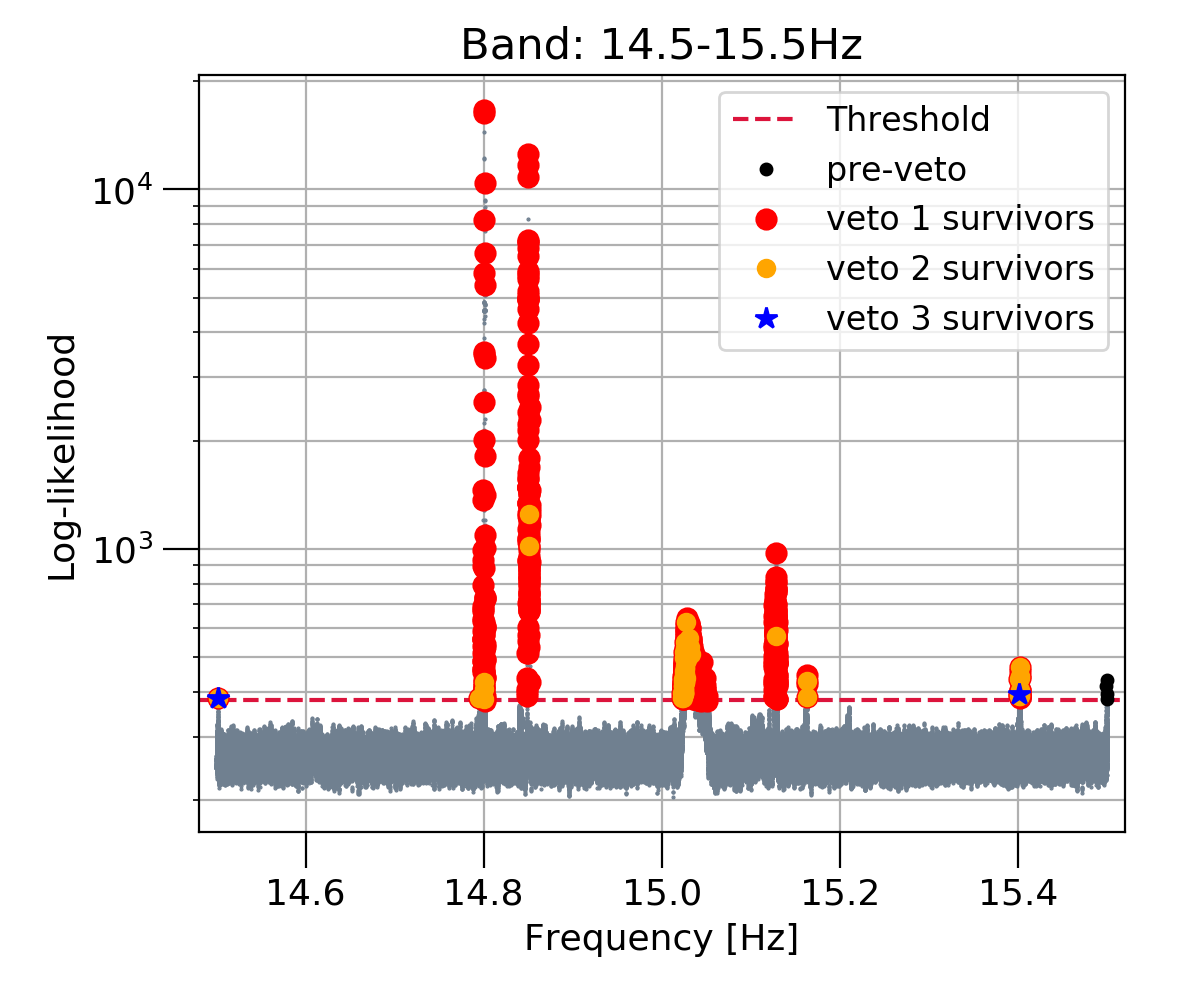}
\end{subfigure}
\begin{subfigure}{0.47\textwidth}
  \centering
  % include second image
  \includegraphics[width=0.9\linewidth]{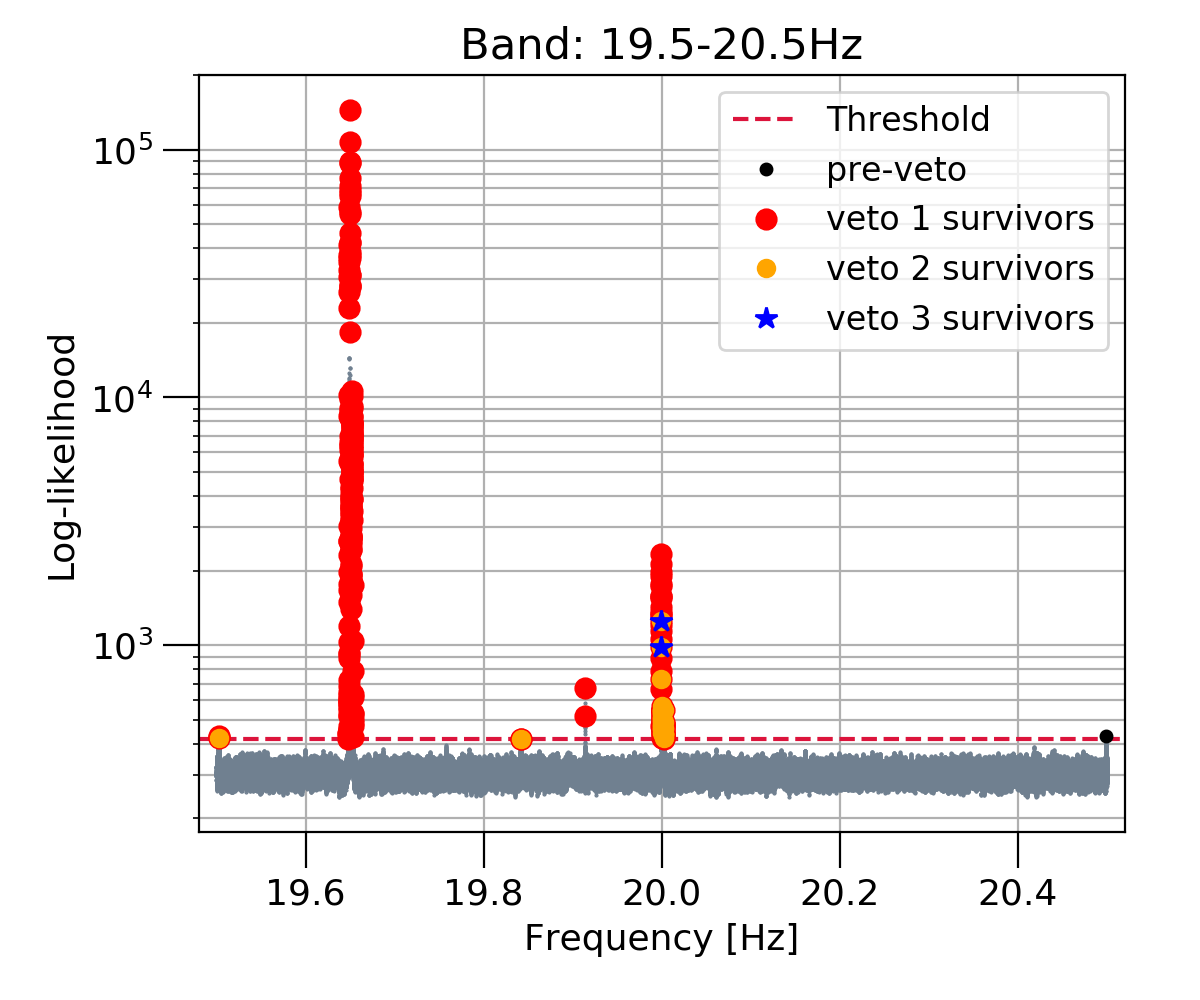}
\end{subfigure}
\begin{subfigure}{0.47\textwidth}
  \centering
  % include second image
  \includegraphics[width=0.9\linewidth]{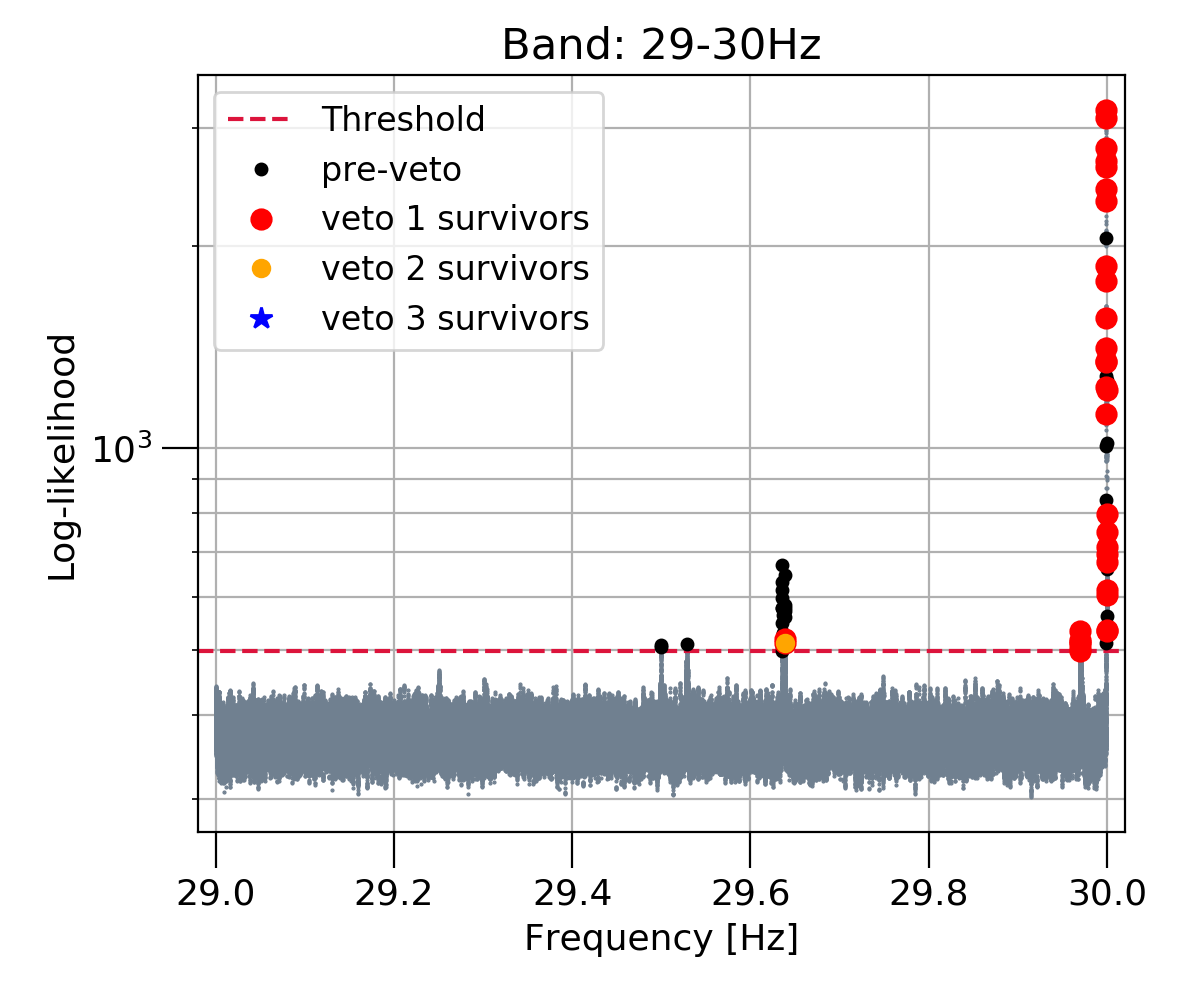}
\end{subfigure}
\caption{\small{Search results for three sub$-$bands of PSR J1831$-$0952, laid out as in Fig.~\ref{Res:PSR_J0534_2200}. There are 1413, 263 and 81 candidates in sub$-$bands corresponding to a signal at $f_*$, $4f_*/3$ and $2f_*$. Only two candidates survive the data quality vetoes at $f_*$ and $4f_*/3$ while none at $2f_*$.}}
\label{Res:PSR_J1831-0952}
\end{figure}

\begin{figure}[h!]
\begin{subfigure}{0.47\textwidth}
  \centering
  % include second image
  \includegraphics[width=0.9\linewidth]{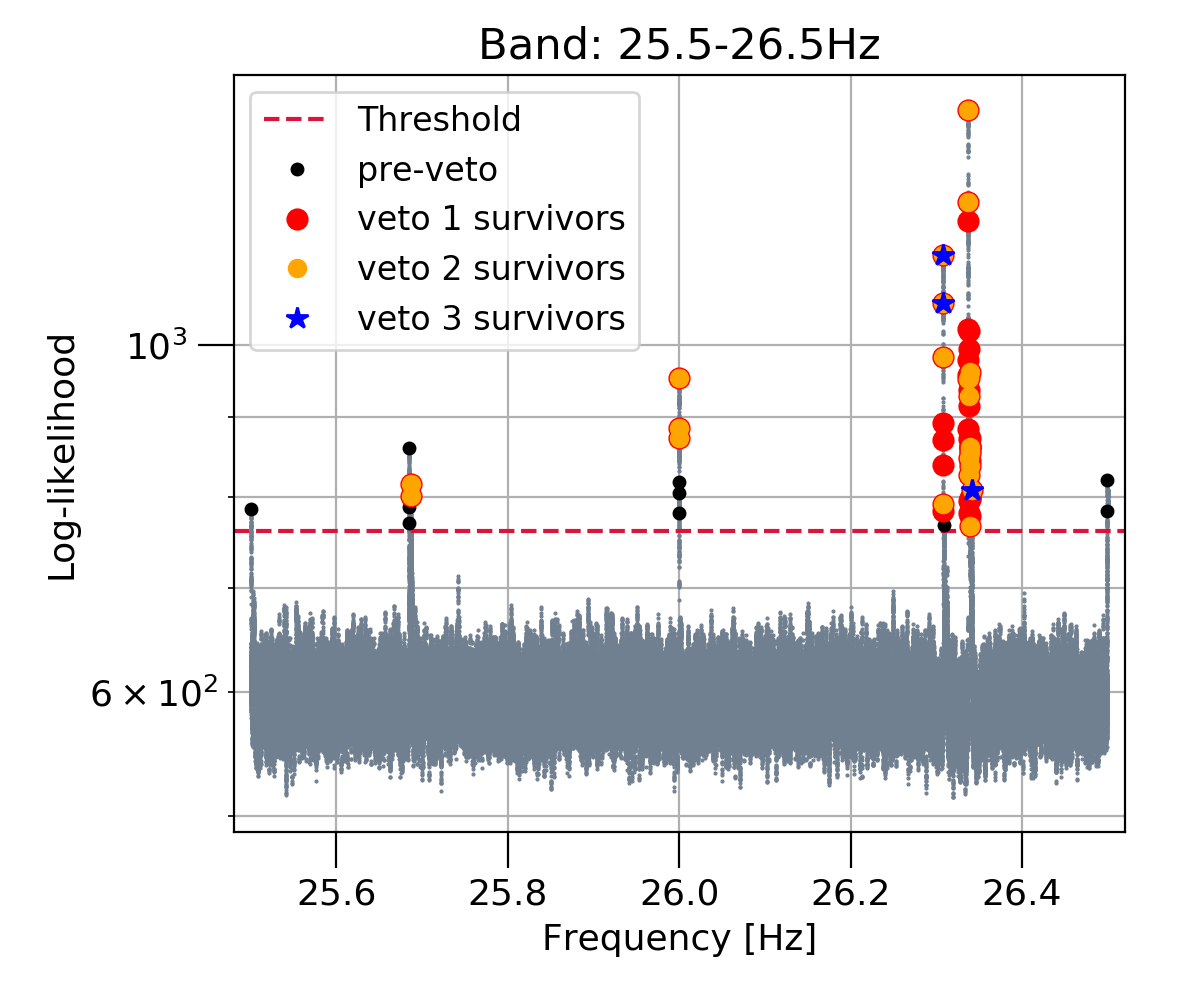}
\end{subfigure}
\begin{subfigure}{0.47\textwidth}
  \centering
  % include first image
  \includegraphics[width=0.9\linewidth]{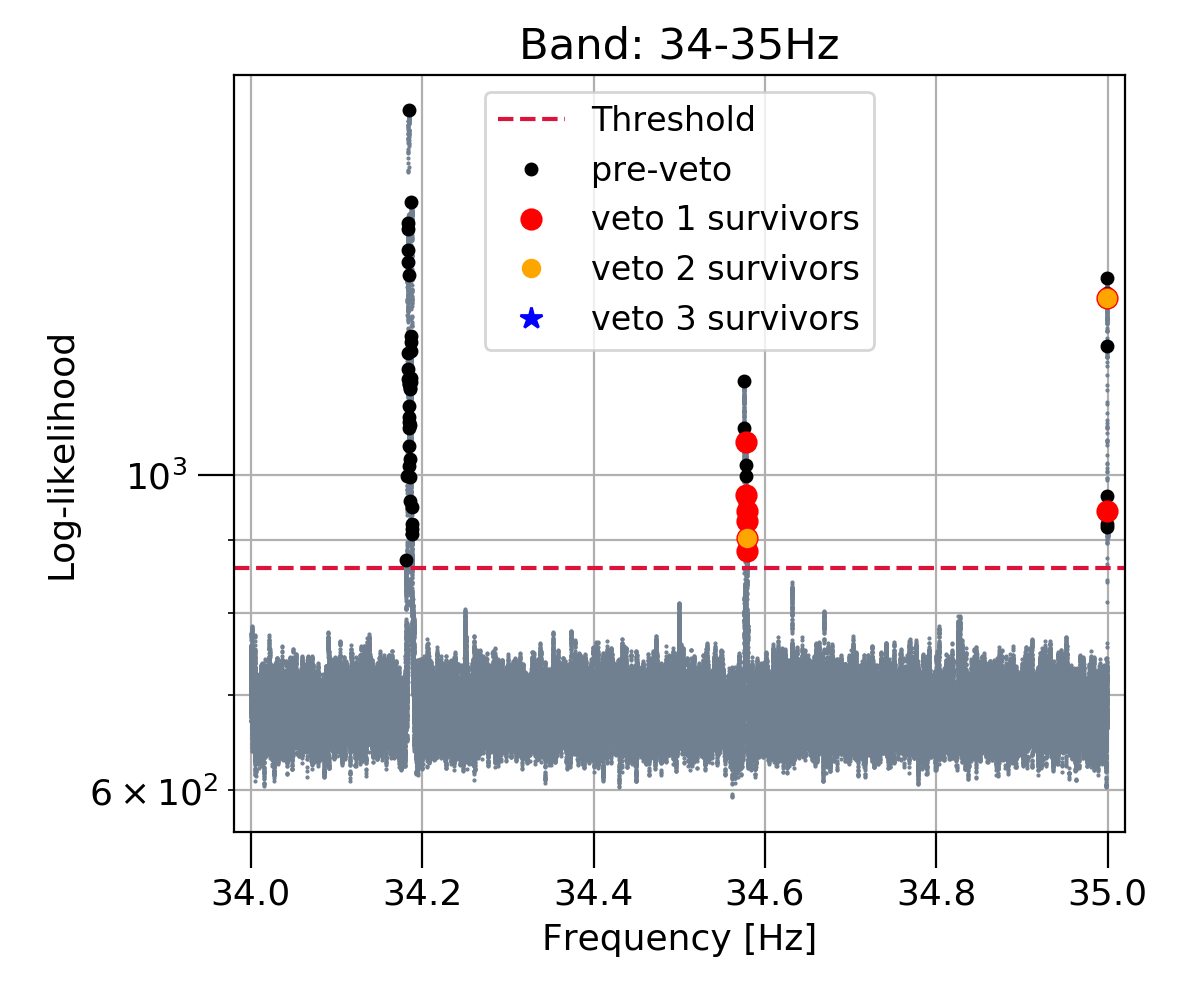}
\end{subfigure}
\begin{subfigure}{0.47\textwidth}
  \centering
  % include first image
  \includegraphics[width=0.9\linewidth]{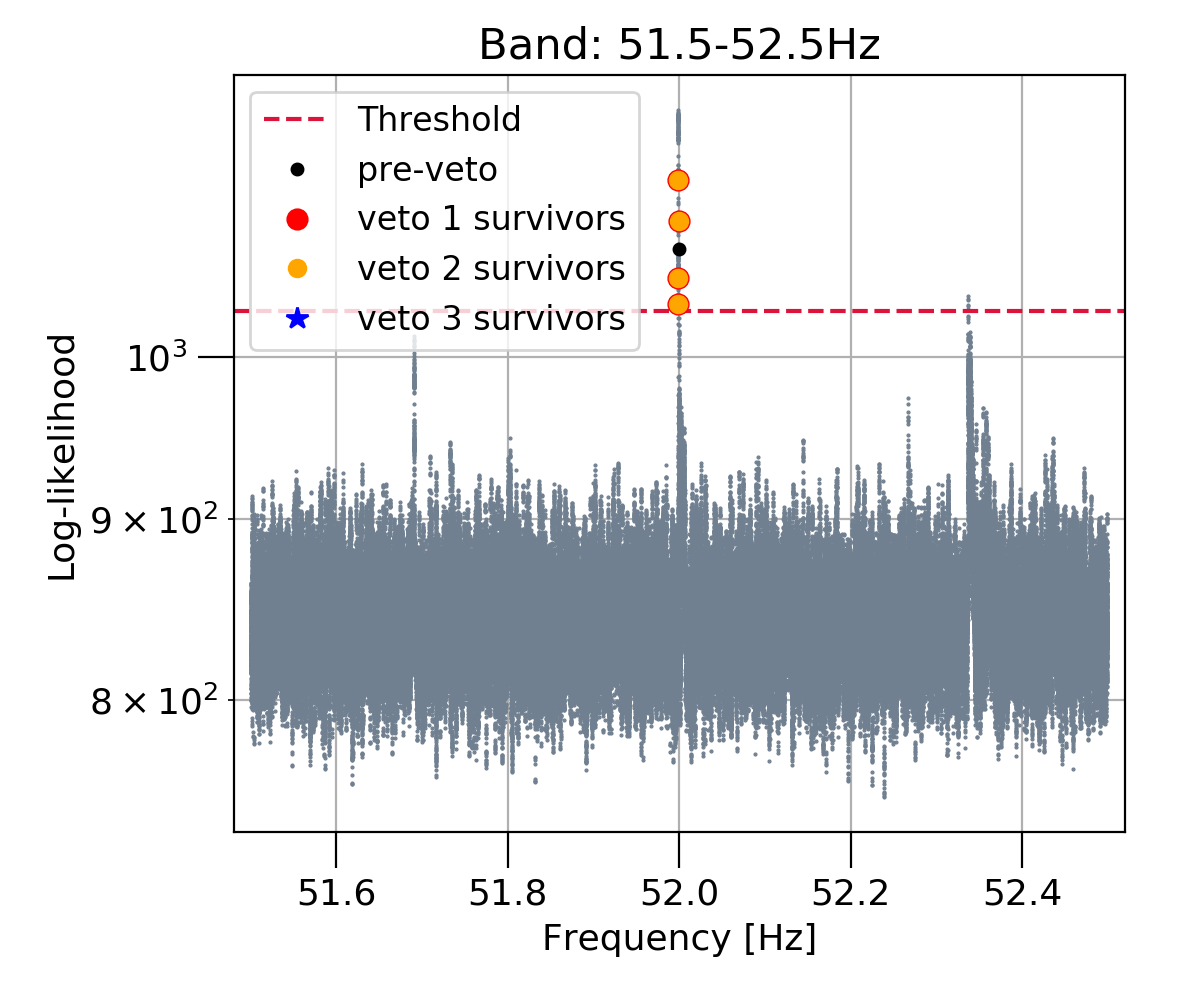}
 \end{subfigure}
\caption{\small{Search results for PSR J1849$-$0001 laid out as in Fig.~\ref{Res:PSR_J0534_2200}. There are 57, 61 and five candidates in sub$-$bands corresponding to signal at $f_*$, $4f_*/3$ and $2f_*$. Only three candidates survive the data quality vetoes at $f_*$ while none survive at $4f_*/3$ or $2f_*$.}}
\label{Res:PSR_J1849-0001}
\end{figure}

Figs.~\ref{Res:PSR_J0534_2200}-\ref{Res:PSR_J1849-0001} show the results for 24 sub$-$bands explored in this study. For each sub$-$band, we plot the log$-$likelihood ($\mathcal{L}$) against the terminating frequency of the associated Viterbi path. Candidates are overlaid along with the outcome at each stage of the data quality vetoes. We discuss the characteristics of veto 3 survivors (indicated by blue stars) in Section \ref{subsec:survivors}.
% ---------------------------------------------------------------
% -----------------------------------------------------------------------------
% -------------------------------------------------------------------------
% -----------------------------------------------------------------------------
% ----------------------------------------------------------------
% ----------------------------------------------------------------
% ----------------------------------------------------------------
% ----------------------------------------------------------------
% ----------------------------------------------------------------
% ----------------------------------------------------------------
\clearpage
% The \nocite command causes all entries in a bibliography to be printed out
% whether or not they are actually referenced in the text. This is appropriate
% for the sample file to show the different styles of references, but authors
% most likely will not want to use it.
%\nocite{*}
\newpage
\nocite{*}
\bibliography{apssamp}% Produces the bibliography via BibTeX.

\end{document}